\documentstyle[12pt,epsfig]{article}
\newcounter{num}

\newcommand{\lsim}{\mathrel{\lower4pt\hbox{$\sim$}}
\hskip-12.5pt\raise1.6pt\hbox{$<$}\;}

\newcommand{\gsim}{\mathrel{\lower4pt\hbox{$\sim$}}
\hskip-12.5pt\raise1.6pt\hbox{$>$}\;}

\begin{document}
\baselineskip 18pt plus 2pt

\noindent \hspace*{10cm}UCRHEP-T201 (August 1997) 

\begin{center}
{\bf CP Violation In Single Top Production And Decay Via $p \bar p \to t
\bar b +X \to W^+ b \bar b +X$ Within The MSSM: A Possible Application
For Measuring $\arg (A_t)$ At Hadron Colliders} 
\vspace{.2in}

%ASf...author format completely chnaged
S. Bar-Shalom \\
\vspace{-5pt}
Department of Physics, University of California at Riverside, Riverside
CA 92521\\ 
\bigskip

D. Atwood \\
\vspace{-5pt}
Theory Group, Thomas Jefferson National Accelerator Facility, Newport
News, VA\ \ 23606\\ 
\vspace{-5pt}
and\\
\vspace{-5pt}
Department of Physics and Astronomy, Iowa State University, Ames, IA\ \
50011\footnote{Address after August 1, 1997.} \\
\bigskip

A. Soni\\
\vspace{-5pt}
Physics Department, Brookhaven Nat.\ Lab., Upton NY 11973, USA.
\vspace{.2in}

%ASf...many changes in the abstract
{\bf Abstract}\\
\end{center}

\noindent CP-nonconserving effects in the reaction $p \bar p \to t \bar b
+X \to W^+ 
b \bar b +X$, driven by the supersymmetric CP-odd phase of the stop trilinear
soft breaking term, $\arg(A_t)$, are studied. We discuss the CP-nonconserving
effects in both production and the associated decay amplitudes of the top.
We find that, within a plausible low energy scenario of the MSSM and keeping
the neutron electric dipole moment below its current limit, a CP-violating
%ASf
cross-section asymmetry as large as  $2-3\%$ can arise 
if some of the parameters lie in a favorable range.
%ASf
A partial rate asymmetry originating only in the top decay $t \to W^+ b$
is  found to be, in general, below the $0.1\%$ level which is somewhat smaller
than previous claims.  For a low $\tan\beta$ of order one 
%ASf
the decay asymmetry
%ASf
can reach at the most $\sim 0.3\%$. This (few) percent level overall
CP-violating 
signal in $p \bar p \to t \bar b +X \to W^+ b \bar b +X$ might be within
the reach of the future 2(4) TeV $p \bar p$ Tevatron collider that may be
able to produce $\sim 10000$($\sim 30000$) such $t\bar b$ events with an
integrated luminosity of 30 fb$^{-1}$. In particular, it may be used to place
an upper bound on $\arg(A_t)$ if indeed ${\rm arg}(\mu) \to 0$, as implied
from the present experimental limit on the neutron electric dipole moment.
%ASf
The partial rate asymmetry in the top decay ( $\sim {\rm few} \times 10^{-3}$ ) 
may also be within the reach of the LHC with $\sim 10^7$ pairs of $t \bar
t$ produced, provided detector systematics are sufficiently small. 
We also show that if the GUT-scale universality of the soft breaking trilinear
$A$ terms is relaxed, then the phases associated with 
$\arg A_u$ and $\arg A_d$ can take values up to $\sim {\rm few} \times
10^{-1}$ even with squarks and gluino masses of several hundred GeV's 
without contradicting
the experimental limit on the neutron electric dipole moment.
\newpage 

\noindent {\bf 1. \underline{Introduction}}\\
 
In the Tevatron $p \bar p$ collider, top quarks will be mainly produced as
pairs of $t \bar t$ via an $s$-channel gluon exchange. Nonetheless, the
subleading 
ElectroWeak (EW) production mechanism of a single top, forms a significant
fraction of the $t \bar t$ pair production. It will therefore be closely
scrutinized in the next runs of the Tevatron \cite{heinson1}. In particular,
the production rate of $t \bar b$ (and the charged conjugate pair) through
an $s$-channel $W$-boson, $p\bar{p} \to W \to t\bar{b}+X$, (the corresponding
partonic reaction is $u \bar d \to W \to t \bar b$) is about 10\% of the
$t \bar t$ production rate \cite{heinson1}. Throughout this paper we will
always refer only to the $s$-channel $W$ exchange when discussing the
reaction  $p\bar{p} \to t\bar{b}+X$. 
       
In a previous work, \cite{tbsusy}, we have shown that a new CP-violating
phase from a Two Higgs Doublet Model (2HDM) of type II can give rise to
CP-violating asymmetries of the order of a few percent in the reaction $p\bar{p}\to
t\bar{b}+X$. In that work we had considered one example of a supersymmetric
(SUSY) mediated, 
one-loop, CP-violating diagram (gluino exchange diagram). The  one SUSY diagram
that was considered in \cite{tbsusy} was found to be p-wave suppressed near
threshold, thus yielding CP asymmetries of the order of 0.1\%. Clearly
a complete study of the possible SUSY-CP effect is required in order to be
able to estimate the true expected magnitude of a potential CP-nonconserving
effect that is driven by new CP-odd phases of low energy SUSY dynamics.

Therefore, in this paper we wish to extend our previous study and explore
the complete CP-violating effect in the process $p\bar{p}\to t\bar{b}+X \to
W^+ b \bar b +X$, induced specifically by the complex soft trilinear parameter
associated with the superpartner of the top, $A_t$, in the Minimal
Supersymmetric 
Standard Model (MSSM). We will separately discuss the SUSY CP-nonconserving
effect in production and decay amplitudes of the top and show how these two
asymmetries combine to give the overall asymmetry in $p\bar{p}\to t\bar{b}+X
\to W^+ b \bar b +X$. 

The experimental limit on the Neutron Electric Dipole Moment (NEDM), $d_n
\leq 1.1 \times 10^{-25}$ e-cm \cite{prd}, places a severe constraint on
the phase of the SUSY Higgs mass parameter $\mu$. In particular, $\arg
(\mu) < {\cal O} (10^{-2})$ \cite{falk,garisto,grossman,oshimo} is essentially
inevitable for a typical SUSY mass scale of the order of a few hundreds GeV.
Therefore, the {\it only significant} SUSY-CP-odd phase, that can potentially
drive notable CP-nonconserving effects in top quark reactions, is the phase
in the stop soft trilinear breaking term $A_t$. As will be shown below,
CP-violating 
effects that can arise from the other $A_{q_i}$ terms (associated with the
lighter quarks $q_i$) are extremely suppressed being proportional to the
masses of the light quarks. In particular, when $m_{q_i} \to 0$ there  is
no mixing between the left and the right squarks. The two mass eigenstates
of the supersymmetric partners of the light quarks are therefore expected
to be nearly degenerate, thus not playing any role in CP-violation effects
at high enough energies. Of  course, this is not the case for the NEDM which
is driven by the slight deviation from degeneracy of the supersymmetric
partners of the $u$ and the $d$ quarks when $\arg(\mu)=0$.         

Moreover, if the phases of $A_u,A_d$ and $A_t$ are not correlated at the
EW scale, which is indeed possible if the GUT-scale universality of the $A$
terms is relaxed, then the experimental limit on the NEDM cannot put any
further constraint on $\arg (A_t)$.  Therefore, it is extremely important to
explore other avenues for constraining $\arg(A_t)$. Of course, the most natural
place to look for CP-nonconserving effects, which are driven by $\arg(A_t)$,
is high energy reactions involving the top quark. Thus, in  the limit $\arg(\mu)
\to 0$, the {\it only\/} important phenomenological CP-violating
SUSY parameter, that enters any CP-odd effect at high energies, resides in
the imaginary part of the ${\tilde t}_L - {\tilde t}_R$ mixing matrix elements
${\rm Im}(Z_t^{1i*}Z_t^{2i})$ (see appendix~A). Being proportional to
$\arg(A_t)$, 
it vanishes for $\arg(A_t) \to 0$. Besides, $\arg(A_t)$ may play
an important role in explaining the observed baryon asymmetry in the universe.
Recently, it was claimed that with  $\arg(\mu) \to 0$, $t$ squarks can mediate
the charge transport mechanism needed to generate the observed baryon asymmetry
even with squark masses $\sim {\rm few} \times{}$GeV, provided that $\arg(A_t)$
is not much suppressed \cite{oshimobarion}.  

In this paper we discuss the CP-odd effects, induced by ${\rm
Im}(Z_t^{1i*}Z_t^{2i}) 
\propto \arg(A_t)$ in the reaction $p\bar{p}\to t\bar{b}+X$ and in the
subsequent top decay $t \to W^+ b$. We find that, with maximum CP-violation
(i.e., $|{\rm Im}(Z_t^{1i*}Z_t^{2i})| = 1/2$), a CP-violating asymmetry in
the cross-section can reach 3\% for squark masses at around 0.5 TeV and
a light stop mass below 100 GeV\null. We also find that, in the same range
of squark masses, a Partial Rate Asymmetry (PRA) effect in $t \to W^+ b$
is predominantly below $0.1\%$ and can be slightly above $0.1\%$ for low
$\tan\beta$. In a narrow window of  the SUSY parameter space, with $\tan\beta
\sim {\cal O}(1)$, it can reach at the most $\sim 0.3\%$.  
This is somewhat less than the estimates in the
literature of a few percent asymmetry in top decay in the MSSM
\cite{cristova}. Therefore, with a predicted few 
percent overall asymmetry in the cross-section and decay, the reaction $p\bar{p}\to
t\bar{b}+X \to W^+ b \bar b +X$ can serve to constrain $\arg(A_t)$ 
in the future runs of the upgraded Tevatron at Fermilab.     

Also, as a by product of our analysis, we found that when the Higgs mass
term, $\mu$, is real at the EW-scale (or has a very small phase) and bearing
theoretical uncertainties associated with the naive quark model approach (see
e.g., \cite{ellis}), the experimental limit on the NEDM can be naturally
accommodated with phases of the soft trilinear breaking $A$ terms at the
order of $\sim {\rm few} \times 10^{-1}$ even with squark and gluino masses
below 500 GeV\null. Therefore, we believe that there is no compelling argument
for  the SUSY CP-odd phases to be necessarily less than or of the order
of $10^{-2}$ for squarks masses below 1 TeV,  contrary to previously and
commonly used claims.   

We wish to emphasize that the analytical formulations given in this paper
hold for the most general low energy realization of the MSSM\null. Moreover,
%ASf
the PRA effect in the top decay $t \to W^+ b$, presented in this paper, 
%is an %independent quantity which 
does  not depend on the specific production mechanism of the $t$ and $\bar t$. 
%ASf
Therefore,
the LHC which may be able to produce $\sim 10^7$ pairs of $t \bar t$ may,
in principle, be able to detect a PRA effect in $t \to W^+ b$ even as small
as $\sim {\rm few} \times 10^{-3}$, provided detector systematics are sufficiently
small. 

The paper is organized as follows: in section~2 we construct a plausible
phenomenological low energy SUSY spectrum. In section~3 we present a complete,
self contained analytical derivation of the CP asymmetry in both production
and decay processes of the  top in a general MSSM framework, keeping all
possible SUSY CP-odd phases. In section~4 we discuss our choice of the low
energy CP phases and its implications on the NEDM\null. In section~5 we
present our numerical results in the limit $\arg(\mu)=0$ and in  section~6
we summarize. The relevant SUSY lagrangian pieces along with a detailed 
derivation of the various SUSY mixing matrix elements is given in appendix~A\null.
In appendix B we list the SUSY CP-violating phases, in the general MSSM 
scenario, that are responsible for the CP-asymmetry in our reaction while
a description of the one-loop integrals is given in appendix~C.\\ 

\noindent {\bf 2. \underline{Low Energy MSSM Phenomenology}}\\

The most general low energy $N=1$ minimal supergravity (SUGRA) $SU(3)\times
SU(2) \times U(1)$ invariant lagrangian  (which apart from gravitational 
interactions, is essentially identical at low energies to a theory with 
softly broken supersymmetry), that consists of three generations of quarks,
two Higgs doublets and the $SU(3)\times SU(2) \times U(1)$ gauge fields,
along with their supersymmetric partners, can be written as
\cite{wise,haber,rosiek}: 

\begin{equation}
{\cal L} = \ \mbox{kinetic terms} + \int d^2\theta W + {\cal L}_{\rm soft}
\ \ , 
\end{equation}

\noindent where $W$ is the superpotential and is given by:

\begin{equation}
W = \epsilon_{ij}(g^{IJ}_U \hat{Q}^i_I \hat{H}^j_2 \hat{U}_J
+g^{IJ}_D \hat{Q}^i_I \hat{H}^j_1 \hat{D}^c_J + 
g^{IJ}_E \hat{L}^i_I \hat{H}^j_1 \hat{R}^c_J+ \mu\hat{H}^i_1\hat{H}^j_2)
\ \ . 
\end{equation}

\noindent $\epsilon_{ij}$ is the antisymmetric tensor with $\epsilon_{12}=1$
and the usual convention was used for the superfields
$\hat{Q},\hat{U},\hat{L},\hat{R}$ and $\hat{H}$ \cite{wise}. The capital
index indicates the generation (i.e., $I,J=1,2$ or 3). 

${\cal L}_{\rm soft}$ consists of the soft breaking terms and can be divided
into three pieces: 

\begin{equation}
{\cal L}_{\rm soft} \equiv {\cal L}_{\rm gaugino} + {\cal L}_{\rm scalar}
+ {\cal L}_{\rm trilinear} \ \ . 
\end{equation}

\noindent These are the soft-supersymmetry breaking gaugino and scalar mass
terms and the trilinear coupling breaking terms. In particular they are given
by \cite{rosiek}: 

\begin{equation}
{\cal L}_{\rm gaugino} = \frac{1}{2} (\tilde{m}_1 \lambda_B \lambda_B +
\tilde{m}_2 \lambda_W^a \lambda_W^a + \tilde{m}_3 \lambda_G^b \lambda_G^b)
\ \ ,  
\end{equation}
\begin{eqnarray}
{\cal L}_{\rm scalar}  &=& -m_{H_1}^2 |H^1_i|^2 -m_{H_2}^2 |H^2_i|^2 - m_L^2
|L^i|^2 -  \nonumber \\ 
&&m_R^2 |R|^2 - m_Q^2 |Q^i|^2 - m_D^2 |D|^2 - m_U^2 |U|^2 \ \ ,
\end{eqnarray}
\begin{eqnarray}
{\cal L}_{\rm trilinear} &=& \epsilon_{ij} (g_U A_U Q^i H^j_2 U +
g_D A_D Q^i H^j_1 D +\nonumber \\
&&g_E A_E L^i H^j_1 R + \mu B H^i_1H^j_2) \ \ .
\end{eqnarray}
%
%
%\begin{equation}
%{\cal L}_{\rm soft} &=& - \frac{1}{2} \sum_a \tilde{m}_a\lambda_a\lambda_a - 
%\frac{1}{2} \sum_{\scriptsize{\matrix{\rm all \cr \rm scalars}}} m_i|\phi_i|^2 + \nonumber \\
%&& + m_g \epsilon_{ij} 
%(A^{IJ}_V g^{IJ}_V Q^i_I H^j_2 V_s +
%A^{IJ}_D g^{IJ}_D Q^i_I H^j_1 D^c_s +\nonumber \\
%&& + A^{IJ}_E g^{IJ}_E L^i_I H^j_1 R_j^c + \mu B H^i_1H^j_2) 
%\end{equation}
%
%

\noindent In the above soft breaking terms we have omitted the family indices
$I$ and $J$. The above scalar fields correspond to the superfields which
were indicated in our notation by a ``hat''. $\lambda_B$,  $\lambda_W^a$
(with $a=1,2$ or 3) and $\lambda_G^b$ (with $b=1,\dots,8$) are the gauge
superpartners of the  U(1), SU(2) and SU(3) gauge bosons, respectively.

The conventional wisdom is to assume a complete universality of the soft
supersymmetric parameters at the GUT scale.  That is, a common scalar mass
$m_0$, a common gaugino mass $M_{1/2}$ and a universal boundary condition
for the trilinear soft breaking terms $A_U=A_D=A_E=A^G$. This gives rise
to an appealing theory with a minimum number of  free parameters, which,
among other interesting features, allows Radiative EW Symmetry Breaking (REWSB)
\cite{sugra}. Such a SUGRA model has only two independent CP-violating phases
at the GUT scale \cite{dugan}. These two physical phases can be chosen as
$\arg(A^G)$ and $\arg(\mu^G)$. Nonetheless, although very attractive, a
complete 
universality of the soft breaking terms at the GUT scale (motivated by a
``would be'' GUT-scale SUGRA model scenario) is a simplifying assumption.
If the universality of the trilinear soft breaking terms mentioned above
is relaxed, REWSB can still occur. Yet, in this less constrained framework,
the $A_t,A_u$ and $A_d$ (the only three soft trilinear breaking terms relevant
for our discussion in this paper) can have different phases at the GUT scale,
and therefore their phases at the EW scale are completely undetermined. As
a consequence, we will show below that, contrary to previous claims, with
no ``fine tuning'' of $\arg(A_u)$ and $\arg(A_d)$, the NEDM can meet its
experimental limit even with squark masses  of a few hundred GeV, leaving
arg($A_t$) large enough to drive significant CP-violating effects in top
quark systems. We will discuss in detail the ``SUSY CP problem'' of the NEDM
in section~4.   

Till now, no supersymmetric particle has been discovered, and in spite of
the very fascinating theoretical features of the GUT-scale SUGRA model, there
is no experimental hint that can indicate the real nature of the underlying,
short distance SUSY model. We therefore wish to keep an open mind and instead
of restricting ourselves to the GUT-scale SUGRA model with complete universality
of the soft breaking terms, our strategy will be to choose a plausible set
of the SUSY parameters at the EW scale subject to the present experimental
limits on the sparticle masses \cite{limits}. Indeed, by relaxing the GUT-scale
universality of only the soft breaking trilinear $A$ terms, almost any low
energy SUSY spectrum can be consistently recovered.  

We now describe the key features and assumptions of our low energy SUSY
scenario:\\ 
\noindent I) Motivated by the strong theoretical prediction of the unification
of the SU(3), SU(2) and U(1) gauge couplings when SUSY particles with a mass
scale around 1 TeV are folded into the Renormalization Group Equations (RGE),
we will follow only one traditional simplifying GUT assumption; that there
is an underlying grand unification. This leads us to have a common gaugino
mass parameter defined at the GUT scale. Then, the difference between the
three low energy gaugino mass parameters comes from  the fact that they undergo
a different renormalization when they evolve from the GUT scale to the EW
scale due to the different gauge structure of their interactions. 
The gaugino masses, at the EW-scale, are then related by \cite{barger}:

\begin{equation}
\frac{3\cos^2\theta_W}{5}\frac{{\tilde m}_1}{\alpha}=\sin^2\theta_W
\frac{{\tilde m}_2}{\alpha}=\frac{{\tilde m}_3}{\alpha_s}~,
\end{equation}

\noindent where $\theta_W$ is the weak mixing angle and ${\tilde m}_3$ is
the low energy gluino mass (from now on, we will refer to the gluino mass
as $m_G$). Thus, once the gluino mass is set at the EW scale, the SU(2) and
U(1) gaugino masses ${\tilde m}_2$ and ${\tilde m}_1$, respectively, are
determined.\\ 
\noindent II) All the squarks except for the light stop are assumed to be
degenerate with a mass $M_s$, while the light stop mass, which we denote
by $m_l$, is chosen subject to $m_l > 50$ GeV \cite{limits}.\\  
III) The mass matrices of the neutralinos, $M_{{\tilde \chi}^0}$, and
charginos, 
$M_{{\tilde \chi}}$, depend on the low energy Higgs mass parameter $\mu$,
the two gaugino masses ${\tilde m}_2$ and ${\tilde m}_1$ (which are resolved
by the gluino mass) and  $\tan\beta$ (see appendix~A).  Therefore, once
$\mu,m_G$ and $\tan\beta$ are set to their low energy values, the four physical
neutralino 
species $m_{{\tilde \chi}_n^0}$ ($n=1-4$) and the two physical chargino species
$m_{{\tilde \chi}_m}$ ($m=1,2$) are extracted by diagonalizing $M_{{\tilde
\chi}^0}$ and $M_{{\tilde \chi}}$. In appendix~A we list the full analytical
prescription for diagonalizing the neutralinos and charginos mixing matrices
in the case that $\mu$ is real. \\ 
IV) We limit the lightest neutralino mass to be above 20 GeV and the lighter
chargino mass to be above 65 GeV \cite{limits}.\\  
V) We will always choose $m_G>200$ GeV within the range $M_s-200~{\rm GeV}
\lsim m_G \lsim M_s+200~{\rm GeV}$ as roughly indicated by recent results
from the Tevatron \cite{limits}.\\ 

With no further assumptions, the low energy SUSY mass spectrum is fully determined
from the only five free parameters $\mu,m_G,M_s,m_l$ and $\tan\beta$. The
possible CP phases, that can arise in this low energy SUSY framework, will
be described in detail in section~4.\\ 

\noindent {\bf 3. \underline{Derivation of the CP Violating Asymmetry}}\\

We now turn to the discussion of the possible CP-violating effects in the
process $p \bar p \to t \bar b+X \to W^+ b \bar b +X$ (i.e., production of
a pair of $t \bar b$ followed by the top decay $t \to b W^+$), driven by
SUSY CP-violating sources \cite{footresonant}. To estimate the possible size
of the effect, we consider the following CP-violating quantity: 

\begin{equation}
{\hat A} \equiv \frac{\hat{\sigma}(\hat{s}) -
 \hat{\bar{\sigma}}(\hat{s})}{\hat{\sigma}(\hat{s})+\hat{\bar{\sigma}}(\hat{s})}
 ~, 
\end{equation}

\noindent where $\hat s = (p_t+p_b)^2$ and $\hat{\sigma}(\hat{s})$ and
$\hat{\bar{\sigma}}(\hat{s})$ 
are the cross-sections for the reactions $u \bar d \to t \bar b \to W^+b
\bar b$ and $\bar u d \to \bar t b \to W^- \bar b b$, respectively.  Folding
in the parton densities in the usual manner \cite{collider} we define the
asymmetry of cross-sections at the level of the  colliding protons as:

\begin{equation}
A \equiv \frac{{\sigma}({s}) -
 {\bar{\sigma}}({s})}{{\sigma}({s})+{\bar{\sigma}}({s})} \label{a0}~.
\end{equation}
 
\noindent Now $s=(p_p+p_{\bar p})^2$ and ${\sigma}({s})$ and
${\bar{\sigma}}({s})$ 
are the cross sections for  $p\bar{p}\to t\bar{b} +X \to W^+ b \bar b +X$
and  $p\bar{p}\to \bar t b+X \to W^- \bar b b +X$, respectively. 

In the narrow width approximation the tree-level cross-section at the quark
level is given by: 

\begin{equation}                              %23
\hat{\sigma_0}(\hat{s}) \approx \frac{\pi\alpha^2 (m^2_t-\hat{s})^2
(m_t^2+2\hat{s})} {24 \sin^4\theta_W\hat{s}^2 (\hat s -m_W^2)^2} \times {\rm
Br}(t \to W^+ b) ~.   
\end{equation} 

\noindent In Fig.~1 we have drawn the tree level diagram for the parton level
reaction $u\bar{d}\to t\bar{b}$.  The possible supersymmetric one loop diagrams
that can violate CP invariance in this reaction are depicted in Fig.~2. Note
that in both Fig.~1 and 2 the subsequent top decay is not explicitly drawn.
The one-loop CP-violating triangle diagrams a-d of Fig.~2 in the production
vertex will enter also the top decay $t \to W^+ b$ with the $t$ and $W^+$
momenta reversed.

To one-loop order in perturbation theory, where the CP-violating virtual
corrections enter only  either the production or the decay vertices
of the top in the overall reaction $p \bar p \to t \bar b +X \to W^+ b \bar
b +X$, and in the narrow width  approximation for the decaying top, the overall
CP asymmetry, $A$, can be broken into:  

\begin{equation}
A = A_P + A_D ~.
\end{equation}

\noindent In this Eq.\  $A_P$ and $A_D$ are the CP asymmetries emanating
from the production and decay of the top, respectively, and are defined
by: 

\begin{eqnarray}
&&A_P \equiv \frac{ \sigma(p \bar p \to t \bar b +X) - \bar\sigma(p \bar p 
\to \bar t b +X) }{ \sigma(p \bar p \to t \bar b+X) + \bar\sigma(p \bar p 
\to \bar t b+X) } \label{ap}~,\\
&&A_D \equiv \frac{ \Gamma(t \to W^+ b) - \bar\Gamma(\bar t \to W^- \bar b) }
{ \Gamma(t \to W^+ b) + \bar\Gamma(\bar t \to W^- \bar b) } \label{ad}~.  
\end{eqnarray}

\noindent The PRA $A_D$, defined in Eq.~\ref{ad}, is an independent quantity
and does not depend on the specific production mechanism of the top. We will
therefore treat the CP-nonconserving asymmetries $A_P$ and $A_D$ separately
keeping in mind that the  total CP-odd effect is simply the sum of the two.
Moreover, it is convenient to further divide the cross-section asymmetry
in $t \bar b$ production, $A_P$, into: 

\begin{equation}
A_P \equiv A_P^{\rm triangle} + A_P^{box} ~,
\end{equation}

\noindent where $A_P^{\rm triangle}$ and $A_P^{box}$ are the CP-violating
cross-section asymmetries which arise from the triangle diagrams $a-d$ and
$a^{\prime}-d^{\prime}$ and box diagrams $e-h$ in Fig.~2, respectively.
Similarly, 
the corresponding parton level asymmetries will be denoted by a ``hat''. \\ 

\noindent \underline{Asymmetry From Triangle Diagrams In The Production
Amplitude - $A_P^{\rm triangle}$}:\\ 

The one-loop $t \bar b$ and $\bar t b$ currents of the production amplitude
can be parameterized as:  

\begin{eqnarray}
J^{\mu(t \bar b)}_ k \equiv i \frac{g_W}{\sqrt 2} \sum_{P=L,R} \bar{u}_t
\left( \frac{{\cal P}_{1(k)}^P p_b^{\mu}}{m_t} + {\cal P}_{2(k)}^P \gamma^{\mu}
\right) Pv_b \label{jtbarb}~, \\ 
J^{\mu(\bar t b)}_ k \equiv  i \frac{g_W}{\sqrt 2} \sum_{P=L,R} \bar{u}_b
\left( \frac{{\bar {\cal P}}_{1(k)}^P p_b^{\mu}}{m_t} + \bar{{\cal P}}_{2(k)}^P
\gamma^{\mu} \right) Pv_t \label{jbartb}~,  
\end{eqnarray}

\noindent where $P=L~{\rm or}~R$ and $L(R) \equiv (1 -(+) \gamma_5)/2$. The
index $k$ indicates the triangle diagram (i.e.\ $k=a,b,c,d,a^{\prime},b^{\prime},c^{\prime}$
or  $d^{\prime}$). ${\cal P}_{1(k)}^{L,R}$ and ${\cal P}_{2(k)}^{L,R}$ defined
in Eqs.~\ref{jtbarb} and \ref{jbartb},   contain the SUSY CP-violating phases
as well as the  absorptive phases of the triangle diagrams, both of which
are needed to render $A_P^{\rm triangle}$ non-zero. Then, in terms of the 
scalar $({\cal P}_{1(k)})$ and vector $({\cal P}_{2(k)})$ form factors, the
parton level cross-section asymmetry ${\hat A}_P^{\rm triangle}$ is given
by: 

\begin{equation}
{\hat A}_ P^{\rm triangle} = \sum_k \left\{ \frac{(x-1)}{2(x+2)} {\rm Re}({\cal
P}_{1(k)}^L + {\bar {\cal P}}_{1(k)}^R) -  {\rm Re}({\cal P}_{2(k)}^L - {\bar
{\cal P}}_{2(k)}^L) \right\} \label{atriangle}~, 
\end{equation}

\noindent where $x \equiv m_t^2/\hat s$ and the sum is carried out over all
triangle diagrams in Fig.~2.  It is easy to show that if one defines:

\begin{eqnarray}
{\cal P}_{1(k)}^L &\sim& e^{i \delta_s^{1(k)}} \times e^{i \delta_w^{1(k)}}
\label{relation1}~,\\ 
{\cal P}_{2(k)}^L &\sim& e^{i \delta_s^{2(k)}} \times e^{i \delta_w^{2(k)}}
\label{relation2}~, 
\end{eqnarray}

\noindent where $\delta_s^{1(k)},\delta_s^{2(k)}$ are the CP-even absorptive
phases (i.e., final state interaction phases) and
$\delta_w^{1(k)},\delta_w^{2(k)}$ are the CP-odd phases associated with each
of the triangle diagrams in Fig.~2, then:  

\begin{eqnarray}
{\bar {\cal P}}_{1(k)}^R &\sim& - e^{i \delta_s^{1(k)}} \times e^{-i
\delta_w^{1(k)}} \label{relation3}~,\\ 
{\bar {\cal P}}_{2(k)}^L &\sim& e^{i \delta_s^{2(k)}} \times e^{-i
\delta_w^{2(k)}} \label{relation4}~. 
\end{eqnarray}

The CP-violating scalar form factors in Eq.~\ref{atriangle}, for each triangle
diagram are then given by: 

\begin{eqnarray}   
{\rm Re}({\cal P}_{1(a)}^L + {\bar {\cal P}}_{1(a)}^R) &=& \frac{8}{3}
\frac{\alpha_s}{\pi} 
m_t m_{\tilde{g}}   {\cal O}_a^1 \left( C^a_0 + C^a_{11} \right) \label{aalr}~, \\
{\rm Re}({\cal P}_{1(b)}^L + {\bar {\cal P}}_{1(b)}^R) &=& - \frac{\alpha}{\pi
\sin^2\theta_W} m_t \left[ m_t {\cal O}_b^1 \left( C^b_{12} + C^b_{23} \right)
- m_{\tilde{\chi}_n^0} {\cal O}_b^2 \left( C^b_0 + C^b_{11} \right) \right]
~, \\ 
{\rm Re}({\cal P}_{1(c)}^L + {\bar {\cal P}}_{1(c)}^R) &=& - \frac{\alpha}{\pi
\sin^2\theta_W} m_t \left[  m_t {\cal O}_c^1 \left( C^c_{12} + C^c_{23} \right)
+ m_{\tilde{\chi}_m} {\cal O}_c^2 C^c_{12} \right. \nonumber \\ 
&& \left. + m_{\tilde{\chi}_n^0} {\cal O}_c^3 \left( C^c_{12} - C^c_{11}
\right) \right] ~, \\ 
{\rm Re}({\cal P}_{1(d)}^L + {\bar {\cal P}}_{1(d)}^R) &=& {\rm Re}({\cal
P}_{1(c)}^L + {\bar {\cal P}}_{1(c)}^R) \left(m_{\tilde{\chi}_n^0} \to -
m_{\tilde{\chi}_m}, m_{\tilde{\chi}_m} \to m_{\tilde{\chi}_n^0}, {\cal O}_c^i
\to  {\cal O}_d^i, C^c_{ij} \to C^d_{ij} \right) ~,
\end{eqnarray}   

\noindent and the CP-violating vector form factors in Eq.~\ref{atriangle}
are: 

\begin{eqnarray}
{\rm Re}({\cal P}_{2(a)}^L - {\bar {\cal P}}_{2(a)}^L) &=& 0 ~, \\ 
{\rm Re}({\cal P}_{2(b)}^L - {\bar {\cal P}}_{2(b)}^L) &=& \frac{\alpha}{\pi
\sin^2\theta_W} {\cal O}_b^1  C^b_{24} ~, \\ 
{\rm Re}({\cal P}_{2(c)}^L - {\bar {\cal P}}_{2(c)}^L) &=& \frac{1}{2}
\frac{\alpha}{\pi 
\sin^2\theta_W} \left[{\cal O}_c^1 \left( (\hat s -m_t^2) C^c_{23} -\hat
s C^c_{22} -2 C^c_{24} -m_t^2 C^c_{12} \right) \right. \nonumber \\
&& \left. - m_t m_{\tilde{\chi}_m} {\cal O}_c^2 C^c_{12} - m_t
m_{\tilde{\chi}_n^0} 
{\cal O}_c^3 \left( C^c_0 + C^c_{12} \right) \right. \nonumber \\   
&& \left. - m_{\tilde{\chi}_m} m_{\tilde{\chi}_n^0} {\cal O}_c^4 C^c_0 \right]
~, \\ 
{\rm Re}({\cal P}_{2(d)}^L - {\bar {\cal P}}_{2(d)}^L) &=& {\rm Re}({\cal
P}_{2(c)}^L - {\bar {\cal P}}_{2(c)}^L) \left(m_{\tilde{\chi}_n^0} \to -
m_{\tilde{\chi}_m}, m_{\tilde{\chi}_m} \to m_{\tilde{\chi}_n^0}, {\cal O}_c^i
\to  {\cal O}_d^i, C^c_{ij} \to C^d_{ij} \right) \label{lastvec}~, \\
{\rm Re}({\cal P}_{2(a^{\prime})}^L - {\bar {\cal P}}_{2(a^{\prime})}^L)
&=& 0 ~, \\ 
{\rm Re}({\cal P}_{2(b^{\prime})}^L - {\bar {\cal P}}_{2(b^{\prime})}^L)
&=& - \frac{\alpha}{\pi \sin^2\theta_W}{\cal O}_{b^{\prime}}^1
C^{b^{\prime}}_{24} ~, \\ 
{\rm Re}({\cal P}_{2(c^{\prime})}^L - {\bar {\cal P}}_{2(c^{\prime})}^L)
&=&  \frac{1}{2} \frac{\alpha}{\pi \sin^2\theta_W} \left[ {\cal
O}_{c^{\prime}}^1 
\left( \hat s (C^{c^{\prime}}_{23} -  C^{c^{\prime}}_{22}) -2
C^{c^{\prime}}_{24} \right) \right. \nonumber \\ 
&& \left. + m_{\tilde{\chi}_m} m_{\tilde{\chi}_n^0} {\cal O}_{c^{\prime}}^2
C^{c^{\prime}}_0 \right] ~, \\ 
{\rm Re}({\cal P}_{2(d^{\prime})}^L - {\bar {\cal P}}_{2(d^{\prime})}^L)
&=&  {\rm Re}({\cal P}_{2(c^{\prime})}^L - {\bar {\cal P}}_{2(c^{\prime})}^L)
\left( m_{\tilde{\chi}_n^0} \to - m_{\tilde{\chi}_m}, m_{\tilde{\chi}_m}
\to m_{\tilde{\chi}_n^0}, {\cal O}_{c^{\prime}}^i \to  {\cal O}_{d^{\prime}}^i,
C^{c^{\prime}}_{ij} \to C^{d^{\prime}}_{ij} \right) \label{cdtagll}~.  
\end{eqnarray}   

\noindent The SUSY CP-weak phases associated with the triangle diagrams,
the ${\cal O}_k^i$'s above, are given in appendix~B and all the corresponding
loop form factors $C^k_0,C^k_{pq}$ ($p=1,2$ and $q=1-4$)  are defined and
given in appendix~C\null. 

Note that for diagrams $a^{\prime},b^{\prime},c^{\prime}$ and $d^{\prime}$
only a vector  form factor in the $u \bar d$ one-loop current contributes
to CP-violation  in the limit $m_u,m_d \to 0$ (in this limit diagram
$a^{\prime}$ contain no CP-violating phase). That is:  

\begin{eqnarray} 
J^{\mu(u \bar d)}_ k \equiv i \frac{g_W}{\sqrt 2} \sum_{P=L,R} \bar{v}_d
{\cal P}_{2(k)}^P \gamma^{\mu} Pu_u ~, \\
J^{\mu(\bar u d)}_ k \equiv  i \frac{g_W}{\sqrt 2} \sum_{P=L,R} \bar{v}_u
\bar{{\cal P}}_{2(k)}^P \gamma^{\mu} Pu_d ~. 
\end{eqnarray}

\noindent and we can therefore consider these diagrams to have the same
structure as diagrams $a,b,c$ and $d$ (see Eqs.~\ref{jtbarb} and  \ref{jbartb}),
taking the vector form factor in the $ud$ current to be the vector form factor
in the $tb$ current. \\  

\noindent \underline{Asymmetry From Box Diagrams In The Production Amplitude
- $A_P^{\rm box}$}:\\ 

There are two kinds of amplitudes for the box diagrams. The amplitudes for
diagrams $e$ and $f$ in Fig.~2 can be written schematically as: 

\begin{equation}   
-i {\cal M}_{k=e,f} \equiv \frac{-g_W^4}{16 \pi^2} \int \frac{d^4q}{i \pi^2}
\left\{ \frac{{\bar u}_t \left[ \sum_{P=L,R} \left( {\cal X}_k^P +  {\cal
X}_k^{{\prime}P} p\hspace{-0.20cm}/_1 \right) P \right]  u_u  {\bar v}_d
\left[ \sum_{P=L,R} \left( {\cal Y}_k^P +  {\cal Y}_k^{{\prime}P}
p\hspace{-0.20cm}/_2 
\right) P \right]  v_b }{{\cal A}_k^1 {\cal A}_k^2 {\cal A}_k^3 {\cal A}_k^4}
\right\} \label{ef}~,          
\end{equation} 

\noindent whereas for Diagrams $g$ and $h$:

\begin{equation}   
-i {\cal M}_{k=g,h} \equiv \frac{-g_W^4}{16 \pi^2}  \int \frac{d^4q}{i \pi^2}
\left\{ \frac{{\bar u}_t \left[ \sum_{P=L,R}  \left( {\cal X}_k^P +  {\cal
X}_k^{{\prime}P} p\hspace{-0.20cm}/_1 \right) P \right] u_d  {\bar u}_b \left[
\sum_{P=L,R} \left( {\cal Y}_k^P +  {\cal Y}_k^{{\prime}P} p\hspace{-0.20cm}/_2
\right) P \right]  u_u}{{\cal A}_k^1 {\cal A}_k^2 {\cal A}_k^3 {\cal A}_k^4}
\right\} \label{gh}~.          
\end{equation} 

\noindent The four vectors $p_1$ and $p_2$, that appear in Eqs.~\ref{ef} and
\ref{gh} above, stand for  either $q$ - the integrated four momentum in the
loop or $p=p_t+p_b$ - the $\hat s$-channel four momentum. Also, the
denominators ${\cal A}_k^i$ of the particles in the loop can be read off
appendix~C (see also \cite{footminus}).     

A straight forward calculation of the interference of diagrams $e,f,g$ and
$h$ with the tree level diagram  yields:

\begin{equation}    
{\hat A}_P^{\rm box} = \frac{3}{2} \frac{\alpha}{\pi \sin^2\theta_W}
\frac{(y-1)}{(1-x)^2 
(2+x)}  \sum_k \frac{1}{{\hat s}^2} \int_{{\hat t}^- = m_t^2 -\hat s}^{{\hat
t}^+ = 0} d{\hat t} \left\{ {\cal B}^1_k({\hat s},{\hat t}) + {\cal
B}^2_k({\hat s},{\hat t}) \right\} ~, 
\end{equation}  

\noindent where $y \equiv m_W^2/\hat s$ and the three Mandelstam variables
at the parton level (i.e., $\hat s,\hat t$ and $\hat u$) are related via
(for $m_b=m_u=m_d=0$) $\hat s +\hat t +\hat u = m_t^2$.  For later use, we
furthermore define the following quantity: 

\begin{equation}   
X \equiv \frac{1}{2} \left[ \hat s \left( \hat s -m_t^2 \right) - \hat t
\left( \hat t -m_t^2 \right) +  \hat u \left( \hat u -m_t^2 \right) \right]
~.  
\end{equation} 

\noindent For diagrams $e$ and $f$, in the limit $m_u=m_d=m_b \to 0$, the
${\cal X}_k^R,{\cal X}_k^{{\prime}R}, {\cal Y}_k^L,{\cal Y}_k^R$ and 
${\cal Y}_k^{{\prime}R}$ terms do not contribute to ${\hat A}_P^{\rm box}$
and we get: 

\begin{eqnarray}   
{\cal B}^1_{k=e,f}({\hat s},{\hat t}) &=& 2 m_t {\rm Im} \left( {\cal X}_k^L
{\cal Y}_k^{{\prime}L} \right) {\rm Im} \left( \int \frac{d^4q}{i \pi^2}
\left\{ \frac{ {\rm Tr} \left[ L p\hspace{-0.20cm}/_u p\hspace{-0.20cm}/_b
(q\hspace{-0.20cm}/-p\hspace{-0.20cm}/)   p\hspace{-0.20cm}/_d \right]}{{\cal
A}_k^1 {\cal A}_k^2 {\cal A}_k^3 {\cal A}_k^4} \right\} \right) \label{b1ef}~,
\\     
{\cal B}^2_{k=e,f}({\hat s},{\hat t}) &=& 2 {\rm Im} \left( {\cal
X}_k^{{\prime}L} 
{\cal Y}_k^{{\prime}L} \right) {\rm Im} \left( \int \frac{d^4q}{i \pi^2}
\left\{ \frac{{\rm Tr} \left[ L p\hspace{-0.20cm}/_t q\hspace{-0.20cm}/
p\hspace{-0.20cm}/_u 
p\hspace{-0.20cm}/_b (q\hspace{-0.20cm}/-p\hspace{-0.20cm}/)  
p\hspace{-0.20cm}/_d \right]}
{{\cal A}_k^1 {\cal A}_k^2 {\cal A}_k^3 {\cal A}_k^4} \right\} \right) ~, 
\end{eqnarray}

\noindent whereas, for diagrams $g$ and $h$ the ${\cal X}_k^L,{\cal
X}_k^{{\prime}L}, {\cal Y}_k^R,{\cal Y}_k^{{\prime}L}$ and 
${\cal Y}_k^{{\prime}R}$ terms vanish, and we have:

\begin{eqnarray}
{\cal B}^1_{k=g,h}({\hat s},{\hat t}) &=& -2 \hat u {\rm Im} \left( {\cal
X}_k^{R} {\cal Y}_k^{L} \right) {\rm Im} \left( \int \frac{d^4q}{i \pi^2}
\left\{ \frac{ {\rm Tr} \left[ L p\hspace{-0.20cm}/_t p\hspace{-0.20cm}/_d
\right]} {{\cal A}_k^1 {\cal A}_k^2 {\cal A}_k^3 {\cal A}_k^4} \right\} \right)
~, \\ 
{\cal B}^2_{k=g,h}({\hat s},{\hat t}) &=& -2 m_t \hat u {\rm Im} \left( {\cal
X}_k^{{\prime}R} {\cal Y}_k^{L} \right) {\rm Im} \left( \int \frac{d^4q}{i
\pi^2} \left\{ \frac{ {\rm Tr} \left[ L q\hspace{-0.20cm}/ p\hspace{-0.20cm}/_d
\right]} {{\cal A}_k^1 {\cal A}_k^2 {\cal A}_k^3 {\cal A}_k^4} \right\} \right)
\label{b2gh} ~. 
\end{eqnarray}

\noindent Eqs.~\ref{b1ef}--\ref{b2gh} then give:

\begin{eqnarray}   
{\cal B}^1_e({\hat s},{\hat t}) &=& -2 X m_t m_{\tilde{\chi}_n^0} {\cal O}_e^1
\left( D^e_0 +D^e_{11} \right) \label{b1e}~, \\ 
{\cal B}^2_e({\hat s},{\hat t}) &=& 2 {\cal O}_e^2 \left\{ X m_t^2 D^e_{12}
+ \left[  \hat u (\hat u -m_t^2) (\hat s -m_t^2) - m_t^2 \hat u \hat t \right]
D^e_{24} \right.  \nonumber \\ 
&& + \left. X (\hat u -m_t^2) D^e_{25} + 2 \hat u (\hat u -m_t^2) D^e_{27}
\right.  \nonumber \\   
&& +\left. (\hat u -m_t^2) \left[ m_t^2 \hat u D^e_{11} + \hat u (\hat s
-m_t^2) D^e_{12} +X D^e_{13} \right]   \right. \nonumber \\  
&& +\left. m_t^2 \left[ \left( X-\hat u (\hat u -m_t^2) \right) D^e_{11}
- \hat u \hat t D^e_{12} \right] \right\} ~,\\ 
{\cal B}^1_f({\hat s},{\hat t}) &=& {\cal B}^1_e({\hat s},{\hat t}) \left(
m_{\tilde{\chi}_n^0} \to m_{\tilde{\chi}_m},D^e_{ij} \to  
D^f_{ij},{\cal O}_e^1 \to {\cal O}_f^1 \right) ~,\\
{\cal B}^2_f({\hat s},{\hat t}) &=& 0 ~, \\ 
{\cal B}^1_g({\hat s},{\hat t}) &=& 2 m_{\tilde{\chi}_n^0} m_{\tilde{\chi}_m}
\hat u (\hat u -m_t^2) {\cal O}_g^1 D^g_0 ~,\\   
{\cal B}^2_g({\hat s},{\hat t}) &=& -2 m_t m_{\tilde{\chi}_m} \hat u {\cal
O}_g^2 \left[ (\hat u -m_t^2) D^g_{11} + \hat tD^e_{12} + \hat s D^e_{13}
\right] ~,\\    
{\cal B}^1_h({\hat s},{\hat t}) &=& {\cal B}^1_g({\hat s},{\hat t}) \left(
D^g_0 \to  D^h_0,{\cal O}_g^1 \to {\cal O}_h^1 \right) ~,\\
{\cal B}^2_h({\hat s},{\hat t}) &=& {\cal B}^2_g({\hat s},{\hat t}) \left(
m_{\tilde{\chi}_m} \to m_{\tilde{\chi}_n^0},D^g_{ij} \to    
 D^h_{ij},{\cal O}_g^2 \to {\cal O}_h^2 \right) \label{b2h} ~.
\end{eqnarray} 

\noindent The SUSY CP-weak phases associated with the box diagrams, the ${\cal
O}_k^i$'s above, are also given in appendix~B and the corresponding four
point one-loop form factors $D^k_0,D^k_{pq}$ ($p=1,2$ and $q=1-7$)  
are defined and given in appendix~C\null. \\

\noindent \underline{PRA From The Decay Amplitude - $A_D$}:\\ 

As mentioned before, there are four potential diagrams that can give rise
to CP-violating PRA in the decay $t \to W^+b$ within the MSSM\null. These are
diagrams a-d in Fig.~2 where it should be understood that the $t$ and $W^+$
momenta are reversed and the $ud$ current is disregarded. It is convenient
to denote these one-loop top decay diagrams with the same alphabetical order,
i.e., $a,b,c$ and $d$, similar to the corresponding production diagrams,
as the SUSY CP-odd phases defined in Eqs.~\ref{aalr}-\ref{lastvec}, the ${\cal
O}_{a,b,c,d}^i$'s, are the same for both production and decay of the top.

We parameterize the $t \to W^+ b$ and $\bar t \to W^- \bar b$ decay vertices
as follows:  

\begin{eqnarray}
J^{\mu(t)}_ k \equiv i \frac{g_W}{\sqrt 2} \sum_{P=L,R} \bar{u}_b \left(
\frac{{\cal D}_{1(k)}^P p_t^{\mu}}{m_t} + {\cal D}_{2(k)}^P \gamma^{\mu}
\right) Pu_t \label{jt}~, \\
J^{\mu(\bar t)}_ k \equiv  i \frac{g_W}{\sqrt 2} \sum_{P=L,R} \bar{v}_t \left(
\frac{{\bar {\cal D}}_{1(k)}^P p_t^{\mu}}{m_t} + \bar{{\cal D}}_{2(k)}^P
\gamma^{\mu} \right) Pv_b \label{jbart}~, 
\end{eqnarray}

\noindent Similar to the production process discussed above,  ${\cal
D}_{1(k)}^{L,R}$ and ${\cal D}_{2(k)}^{L,R}$ defined in Eqs.~\ref{jt} and
\ref{jbart}, contain 
the CP-violating odd phase as well as the  absorptive phases of the decay
diagram $k$, $k=a-d$. Then, in terms of the  scalar $({\cal D}_{1(k)})$ and
vector $({\cal D}_{2(k)})$ form factors, the  PRA, $A_D$, for the decay process
is given by: 

\begin{equation}
A_ D = \sum_k \left\{ \frac{(x-1)}{2(x+2)} {\rm Re}({\cal D}_{1(k)}^R + {\bar
{\cal D}}_{1(k)}^L) +  {\rm Re}({\cal D}_{2(k)}^L - {\bar {\cal D}}_{2(k)}^L)
\right\} \label{adecay}~. 
\end{equation}

\noindent Now $x \equiv m_t^2/m_W^2$ and the sum is carried out over all
decay diagrams in Fig.~2  (i.e., $k=a,b,c$ and $d$). The relations between
${\cal D}_{1(k)}^{R}$ and ${\bar {\cal D}}_{1(k)}^L$ and between ${\cal
D}_{2(k)}^{L}$ 
and ${\bar {\cal D}}_{2(k)}^L$ are the same as the relation between  ${\cal
P}_{1(k)}^{L}$ and ${\bar {\cal P}}_{1(k)}^R$ and between ${\cal P}_{2(k)}^{L}$
and ${\bar {\cal P}}_{2(k)}^L$, respectively, as is given in 
Eqs.~\ref{relation1}--\ref{relation4}.

For the scalar form factors in Eq.~\ref{adecay} we get:

\begin{eqnarray}   
{\rm Re}({\cal D}_{1(a)}^R + {\bar {\cal D}}_{1(a)}^L) &=& - \frac{8}{3}
\frac{\alpha_s}{\pi} m_t m_{\tilde{g}}   {\cal O}_a^1 C^a_{12} \label{aalrd}~,
\\ 
{\rm Re}({\cal D}_{1(b)}^R + {\bar {\cal D}}_{1(b)}^L) &=& - \frac{\alpha}{\pi
\sin^2\theta_W} m_t \left[m_t {\cal O}_b^1 \left( C^b_{22} - C^b_{23} \right)
+ m_{\tilde{\chi}_n^0} {\cal O}_b^2 C^b_{12} \right] ~, \\ 
{\rm Re}({\cal D}_{1(c)}^R + {\bar {\cal D}}_{1(c)}^L) &=& \frac{\alpha}{\pi
\sin^2\theta_W} m_t \left[m_t {\cal O}_c^1 \left( C^c_{23} - C^c_{22} \right)
- m_{\tilde{\chi}_m} {\cal O}_c^2 \left( C^c_{11} - C^c_{12} \right)
\right. \nonumber \\ 
&& \left. + m_{\tilde{\chi}_n^0} {\cal O}_c^3 \left( C^c_{0} + C^c_{11} \right)
\right] ~, \\ 
{\rm Re}({\cal D}_{1(d)}^R + {\bar {\cal D}}_{1(d)}^L) &=& {\rm Re}({\cal
D}_{1(c)}^R + {\bar {\cal D}}_{1(c)}^L) \left(m_{\tilde{\chi}_n^0} \to -
m_{\tilde{\chi}_m}, m_{\tilde{\chi}_m} \to m_{\tilde{\chi}_n^0}, {\cal O}_c^i
\to  {\cal O}_d^i, C^c_{ij} \to C^d_{ij} \right) ~,
\end{eqnarray}   

\noindent while the vector form factors in Eq.~\ref{adecay} are given
by:

\begin{eqnarray}
{\rm Re}({\cal D}_{2(a)}^L - {\bar {\cal D}}_{2(a)}^L) &=& 0 ~, \\
{\rm Re}({\cal D}_{2(b)}^L - {\bar {\cal D}}_{2(b)}^L) &=& - \frac{\alpha}{\pi
\sin^2\theta_W} {\cal O}_b^1 C^b_{24} ~, \\  
{\rm Re}({\cal D}_{2(c)}^L - {\bar {\cal D}}_{2(c)}^L) &=& \frac{1}{2}
\frac{\alpha}{\pi 
\sin^2\theta_W} \left[{\cal O}_c^1 \left( m_t^2 \left( C^c_{22} - C^c_{23}
\right) +  m_W^2 \left( C^c_{11} + C^c_{22} - C^c_{12} - C^c_{23} \right)
+ 2 C^c_{24} \right) \right. \nonumber \\
&& \left. + m_t m_{\tilde{\chi}_m} {\cal O}_c^2 \left( C^c_{11} - C^c_{12}
\right)   - m_t m_{\tilde{\chi}_n^0} {\cal O}_c^3
\left( C^c_0 + C^c_{11} - C^c_{12} \right) \right. \nonumber \\   
&& \left. - m_{\tilde{\chi}_m} m_{\tilde{\chi}_n^0} {\cal O}_c^4 C^c_0 \right]
~, \\ 
{\rm Re}({\cal D}_{2(d)}^L - {\bar {\cal D}}_{2(d)}^L) &=& {\rm Re}({\cal
D}_{2(c)}^L - {\bar {\cal D}}_{2(c)}^L) \left(m_{\tilde{\chi}_n^0} \to -
m_{\tilde{\chi}_m}, m_{\tilde{\chi}_m} \to m_{\tilde{\chi}_n^0}, {\cal O}_c^i
\to  {\cal O}_d^i, C^c_{ij} \to C^d_{ij} \right) \label{dalrd}~.
\end{eqnarray}   

\noindent As mentioned above, the SUSY CP-weak phases for the decay diagrams,
the ${\cal O}_k^i$'s above, are the same as those for the production triangle
diagrams $a$--$d$ in Fig.~2 and are given in appendix~B\null. Although the
same notation (as in the production case) for the corresponding loop form
factors $C^k_0,C^k_{pq}$ is used in the decay case at hand, these form factors
are separately defined and given in appendix~C\null. \\ 

\noindent{\bf 4. \underline{CP Phases of The Low Energy MSSM and the Neutron
EDM}}\\ 

Before presenting our numerical results, we wish to clarify our approach
with regard to the CP-violating sector of the low energy MSSM and their effect
on the NEDM\null.  Disregarding for now the possible phases at the GUT scale,
we assume for simplicity that at the EW scale, all CP-violation (apart from
the usual SM phases) resides in the complex trilinear soft breaking terms,
the $A_f$'s. For definiteness, we assume $\arg(\mu)=0$ motivated by analyzing
the NEDM, which suggests that $\arg(\mu) < {\cal O} (10^{-2})$ is required
for the NEDM to satisfy the experimental bound for squark masses of a few
hundred GeV (which we are assuming throughout the paper)
\cite{falk,garisto,grossman,oshimo}. 

In fact, from a phenomenological point of view, this is certainly a plausible
scenario that can emanate from a GUT-scale SUSY model in which the universality
of the $A$ terms is relaxed to give arbitrary phases $\arg(A_f^G)$ at the
GUT-scale. In this regard, we take note of a recent very interesting work
by Garisto and Wells \cite{garisto}. They obtained severe constraints on
the low energy phases of $A_t$ and $\mu$ by deriving relations between the
two. For that they use the complete set of RGE involving the complex parameters
of a GUT-scale SUSY model with and without universal $A$ terms and with some
definite 
boundary conditions at the GUT-scale (for example $\arg(\mu^G)=\arg(A_f^G)=0$).
However, taking non-zero phases at the GUT-scale, the constraints obtained
in \cite{garisto} may not hold as ,in general, assuming arbitrary phases at the GUT-scale one can practically make no prediction on the corresponding phases at the EW-scale.
Our approach, the ${\rm EW} \to {\rm GUT}$
approach, will be to assume a set of SUSY phases at the EW-scale, subject
to existing experimental data, which implicitly assumes arbitrary phases
at the scale in which the soft breaking terms are generated. With only the
low energy phases of the various $A_f$ terms, if all the squark masses except
for the stops are degenerate with a mass  $M_s$, then only $\arg(A_t)$
contributes to CP-violation in the reaction $p \bar p \to t \bar b +X \to
W^+ b \bar b +X$. Any  CP-odd phase from the trilinear soft breaking terms
of the other squarks will enter into the asymmetry $A$, defined in
Eq.~\ref{a0}, 
with a suppression factor of $m_q^2/m_t^2$ ($q$ stands for any quark but
the top) even if the squarks masses were not taken exactly degenerate. One
power of  $m_q/m_t$ comes from the Dirac algebra when evaluating the squared
matrix elements, if $m_q$ is not neglected. Another $m_q/m_t$ comes from
the squarks mixing matrices: as can be seen from appendix~A\null. The
CP-violating quantity that arises from ${\tilde q}_L - {\tilde q}_R$ mixing
(i.e., ${\rm Im} (Z_q^{1i*} Z_q^{2i})$) is proportional to the SM quark masses,
thus for $m_q/A_q \to 0$ there will be no mixing in the sector of the
supersymmetric partners of the light quarks. 

In appendix~A we list the Feynman rules needed for calculating the CP-odd
effect in the reaction $u\bar d \to t\bar b \to W^+ b \bar b$. It is then
obvious, that in principle, all the CP-violating vertices of the MSSM arise
from diagonalization of the complex mass matrices $Z_u, \ Z_d, \ Z_N, \ Z^-$
and $Z^+$. However, with the assumption of $\arg(\mu)=0$, it turns out that
$Z_N, \ Z^-$ and $Z^+$ are real. In particular, assuming the universality
of the gaugino masses at the GUT-scale, the phase of the common gaugino mass,
$\arg(M_{1/2})$, can be set to zero by a phase rotation \cite{rrotation},
thus leaving the ${\tilde m}_i$ $i=1-3$ ``phaseless'' at any scale.  
We therefore have all CP-violation arising from ${\tilde t}_L - {\tilde t}_R$
mixing (i.e., from $\arg(A_t)$); in particular from (see appendix~A): 

\begin{equation}
{\rm Im} (\xi_t^i) \equiv {\rm Im}(Z_t^{1i*}Z_t^{2i}) = \frac{(-1)^{i-1}}{2}
\sin2\theta_t \sin\beta_t \label{xit}~. 
\end{equation}      

We will choose maximal CP-violation in the sense that ${\rm Im} (\xi_t^i)=
(-1)^{i-1}/2$ thus presenting all our numerical results in units of $
\sin2\theta_t 
\sin\beta_t$. With no further assumptions, what is left to be considered
is the NEDM for a chosen set of the free parameters of the above
phenomenological low energy MSSM\null. 
 
The EDM of the neutron, $d_n$, is presumably one of the most important
phenomenological 
problems associated with SUSY models especially with regard to CP violation.
With a low energy MSSM that originates from a GUT-scale SUGRA model (with
complete universality of the soft breaking terms), keeping $d_n$ within its
allowed experimental value (i.e., $d_n \leq 1.1\times 10^{-25}$  e-cm
\cite{prd}) requires the ``fine tuning'' of the SUSY phases to be less then
or of the order of $10^{-2}-10^{-3}$ for SUSY particle masses at around
several hundred GeV's \cite{garisto}. 

When $\arg(\mu)=0$, the leading contribution to a light quark EDM comes from
gluino exchange, which, with the approximation of degenerate $\tilde u$ and
$\tilde d$ squark masses (which we will denote by $m_{\tilde q}$), can be
written as \cite{garisto,oshimo}:

\begin{equation}
d_q(G)=\frac{2 \alpha_s}{3\pi} Q_qe m_q \frac{|A_q|\sin\alpha_q}{m_{\tilde
q}^3} {\sqrt r} K(r) \label{qedm} ~, 
\end{equation}  

\noindent where $m_q$($m_{\tilde q}$) is the quark(squark) mass and $Q_q$
is its charge. Also, $r \equiv m_G^2/m_{\tilde q}^2$ and $K(r)$ is given
by \cite{oshimo,footnedm}: 

\begin{equation}
K(r)= \frac{1}{(r-1)^3}\left( \frac{1}{2}+\frac{5}{2}r + \frac{r(2+r)}{1-r}{\rm
ln}r \right) \label{kr}~. 
\end{equation} 

\noindent $A_q$ is the complex trilinear soft breaking term at the EW scale
associated with the squark $\tilde q$, and we have defined $A_q=|A_q|e^{i
\alpha_q}$.  Then, within the naive Quark Model, the NEDM can be obtained by
relating it to the $u$ and $d$ quarks EDMs (i.e., $d_u$ and $d_d$, respectively)
by $d_n=(4d_d-d_u)/3$. 

We stress again that with no universal boundary conditions for the soft
trilinear terms and their phases at the GUT scale (i.e.,
$\arg(A_U)=\arg(A_D)=\arg(A_E)=\arg(A^G)$), 
there is no {\it a-priori\/} reason to believe that the low energy phases
associated with the different $A_f$ soft breaking terms are related at the
EW scale. We therefore consider $\arg(A_u)$ and $\arg(A_d)$ to be free
parameters 
of the model no matter what $\arg(A_t)$ is.  In Figs.~3a and 3b we have plotted
the allowed regions in the $\sin\alpha_u - \sin\alpha_d$ plane for $|d_n|$
not to exceed $1.1\times 10^{-25}$ e-cm (the present experimental limit)
and $3\times 10^{-25}$ e-cm, respectively. In calculating $d_n$ we assumed
that the above naive Quark Model relation holds. Although there is no doubt
that it can serve as a good approximation for an order of magnitude estimate
it may still deviate from the true theoretical value which involves
uncertainties in the calculation  of the corresponding hadronic matrix elements
(see \cite{grossman} and references therein). Note also that it was recently
argued that the naive quark model overestimates the NEDM, as the strange
quark may carry an appreciable fraction of the neutron spin which  can partly
screen the contributions to the NEDM coming from the $u$ and the $d$ quarks
\cite{ellis}. To be on the safe side, we therefore slightly relax the
theoretical limit on $d_n$ in Fig.~3b to be $3\times 10^{-25}$ e-cm.   

We have used, for these plots, $m_{\tilde d}=m_{\tilde u}=M_s=400$ GeV,
$m_G=500$ 
GeV and for simplicity we also took $|A_u|=|A_d|=M_s$ ($M_s$ is the only
high SUSY mass scale associated with the squarks sector in our low energy
MSSM\null. Therefore it is only  natural to choose the mass scale of the
soft breaking terms according to $M_s$). Also, we took the current quarks
masses as $m_d=10$ MeV, $m_u=5$ MeV and $\alpha_s(m_Z)=0.118$.   

From Fig.~3a and in particular Fig.~3b, it is evident that $M_s=400$ GeV
and $m_G=500$ GeV can be safely assumed, leaving ``enough room'' in the
$\sin\alpha_u 
- \sin\alpha_d$ plane for $|d_n|$ not to exceed $1.1-3 \times 10^{-25}$ e-cm.
In particular, while $\sin\alpha_u$ is basically not constrained, depending
on $\sin\alpha_u$, $-0.35 \lsim \sin\alpha_d \lsim 0.35$ is needed for $|d_n|
< 1.1 \times 10^{-25}$ e-cm and $-0.55 \lsim \sin\alpha_d \lsim 0.55$ is
needed for $|d_n| < 3 \times 10^{-25}$  e-cm. Moreover, varying $m_G$ between
250 GeV to 650 GeV  (we will vary $m_G$ in this range when discussing the
numerical results below) almost has no effect on the allowed areas in the
$\sin\alpha_u 
- \sin\alpha_d$ plane that are shown in Figs.~3a and 3b. That is, keeping
$M_s=400$ GeV and lowering $m_G$ down to 250 GeV, very slightly shrinks the
dark areas in Figs.~3a and 3b, whereas, increasing $m_G$ up to 650 GeV slightly
widens them. Of course, $d_n$ strongly depends on the scalar mass $M_s$ -
increasing $M_s$ enlarges the allowed regions in Figs.~3a and 3b as expected
from Eq.~\ref{qedm}. It is also very interesting to note from Fig.~3a that,
in some instances, for a cancellation between the $u$ and $d$ quarks
contributions to apply, $\sin\alpha_u,\sin\alpha_d > 0.1$ is essential rather
than being just possible. For example, with $|\sin\alpha_u| \gsim 0.75$,
$|\sin\alpha_d| 
\gsim 0.1$ is required in order to keep $d_n$ below its experimental limit.

We can therefore conclude that CP-odd phases in the $A$ terms at the order
of ${\rm few} \times  10^{-1}$ can be accommodated without too much difficulty
with the existing experimental constraint on the NEDM even for typical SUSY
masses of $\lsim 500$ GeV\null. Therefore, somewhat in contrast to the commonly
held viewpoint we do not find that a ``fine-tuning''
at the level of $10^{-2}$ is necessarily 
required for the SUSY CP-phases nor for the
squark masses. \\ 
\bigskip
\bigskip
\bigskip

\noindent {\bf 5.\underline{ Numerical Results}}\\

We now turn to the discussion of our main numerical results. We first focus
on CP-violating effects in the production amplitude which gives rise to the
cross-section asymmetry, $A_P$, and then we will present our results for
the PRA,   
%ASf
%effect, 
$A_D$, associated with the top decay $t \to W^+b$ and its
conjugate one. Instead of using the asymmetry $A_P$ defined in Eq.~\ref{a0},
after folding in the parton luminosities in the usual manner (see
\cite{collider}), 
we define a Partially Integrated Cross-section Asymmetry (PICA) with respect
to the variable $\tau$ or equivalently  to $\hat s$ ($\hat s=\tau s$). Thus: 

\begin{equation}
A_P^{\rm PICA} \equiv \frac{ \int_{{\tau}_{-}}^{\tau_+} \left[ \frac{d{\cal
L}_{ud}(\tau)}{d\tau} \left( {\hat \sigma}({\hat s}=\tau s) - {\hat {\bar
\sigma}}({\hat s}=\tau s) \right) \right] d\tau}{ \int_{{\tau}_{-}}^{\tau_+}
\left[ \frac{d{\cal L}_{ud}(\tau)}{d\tau} \left( {\hat \sigma}({\hat s}=\tau
s) + {\hat {\bar \sigma}}({\hat s}=\tau s) \right) \right] d\tau} \label{a0dif}
~, 
\end{equation}  
    
\noindent where we have introduced the ``parton luminosity'' ${d{\cal
L}_{ud}(\tau)}/{d\tau}$ 
\cite{collider}. We also define the invariant mass of the $tb$ system to
be $m_{tb} \equiv \sqrt{\hat s}$. Then, in the numerical evaluation of the
asymmetry $A_P^{\rm PICA}$ we choose $\tau_{-}$ and $\tau_{+}$ according
to $m_{tb}^-=m_t$ ($m_b=0$ is assumed through out the paper) and $m_{tb}^+=350$
GeV, respectively. In the Tevatron, one of the  problematic ``backgrounds''
to the reaction $p \bar p \to t \bar b +X$ is $t\bar t$ production
\cite{stelzer1}. 
In principle, the $t\bar t$ background can be eliminated by imposing the
above naive upper cut on $m_{tb}$, which is below the $t\bar t$ production
threshold (i.e., $m_{tb}^+=350$ GeV $\lsim 2m_t$). Although the actual
experimental 
cuts that will be made in order to remove the $t\bar t$ ``background'' in
the Tevatron (when studying the $t\bar b$ final state) might be more involved,
this naive cut on $m_{tb}$ serves our purpose. A detailed discussion of the
exact experimental cuts for the $t\bar b$ final state is beyond the scope
of this paper but can be found in \cite{heinson1,stelzer1}. Note also that
this upper cut on $m_{tb}$ has practically no effect on the cross-section
for $p \bar p \to t \bar b +X$ (or equivalently the available number of $t
\bar b$ events in the future 2 TeV Tevatron),  as most of the $t\bar b$ pairs
will be produced with an invariant mass of $m_{tb} \lsim 350$ GeV\null. In
particular, 
the fully integrated cross-section for $p \bar p \to t \bar b +X$ is (i.e.,
up to $m_{tb}=2$ TeV) $\sigma(s=2~ {\rm TeV})\approx 360~[{\rm fb}]$
\cite{heinson1}, 
out of which $\sim 80\%$ of the $t \bar b$ pairs are produced with an invariant
mass $m_{tb} \lsim 350$ GeV\null.   

For the numerical evaluation of the CP-violating asymmetry $A_P^{{\rm PICA}}$,
we take $M_s=400$ GeV\null. We also vary the gluino mass between $250~{\rm GeV}<
m_G<650~{\rm GeV}$ (as was emphasized in the previous section, these masses
of the degenerate squarks  and the gluino do not contradict the existing
experimental upper limit on the NEDM). We choose two representative values
for $\tan\beta$; $\tan\beta=1.5$ and $\tan\beta=35$ which correspond to a
%ASf
low and a high $\tan\beta$ scenario, respectively. The low energy Higgs mass
parameter $\mu$ is varied between -400 to +400 GeV, whereas, the mass of
%ASf
the light stop ($m_l$) is varied between 50 to 400 GeV\null. 
As mentioned above,
we also choose maximal CP-violation, driven by ${\tilde t}_L - {\tilde t}_R$
mixing, in the sense that $\xi_t^i=(-1)^{1-i}/2$ (see Eq.~\ref{xit}), whereas,
we take $\arg(\mu)=0$. Therefore the asymmetries are always given in
units  of $\sin2\theta_t \sin\beta_t$. 

The consequences of this low energy MSSM scenario (with the above chosen
mass spectrum and CP-odd phases) with regard to the various diagrams that
are depicted in Fig.~2 are: \\ 
I) The triangle diagrams $2c,2b^{\prime},2c^{\prime}$ and $2d^{\prime}$ as
well as the box diagrams $2f$ and $2h$ do not acquire any CP-violating phase
in the limit $\arg(\mu)=0$.\\ 
II) Diagram $2a^{\prime}$ does not have any CP-violating phase in the limit
$m_u,m_d \to 0$.\\ 
III) Diagram $2a$ does not have an absorptive cut for $m_{tb}<350$
GeV\null.\\ 
IV) Diagram $2b$ is $p$-wave suppressed near threshold. Thus its contribution
to $A_P^{{\rm PICA}}$ is negligible compared to the other diagrams.\\   
   
With the above points I--IV, the study of CP-violation in this reaction
simplifies to a large extent.  That is, we are left with only three CP-violating
diagrams that can 
%ASf
make important
contributions to $A_P^{{\rm PICA}}$. These are: diagram $2d,2e$ and
%ASf
$2g$ out of which diagram $2d$ is expected to be the most important one.
In particular, in the range where the asymmetry exceeds the $1\%$ level,
diagram $2d$ is responsible for $\sim 90\%$ of the total CP-odd effect in
$A_P^{{\rm PICA}}$.Moreover, it is 
 important to note that while the CP-violating effects from
diagrams 2e and 2g strongly depend on the heavy scalars mass $M_s$ (through
two scalar exchanges in the loop), diagram 2d is less sensitive to $M_s$
having only one scalar exchange  in the loop. In particular, as $M_s$ increases,
diagram 2d becomes relatively more dominant as far as CP is concerned (Of
course, in general, increasing the SUSY scale $M_s$ causes the asymmetry
to drop for 
diagrams 2d as well as for diagrams 2e and 2g). It is worth mentioning that,
in that range where the asymmetry is above the percent level mainly due to
the CP-effects coming from diagram $2d$, the CP-violating effect driven by
diagrams $2e$ and $2g$ has the same relative sign to that of diagram 2d,
thus yielding a slightly bigger overall asymmetry. Nonetheless, for 
simplicity, we will present all our numerical results only for diagram 2d.
In any case,  we are only interested in a rough estimate of the CP-violating
effect 
in the reaction $p\bar p \to t \bar b+X$ and not in the exact numbers, as
many simplifying assumptions had to be made along the way. 
 
In Figs.~4a and 4b we have plotted $A_P^{{\rm PICA}}$ for $M_s=400$ GeV,
$m_l=50$ GeV and for three values of $m_G$ as a function of the low energy
Higgs mass parameter $\mu$, for $\tan\beta=1.5$ and $\tan\beta=35$, respectively
\cite{footlines,footcuts}.  
%ASf
We see from Fig.~4a that in the low $\tan\beta$ scenario and a gluino mass
between $350~ 
{\rm GeV}\lsim m_G \lsim 550$ GeV, moderately negative values of $\mu$, 
%ASf
lying in a narrow range,
$-100~
{\rm GeV}\lsim \mu \lsim -70$  GeV, are required for the asymmetry
to be above 2\%. In the best case, for $m_G=450$ GeV the asymmetry peaks
at around $\mu \approx -90$ GeV, reaching $\sim 2.75\%$.    
A positive $\mu$ around 140(240) GeV may also give rise to a $\sim 2(1)\%$ 
%ASf
asymmetry for a gluino mass $m_G \approx 550$ GeV\null. 
Also, for $\mu \lsim -200$ GeV and $\mu \gsim 350$ GeV 
the asymmetry drops below the 0.25\% level. 
%ASf
From Fig.~4b (high $\tan(\beta$) we see that around 
$|\mu| \sim 110$ GeV the asymmetry can also reach the $2\%$ level for $ 400~{\rm
GeV} \lsim m_G \lsim 575$ GeV (see also Fig.~6b). Here, the asymmetry drops
below the 0.25\% level for $|\mu| \gsim 250$ GeV\null. 
Notice that, in the high $\tan\beta$ scenario, $A_P^{{\rm PICA}}$ is almost
insensitive to the sign of $\mu$. This happens mainly due to the fact that,
for  high values of $\tan\beta$,
the charginos and neutralinos masses are almost independent of the sign of
$\mu$ as can be also seen in Table~1. The only terms in the chargino and
neutralino masses which linearly depend on $\mu$ are proportional to $\sin2\beta$
which is of the order of ${\rm few} \times 10^{-2}$ for $\tan\beta \gsim
30$ (see Eq.~A.22 and A.25--A.32).

Before discussing the results in Figs.~5--11 we remark that, in what follows,
we choose several representative values of $\mu$; some of them maximize
CP-effect in the production amplitude and some maximize the CP-violating
PRA effect in the decay $t\to bW$ to be discussed later. In particular, for
$\tan\beta =1.5$ we always take the values $\mu=-70,-90,-130,140$ and 240
GeV, while for $\tan\beta =35$ we choose $\mu=-110,-170,-190,110$ and 170
GeV, although, in some cases the combination of $\left\{ \tan\beta,\mu,m_G
\right\}$ for some of those values is forbidden according to our criteria
(see section~2). It is useful to keep track of the charginos and neutralinos
masses for a given set of $\left\{ \tan\beta,\mu,m_G \right\}$, especially
due to the various absorptive thresholds that can emanate from the loop integral
of diagram 2d \cite{footcuts}. Therefore, for the reader's convenience, we
give in Table~1 the masses of the charginos and neutralinos for various sets
of $\left\{ \tan\beta,\mu,m_G \right\}$ that are repeatedly being used throughout
this analysis.    

In Figs.~5a and 5b we show the dependence of $A_P^{{\rm PICA}}$ on $m_l$,
the mass of the light stop, for various values of $\mu$, $M_s=400$ GeV and
for $\tan\beta=1.5$ and $\tan\beta=35$, respectively.
However, the general behavior that is depicted
in Figs.~5a and 5b holds for any value of $\mu$. That is, as expected, when
$m_l$ is increased the asymmetry falls till it totally vanishes for $m_l=400$
GeV in which case the two stop species are degenerate. It is evident from
Figs.~5a and 5b that, in general, $m_l \lsim 75$ GeV is needed for the asymmetry
to be above $\sim 1\%$, although in some cases, for example when $\mu= -70$
GeV and $\tan\beta=1.5$, 
 a 1\% asymmetry can arise even with $m_l \gsim 100$ GeV\null.    

The dependence of $A_P^{{\rm PICA}}$ on the gluino mass $m_G$ is shown in
Figs.~6a and 6b for $\tan\beta=1.5$ and $\tan\beta=35$, respectively. Here
again, we took $M_s=400$ GeV and $m_l=50$ GeV and choose the same
 representative values for $\mu$. We see that for a small $\tan\beta$ and
 a negative $\mu$ around $-100$ GeV, $A_P^{{\rm PICA}} \gsim 2.5\%$ for $350~
 {\rm GeV}\lsim m_G \lsim 550$ GeV and peaks at around $m_G \sim 400$ GeV\null.
Note that even a 
%ASf
%very 
heavy gluino, 
%ASf
e.g., $m_G \approx 650$ GeV, can give rise to a $\gsim 2\%$ asymmetry if
$\mu \approx -90$ GeV and $\tan\beta$ is of order one.   
We see from Fig.~6b that for large $\tan\beta$ and for $|\mu|=110$ GeV,
$A_P^{{\rm PICA}} \sim 2\%$ for $400~ {\rm GeV}\lsim m_G \lsim 550$ GeV and 
peaks at $m_G \sim 450$ GeV\null. In both the low and high $\tan\beta$ scenarios,
with a ``light'' 
gluino, $m_G \lsim 300$ GeV, the cross-section asymmetry is below $1\%$.

Figs.~7a and 7b show the dependence of $A_P^{{\rm PICA}}$ on $\tan\beta$
for $m_G=350$ and 550 GeV, respectively. Evidently, as far as CP is concerned,
%ASf 
a low $\tan\beta$ is better then a high one. We see that as one goes to $\tan\beta
\gsim 10$, the asymmetry is almost insensitive to $\tan\beta$. This also
holds for high values up to $\tan\beta=65$ which are not shown in Figs.~7a
and 7b. 
      
Until now we were not interested in the overall sign of the cross-section
asymmetry $A_P^{{\rm PICA}}$. Including the PRA CP-violating effect,
$A_D$, coming from the decaying top, the relative sign between $A_P^{{\rm
PICA}}$ and $A_D$ becomes important as the total CP-violating effect is the
sum of the two i.e., $A=A_P^{{\rm PICA}}+A_D$. However, we will show below
that, in most instances, $A_D$ is smaller than $A_P^{{\rm PICA}}$ by more
than an order of magnitude and, therefore, its relevance to the CP-effect
in $p \bar p \to t \bar b +X \to  W^+ b \bar b +X$ is negligible.   

Let us now discuss the PRA effect, $A_D$. Once again, we will use our low
energy MSSM scenario (described in detail in section~2 and 4) to
estimate the CP-odd effect. Although, as was mentioned before, the PRA $A_D$
%ASf
%is an independent quantity which 
does not depend on the specific production
mechanisms of the $t$ and $\bar t$, in most 
%ASf
%of the 
cases we will evaluate its magnitude within
%ASf
the same ranges of the SUSY free parameter space as was taken in Figs.~4--7.
%ASf
This should enable the reader to easily extract the overall asymmetry emanating
from both production and decay processes discussed in this paper. 
However, as was mentioned above, throughout most of the range of our SUSY
parameter space, $A_D$ is more than one order of magnitude smaller than $A_P^{{\rm
PICA}}$, therefore,   the relative sign between $A_D$ and $A_P^{{\rm PICA}}$
is essentially irrelevant. Note that
the consequences of the low energy MSSM framework on the various diagrams
(i.e., the ``reversed''
diagrams $a-d$ in Fig.~2) that can potentially contribute
to $A_D$ are: \\ 
I) For $m_G > 250$ GeV diagram $2a$ does not have the needed absorptive cut
and thus, does not contribute to $A_D$.\\ 
%ASf
II) As in the case of production, diagram $2c$ does not have a CP-violating
phase in the limit $\arg(\mu)=0$. 
We are therefore left with only two diagrams that can contribute to $A_D$.
These are: diagram $2b$ and $2d$, where the leading contribution to $A_D$
again comes from diagram $2d$.  

%ASf
Our main results for $A_D$ are shown in Figs.~8--11. A quick look at
Figs.~8--11 reveals that,  
%ASf
throughout a large portion of the SUSY parameter space discussed here, 
$A_D$ tends to be below $0.1\%$, although for some specific values 
of the parameters it can approach $0.3\%$.
%with our chosen range of the SUSY free parameter
%space, $|A_D|<0.3\%$, where in fact it is below $0.1\%$ throughout a large
%portion of the SUSY parameter space discussed here. 
The asymmetry we find is therefore somewhat small compared
to the estimates of Grzadkowski and Keung (GK) and 
of Christova and Fabbrichesi (CF) in \cite{cristova}. In the GK limit only
the gluino exchange of diagram 2a was considered in which case $m_G \lsim
120$ GeV is required in order to have the necessary absorptive cut (when
$m_l=50$ GeV).  In the
best case, GK found a $\sim 1\%$ asymmetry for $m_G=m_{\tilde b}=100$ GeV
\cite{footgk}. On the other hand, in the CF limit, numerical results were
given only for the neutralino exchange diagram (i.e., diagram 2b) wherein
the CP-phase was chosen to be proportional to $\arg(\mu)$ and maximal
CP-violation with regard to $\arg(\mu)$ was taken \cite{footeinhorn}. In
the best case, CF found a $\sim 2\%$ asymmetry for $m_{\tilde b}=100$ GeV\null.
However, each of those largish PRA asymmetries, reported by GK and CF in
\cite{cristova}, suffer from two drawbacks. ($i$) For the GK limit, $m_G
\lsim 120$ GeV is now essentially  disallowed by the current experimental
bounds. ($ii$) For the CF limit, $\arg(\mu) \gsim 10^{-2}$ is an unnatural
choice bearing the stringent constraints on this phase coming from the
experimental 
limits on the NEDM as discussed in section~4. ($iii$) For both the GK and
CF limits, the large asymmetry arises once the masses of the superpartners
of the light quarks are set to 100 GeV\null. Again, this is a rather
unnatural choice  as it is theoretically very hard, if at all possible, to
meet the NEDM experimental limits when the masses of the squarks (except
for  the lighter stop) are at the order of 100 GeV\null. Instead, we discuss
below the PRA effect, $A_D$, including all possible diagrams and subject
to the available experimental bounds on both the NEDM and the supersymmetric
spectrum. In particular, we again take $\arg(\mu)=0$, $m_{\tilde q}=M_s=400$
GeV, $m_l  > 50$ GeV,  $m_G > 250$ GeV,
the LSP mass to be above 20 GeV and the mass of the lighter
chargino to be above 65 GeV (see sections~2 and 4) \cite{footlines}. 

As can be seen from Figs.~8a, 9a and 10a, in the low $\tan\beta$ case (i.e.,
$\tan\beta=1.5$) and for negative values of $\mu$ in the range $-90~{\rm
GeV} \lsim \mu \lsim -30~{\rm GeV}$,  
%ASf
$A_D$ can be about 0.1--0.2\% provided the gluino mass lies in the
range $300~{\rm GeV} \lsim m_G \lsim 550~{\rm GeV}$ and the lighter stop
mass $(m_\ell)$ is  
%ASf
between $\sim 50$ and $\sim 70$ GeV\null. 
%ASf
In fact, $A_D$ can even reach $\sim 0.3\%$ 
if in addition to $m_l$ being in that narrow range we also have
$m_G \approx 320$ GeV\null. As $\mu$ becomes positive, at $\mu \approx 150$ GeV,
$A_D$ tends to grow with the gluino mass. In particular, we find for example
that for $\mu \approx 140$ GeV and $m_G \gsim 600$ GeV, $A_D \sim 0.2\%$
becomes possible.
The results for the high $\tan\beta$ case (i.e., $\tan\beta=35$) are shown
in Figs.~8b, 9b and 10b. Evidently, with a high $\tan\beta$, $|A_D|<0.1\%$
throughout all the range of our SUSY parameter space.   

The dependence of $A_D$ on the lighter stop mass, $m_l$, is shown in Figs.~9a
and 9b. As expected, $A_D$ drops as $m_l$ is increased and vanishes for $m_l
> m_t -m_{\rm LSP}$ ($m_{\rm LSP}$ is the LSP mass), for which case there
is no absorptive cut in any of the diagrams $a-d$ in Fig.~2 \cite{footcuts}. 

Finally, in Figs.~11a and 11b we evaluate $A_D$ as a function of $\tan\beta$
for $m_G=350$ and 550 GeV, respectively. We can see that, here also, a small
$\tan\beta$ gives rise to a bigger asymmetry and, as $\tan\beta$ grows above
$\sim 10$, $A_D$ becomes  insensitive to it. In the best cases, for $\tan\beta
\sim {\cal O} (1)$, $|A_D| \sim 0.3\%$ become possible for both $m_G=350$ and 550 GeV\null.         

Before summarizing, we wish to emphasize that the decay asymmetry $A_D$ is
independent of the production mechanism. Therefore, the LHC with its large
$(\sim10^7)$ production rate for $t\bar t$ pairs could be a good place to
search for $A_D$. To probe such a PRA signal
in the top decay $t \to b W$, of the order of a few tenths of a percent, will
naively require $\sim 10^6$ $t \bar t$ pairs when
 no efficiency factors are taken into account. Note that such a measurement
 requires only the detection of the charge of the top for which the systematic
 errors may be kept relatively small. Even if we take an efficiency overall
 factor of $\sim 0.1$, then with $\sim 10^7$ $t \bar t$ pairs the reach of
 the LHC will be around $A_D\gsim0.3\%$. Although, as we discussed in the
 preceding pages, in our SUSY study it appears difficult to attain that large
 a decay asymmetry, the experimental search would still be worthwhile.
 \\     
\noindent{\bf 6. \underline{Summary and Conclusions}}\\

%ASf
To summarize, we found that within a portion of a plausible MSSM low
energy parameter space, a CP-violating cross-section asymmetry in the reaction
$p \bar p \to t \bar b +X$ can be at the level of a few percent. Furthermore,
a CP-violating PRA effect in the subsequent top decay cannot exceed the 0.3\%
level throughout our chosen range of the MSSM free parameter space. Therefore,
the CP-violating asymmetry in this reaction arises predominantly from the
production vertex.  These asymmetries, in production
and decay of the top, are driven by the complex entry of the soft trilinear
breaking term associated with the top, $\arg(A_t)$.  In particular, we have
shown that for scalar SUSY masses as well as a gluino mass of the order of
0.5 TeV, and with a Higgs mass parameter, $\mu$,
at the range of $-250~{\rm
GeV} \lsim \mu \lsim 250~ {\rm GeV}$ a CP-violating signal above the percent
level  might indeed arise in the reaction $p \bar p \to t \bar b +X \to W^+
b \bar b +X$, provided that the light stop particle have a mass below 100
GeV\null.  

We have also shown that this phenomenologically
 acceptable low energy MSSM
scenario does not contradict the experimental limit on the NEDM (i.e., $d_n
< 1.1\times 10^{-25}$ e-cm). In particular, in our low energy MSSM framework,
SUSY CP-odd phases in the trilinear soft breaking terms ($A_f$'s) at
the order of ${\rm few} \times 10^{-1}$ are allowed even with squark masses
below 500 GeV\null. The theoretical uncertainties involved in calculating
the NEDM may even strengthen this statement. In this respect our findings
are somewhat different from what is commonly claimed in the literature. 

%ASf... the para below has many changes
In the future 2 TeV $p \bar p$ Tevatron collider, the cross-section for $p
\bar p \to t \bar b +X$ is expected to be about 300 (fb) if a cut of
$m_{tb}<350$ 
GeV is applied on the invariant mass of the $t \bar b$. Therefore with an
integrated luminosity of ${\cal L} = 30$ (fb)$^{-1}$ \cite{heinson1,stelzer1}
an asymmetry $A_{\rm PICA} \sim 3\%$, which was found here
in the best cases, can be naively detected 
with a statistical significance of $3\sigma$. We must caution, however, that
the values of the asymmetry as large as 3\% appear attainable only for some
specific range for some of the input parameters. Note also such a detection
at the Tevatron will require
%ASf
the identification of all $t\bar b$ pairs , which, in principle,
can be achieved only if the top can be reconstructed even when the $W$ decays
hadronically. Therefore, at the future 2 TeV $p \bar p$ collider, a percent
level CP-violating signal in the reaction $p \bar p \to t \bar b +X \to W^+
b \bar b +X$ may  become accessible. We have also emphasized that the PRA
effect in the decay $t \to b W$ may be measurable at the LHC\null. 

%ASf
Note that the above line of reasoning can be reversed. That is, one can ask: what
limit can be obtained on $\arg(A_t)$ by studying CP-violation in the reaction
$p \bar p \to t \bar b +X \to W^+ b \bar b +X$ at the next runs of the Tevatron or in the decay $t \to b W$ at the LHC?
We found that in the best case, the 2 TeV Tevatron with ${\cal L} = 30$ (fb)$^{-1}$
will be marginally sensitive to CP-odd effects coming from $\arg(A_t)$ and,
taking into account the various efficiency factors for such a detection,
perhaps a $\sim 1-2$-sigma bound on $\arg(A_t)$ will be feasible.
However, note that in a possible Tevatron upgrade with center of mass energy
of 4 TeV \cite{heinson1}, the production rate of $t \bar b$ becomes three
times bigger and, therefore, in such an upgraded version of the Tevatron a
more stringent upper limit may indeed be placed on $\arg(A_t)$ through the
study of the $t \bar b$ cross-section as suggested here. The LHC may also
be sensitive to $\arg(A_t)$ through a study of a PRA effect in the top decay
%ASf
$t \to b W$ if the SUSY parameter space lies in the window wherein the PRA
effect reaches its maximum values i.e., 0.1--0.3\%. In such a scenario a statistically
significant limit on $\arg(A_t)$ may be within its reach.    
      
In closing, we wish to remark that it will be useful to explore the SUSY
mediated CP-violating effects that can emanate in the $W$-gluon fusion
subprocess 
which contributes to the same final state (i.e., $W g \to t \bar b d$) and
which has a comparable production rate to that of the simple $u \bar d \to
t \bar b$ in the 2 TeV Tevatron. Unlike in the case of the 2HDM \cite{tbsusy},
where, to one-loop order, no CP-violating corrections enter the $W$-gluon
subprocess, within the MSSM, various one-loop triangle and box corrections
can give rise to CP-nonconservation also in the $W$-gluon fusion subprocess.
However, it is also important to emphasize that due to the extra light jet in
$Wg \to t \bar b d$ and the different kinematics, in principle, the $t \bar
b$ pairs originating from the $W$-gluon fusion can be distinguished
from those produced via the $u \bar d$ fusion \cite{heinsonp}.  

\begin{center}
{\bf Acknowledgments}\\
\end{center}

We thank Gad Eilam for discussion and advice. 
We must also thank Galit
Eyal for checking many of the calculations. S.B. also thanks Ann Heinson for discussions. We acknowledge partial support
from U.S. DOE contract numbers DE-AC02-76CH00016(BNL), 
DC-AC05-84ER40150(Jefferson Lab) and DE-FG03-94ER40837(UCR).

\pagebreak
  
\begin{center}
{\bf Appendix A}\\
\end{center}

In this appendix we list the relevant pieces of the SUSY lagrangian written
in terms of the mass eigenstates. We also give the analytical expressions
for all the corresponding mass mixing matrices $Z_f, \ Z_N, \ Z^+, \ Z^-$
and the masses of the supersymmetric particles which are functions of these
matrix elements. 

The SUSY lagrangian pieces are \cite{rosiek}:
\setcounter{num}{1}
\setcounter{equation}{0}
\def\theequation{\Alph{num}.\arabic{equation}}

\begin{eqnarray}
{\cal L}_{\tilde{u}_i\tilde{d}_jW} &=& - \frac{ig}{\sqrt{2}}   %8
Z^{1j}_d Z^{1i}_u V^{ud} 
(\tilde{d}^+_j \stackrel{\leftrightarrow}{\partial^\mu} \tilde{u}_i^+)W^-_\mu
+ H.c. \label{feynman1} ~,\\ 
{\cal L}_{\tilde{u}_i d \tilde{\chi}_m} &=& \tilde{u}_i^+ \bar{d}   %9
\left\{\left[-g Z^{1i}_u Z^{+*}_{1m} + \frac{\sqrt{2}m_u}{v_2}
Z^{2i}_u Z^{+*}_{2m}\right] \right.R + \nonumber \\
&&\left. \frac{\sqrt{2}m_d}{v_1} Z^{1i}_u Z^-_{2m} L\right\} V^{ud}
\tilde{\chi}_m^c + H.c. ~,\\ 
{\cal L}_{\tilde{d}_j u \tilde{\chi}_m} &=&                          %10
- \tilde{d}_j^+ \bar{\tilde{\chi}}_m\left\{\left[ gZ^{1j}_d Z^-_{1m} -
\frac{\sqrt{2}m_d}{v_1} Z^{2j}_d Z^-_{2m}\right]L-\right. \nonumber \\
&&\left. \frac{\sqrt{2}m_u}{v_2} Z^{1j}_d Z^{+*}_{2m} R \right\} V^{ud} u
+ H.c. ~,\\ 
{\cal L}_{\tilde{u}_i u\tilde{\chi}_n^0} &=&   \tilde{u}^-_i
\bar{\tilde{\chi}}^0_n 
%11 
\left\{\left[-\frac{g}{\sqrt{2}} Z^{1i*}_u L^+ -
\frac{\sqrt{2}m_u}{v_2} Z^{2i*}_u Z^{4n}_N\right]L+\right. \nonumber \\
&&\left. \left[\frac{2\sqrt{2}}{3} g \tan \theta_W Z^{2i*}_u Z^{1n*}_N -
\frac{\sqrt{2}m_u}{v_2}Z^{1i*}_u Z^{4n*}_N\right]R\right\}u + H.c. ~,\\
{\cal L}_{\tilde{d}_j d\tilde{\chi}_n^0} &=&         %12
\tilde{d}_j^+ \bar{\tilde{\chi}}^0_n \left\{\left[-\frac{g}{\sqrt{2}}Z^{1j}_d
L^- - \frac{\sqrt{2}m_d}{v_1} Z^{2j}_d Z^{3n}_N\right]L+\right. \nonumber \\
&&\left. \left[-\frac{\sqrt{2}}{3} g \tan \theta_W Z^{2j}_d Z^{1n*}_N -
\frac{\sqrt{2}m_d}{v_1}Z^{1j}_d Z^{3n*}_N\right]R\right\}d+ H.c. ~,\\
{\cal L}_{\tilde{u}_i u\tilde{g}} &=& \sqrt{2}g_s \tilde{u}_i^-
T^a\bar{\tilde{g}}^a 
%13 
\left[-Z_u^{1i*} L + Z^{2i*}_u R\right] u + H.c.  ~,\\
{\cal L}_{\tilde{d}_j d\tilde{g}} &=& \sqrt{2}g_s \tilde{d}_j^+
T^a\bar{\tilde{g}}^a 
%14 
\left[-Z_d^{1j} L + Z^{2j}_d R\right] d + H.c.  ~,\\
{\cal L}_{W \tilde{\chi}_m\tilde{\chi}^0_n} &=& g \bar{\tilde{\chi}}_m \gamma^\mu
%15 
\left\{ K^- L + K^+ R \right\} \tilde{\chi}^0_n W^+_\mu + H.c. \label{feynman2}
~, 
\end{eqnarray}

\noindent where $L(R) = \frac{1}{2}(1-(+)\gamma_5)$ and the $\tilde{u}_i$
and $u$ ($\tilde{d}_j$ and $d$) stand for up squark and up quark (down squark
and down quark), respectively. Also $\tilde{\chi}_m, \ \tilde{\chi}^0_n$
and $\tilde{g}$ are the charginos, neutralinos and gluinos respectively.
We have also defined:

\begin{eqnarray} 
L^{\pm} &\equiv& \frac{1}{3} \tan\theta_W Z_N^{1n} \pm Z_N^{2n} \label{lpm}
~, \\ 
K^+ &\equiv& Z_N^{2n*} Z_{1m}^- + \frac{1}{\sqrt 2} Z_N^{3n*} Z_{2m}^-
\label{kp} 
~, \\ 
K^- &\equiv& Z_N^{2n} Z_{1m}^{+*} - \frac{1}{\sqrt 2} Z_N^{4n} Z_{2m}^{+*}
\label{km} ~, 
\end{eqnarray}  

\noindent and the mixing matrices $Z_u, \ Z_d, \ Z_N, \ Z^-$ and $Z^+$ are
given below. 

Let us define the following diagonalizing mass matrices 
\cite{oshimo,haber,rosiek,barger,shafik,gunion}:

\begin{eqnarray}
&&Z^+_f M^2_{\tilde{f}} Z_f
= {\rm diag}\left(m^2_{\tilde{f}_1}, m^2_{\tilde{f}_2}\right) \ \ ,\\
&&(Z^-)^TM_{\tilde{\chi}} Z^+
= {\rm diag}\left(m_{\tilde{\chi}_1}, m_{\tilde{\chi}_2}\right) \ \ ,\\
&&Z_N^T M_{\tilde{\chi}^0} Z_N
= {\rm diag}\left(m_{\tilde{\chi}^0_1}, m_{\tilde{\chi}^0_2},
m_{\tilde{\chi}^0_3}, m_{\tilde{\chi}_4^0}\right) \ \ ,  %A3
\end{eqnarray}

\noindent where $M^2_{\tilde{f}}$ is the mass squared matrix of the scalar
partners of a fermion. $M_{\tilde{\chi}}$ and $M_{\tilde{\chi}^0}$ are the
mass matrices of the charginos and neutralinos, respectively.

$M^2_{\tilde{f}}$ is then given by \cite{oshimo,haber,rosiek}:

\begin{equation}
M^2_{\tilde{f}} = \left[\matrix{
m^2_f - \cos 2 \beta(T_{3f} - Q_f \sin^2 \theta_W)M_Z^2 + m^2_{\tilde{f}_L}
& - m_f(R_f\mu + A_f^*) \cr
- m_f(R_f\mu^* + A_f)
& m^2_f - \cos 2 \beta Q_f \sin^2 \theta_WM_Z^2 + m^2_{\tilde{f}_R} }\right]
\ , \ \  
\end{equation}

\noindent where $m_f$ is the mass of the fermion $f$, $Q_f$ its electric
charge and $T_{3f}$ the third component of the weak isospin of a left-handed
fermion $f$. $m^2_{\tilde{f}_L} (m^2_{\tilde{f}_R})$ is the low energy mass
squared parameter for the left(right) sfermion $\tilde{f}_L(\tilde{f}_R)$.
$R_f = \cot\beta(\tan\beta)$ for $T_{3f}=\frac{1}{2}(-\frac{1}{2})$ where
$\tan\beta = v_2/v_1$. 

$M_{\tilde{\chi}}$ and $M_{\tilde{\chi}^0}$ are given by 
\cite{oshimo,haber,rosiek,barger,gunion}:

\begin{equation}
M_{\tilde{\chi}} = \left[\matrix{
\tilde{m}_2 & \sqrt{2} M_W \sin\beta \cr
\sqrt{2} M_W \cos\beta & \mu } \right] \ \ ,
\end{equation}
\begin{equation}
M_{\tilde{\chi}^0} = \left[\matrix{
\tilde{m}_1   & 0 & -M_Z\cos\beta\sin\theta_W & M_Z\sin\beta\sin\theta_W \cr
0 & \tilde{m}_2 & M_Z\cos\beta\cos\theta_W & -M_Z\sin\beta\cos\theta_W \cr
-M_Z\cos\beta\sin\theta_W & M_Z\cos\beta\cos\theta_W & 0 & -\mu\cr
M_Z\sin\beta\sin\theta_W & -M_Z\sin\beta\cos\theta_W & -\mu & 0 } \right] \ , \ \ 
\end{equation}

\noindent where $\tilde{m}_1$($\tilde{m}_2$) is the mass parameter for the
$U(1)$($SU(2)$) gaugino. 

After diagonalizing these mass matrices according to Eq.~A.12--A.14, the
corresponding masses of the supersymmetric partners of the fermions, the
charginos and the neutralinos are given by: 

\begin{eqnarray}
&&M^2_{\tilde{f}_1(\tilde{f}_2)}
= \frac{a_f-(+)\sqrt{b_f^2 + c_f^2} }{2} \ , \\
&&a_f = 2m^2_f -\cos2\beta T_{3f}M_Z^2+m^2_{\tilde{f}_L} + m^2_{\tilde{f}_R}
\ , \\
&&b_f = \cos2\beta(2Q_f\sin^2\theta_W - T_{3f}) M^2_Z + m^2_{\tilde{f}_L}
- m^2_{\tilde{f}_R} \ , \\ 
&&c_f = - 2m_f\left| R_f\mu + A^*_f\right| \ ,
\end{eqnarray}
\begin{eqnarray}
M^2_{\tilde{\chi}_1(\tilde{\chi}_2)} &=& \frac{1}{2} \left[
\tilde{m}^2_2 + \mu^2 + 2M_W^2 -(+)
\left[(\tilde{m}^2_2-\mu^2)^2+4M_W^4\cos^22\beta +\right.\right. \nonumber \\
&+& 4M_W^2
\left.\left.(\tilde{m}^2_2 + \mu^2 +
2\tilde{m}_2\mu\sin2\beta)\right]^{\frac{1}{2}}\right] \ \ . 
\end{eqnarray}

\noindent In Eq.~A.14 we have defined $Z_N$ such that the elements of the
diagonal neutralinos mass matrix are real and non-negative.  It is sometimes
more convenient to allow for negative entries for the $m_{\tilde{\chi}^0_i}$
which implies the change of Eq.~A.14 to \cite{shafik,gunion}: %%%1096 

\begin{equation}
NM_{\tilde{\chi}_0}N^{-1} = {\rm diag}(\epsilon_1m_{\tilde{\chi}^0_1}, \
\epsilon_2m_{\tilde{\chi}^0_2}, \ \epsilon_3m_{\tilde{\chi}^0_3}, \
\epsilon_4m_{\tilde{\chi}^0_4}) \ , 
\end{equation}

\noindent where $N$ is a real matrix.  While the $m_{\tilde{\chi}^0_i}$ are
always positive, the $\epsilon_i$'s are either $\pm 1$.  With this substitution,
in the Feynman rules one has to use the relation \cite{shafik,gunion} (note
the slight difference between our notation for $Z_N$ and the one used in
\cite{shafik}): 

\begin{equation}
Z_{N_{ij}} = \left(\sqrt{\epsilon_j}\right)^*N_{ji} \ ,
\end{equation}

\noindent The positive real masses of the neutralinos are then given
by:

\begin{eqnarray}
&& \epsilon_1m_{\tilde{\chi}^0_1} = - a_{\tilde{\chi}} +
\sqrt{b_{\tilde{\chi}}+c_{\tilde{\chi}}} + d_{\tilde{\chi}} \ , \\ 
&& \epsilon_2m_{\tilde{\chi}^0_2} =  a_{\tilde{\chi}} -
\sqrt{b_{\tilde{\chi}}-c_{\tilde{\chi}}} + d_{\tilde{\chi}} \ , \\ 
&& \epsilon_3m_{\tilde{\chi}^0_3} = - a_{\tilde{\chi}} -
\sqrt{b_{\tilde{\chi}}+c_{\tilde{\chi}}} + d_{\tilde{\chi}} \ , \\ 
&& \epsilon_4m_{\tilde{\chi}^0_2} =  a_{\tilde{\chi}} +
\sqrt{b_{\tilde{\chi}}-c_{\tilde{\chi}}} + d_{\tilde{\chi}} \ , 
\end{eqnarray}

\noindent where:

\begin{eqnarray}
&&a_{\tilde{\chi}} = \left(\frac{1}{2}a-\frac{1}{6} C_2\right)^{\frac{1}{2}}
\ , \\ 
&&b_{\tilde{\chi}} = -\frac{1}{2}a-\frac{1}{3} C_2 \ , \\
&&c_{\tilde{\chi}} = \frac{C_3}{(8a-\frac{8}{3}C_2)^{\frac{1}{2}}} \ , \\
&&d_{\tilde{\chi}} = \frac{1}{4}(\tilde{m}_1+\tilde{m}_2) \ ,
\end{eqnarray}

\noindent and
$a, \ C_2$ and $C_3$ are defined in \cite{barger} with the
substitution $\mu\to - \mu$.

The mass mixing matrices are then given by \cite{oshimo,footmixing}:

\begin{equation}
Z_f = \left[ \matrix {\cos \theta_f & - e^{-i\beta_f}  \sin\theta_f \cr
 e^{i\beta_f}  \sin\theta_f & \cos\theta_f } \right] \ ,
\end{equation}

\noindent where:

\begin{equation}
\tan\theta_f = \frac{c_f}{b_f} \ .
\end{equation}

\noindent When $\mu$ is chosen as real, the phase $\beta_f$ is given by:

\begin{equation}
\tan\beta_f = \frac{-|A_f|\sin\theta_{A_f}}{\mu R_f + |A_f|\cos\theta_{A_f}}
\ , 
%\left\{ \matrix{
%\frac{-|A_f|\sin\theta_{A_f}}{\mu R_f + |A_f|\cos\theta_{A_f}}
%& (T_{3f} = \frac{1}{2})   \cr & \cr
%\frac{-A_fm_0\sin\theta_{Af}}{\mu\tan\beta + |A_f|m_0\cos\theta_{Af}}
%& (T_{3f} =-\frac{1}{2})  } \right.
\end{equation}

\noindent where $\theta_{A_f} = \arg (A_f)$.

The charginos mixing matrices are given by \cite{barger,oshimo}:

\begin{equation}
Z^-=O_-^{\rm T}, \ Z^+ =
\left\{ \matrix{(O_+)^{-1} \ \ \ if \ \ \ \det (M_{\tilde{\chi}})\geq 0 \cr ~~~ \cr
(\sigma_3 O_+)^{-1} \ \ \ if \ \ \ \det (M_{\tilde{\chi}}) < 0 } \right. \ ,
\end{equation}

\noindent where:

\begin{equation}
O_\pm = \left(\matrix{\cos\phi_\pm & \sin\phi_\pm \cr
-\sin\phi_\pm & \cos\phi_\pm } \right) \ ,
\end{equation}

\noindent and:

\begin{equation}
\tan2\phi_- = 2\sqrt{2}M_W \frac{\mu\sin\beta+\tilde{m}_2\cos\beta}
{m^2_2-\mu^2 - 2M_W^2\cos2\beta} \ ,
\end{equation}
\begin{equation}
\tan2\phi_+ = 2\sqrt{2}M_W \frac{\mu\cos\beta+\tilde{m}_2\sin\beta}
{m^2_2-\mu^2 + 2M_W^2\cos2\beta} \ .
\end{equation}

\noindent The neutralinos mixing matrix elements are given by
\cite{barger}:

\begin{eqnarray}
&&\frac{N_{i2}}{N_{i1}} = \frac{1}{\tan\theta_W} \times
\frac{\tilde{m}_1 - \epsilon_im_{\tilde{\chi}^0_i}}
{\tilde{m}_2 - \epsilon_im_{\tilde{\chi}^0_i}} \ , \\ %[0.25cm]
&&\frac{N_{i3}}{N_{i1}} = \frac{\mu[\tilde{m}_2-\epsilon_im_{\tilde{\chi}^0_i}]
[\tilde{m}_1-\epsilon_im_{\tilde{\chi}^0_i}] -
M_Z^2\sin\beta\cos\beta[(\tilde{m}_1-\tilde{m}_2)\cos^2\theta_W +\tilde{m}_2
-\epsilon_im_{\tilde{\chi}^0_i}]}
{M_Z [\tilde{m}_2 - \epsilon_im_{\tilde{\chi}^0_i}]\sin\theta_W[\mu\cos\beta
+ \epsilon_im_{\tilde{\chi}^0_i}\sin\beta]} \ , \nonumber \\ %[0.25cm]
&& ~~ \\
&&\frac{N_{i4}}{N_{i1}} = \frac{-\epsilon_im_{\tilde{\chi}^0_i}
[\tilde{m}_2-\epsilon_im_{\tilde{\chi}^0_i}]
[\tilde{m}_1-\epsilon_im_{\tilde{\chi}^0_i}] -
M_Z^2\cos^2\beta[(\tilde{m}_1-\tilde{m}_2)\cos^2\theta_W +\tilde{m}_2
-\epsilon_im_{\tilde{\chi}^0_i}]}
{M_Z [\tilde{m}_2 - \epsilon_im_{\tilde{\chi}^0_i}]\sin\theta_W[\mu\cos\beta
+ \epsilon_im_{\tilde{\chi}^0_i}\sin\beta]} \ , \nonumber \\ %[0.25cm]
&& ~~ \\
&&N_{i1} = \left[1+\left(\frac{N_{i2}}{N_{i1}}\right)^2 +
\left(\frac{N_{i3}}{N_{i1}}\right)^2 +
\left(\frac{N_{i4}}{N_{i1}}\right)^2 \right]^{-\frac{1}{2}} \ ,
\end{eqnarray}

\noindent and the relation between $Z_N$ and $N$ is given by Eq.~A.24.
\pagebreak

\begin{center}
{\bf Appendix B}\\
\end{center}

In this appendix we list the imaginary parts which arise from the SUSY
couplings and were defined by the ${\cal O}_k^i$'s in 
Eqs.~\ref{aalr}-\ref{cdtagll}, Eqs.~\ref{b1e}-\ref{b2h} and
Eqs.~\ref{aalrd}-\ref{dalrd} ($k$ stands for the corresponding
diagram in Fig.~2,
i.e.\ $k=a,b,c,d,a^{\prime},b^{\prime},c^{\prime},d^{\prime},e,f,g,h$). 
These contain  the necessary CP-odd SUSY phases
needed for both the production cross-section asymmetry, $A_P$, and the decay
PRA in the top decay, $A_D$.   

\setcounter{num}{2}
\setcounter{equation}{0}
\def\theequation{\Alph{num}.\arabic{equation}}

\begin{eqnarray} 
{\cal O}_a^1 &\equiv& |Z_b^{1j}|^2 {\rm Im} \left\{ \xi_t^i \right\} ~, \\
{\cal O}_b^1 &\equiv& - \frac{1}{2} |Z_b^{1j}|^2 {\rm Im} \left\{ L^- \left(
|Z_t^{1i}|^2 L^{+*} +  
\frac{m_t}{M_W \sin\beta} Z_N^{4n*} \xi_t^i \right) \right\} ~, \\
{\cal O}_b^2 &\equiv& \frac{1}{2} |Z_b^{1j}|^2 {\rm Im} \left\{ L^- \left(
\frac{4}{3} \tan\theta_W Z_N^{1n} \xi_t^i -  
\frac{m_t}{M_W \sin\beta} |Z_t^{1i}|^2 Z_N^{4n} \right) \right\} ~, \\
{\cal O}_c^1 &\equiv& |Z_b^{1j}|^2 {\rm Im} \left\{ Z_{1m}^{-*} L^- K^+
\right\} ~, \\ 
{\cal O}_c^2 &\equiv& - \frac{1}{\sqrt 2} \frac{m_t}{M_W \sin\beta}
|Z_b^{1j}|^2 {\rm Im} \left\{   Z_{2m}^{+} L^- K^+ \right\} ~, \\
{\cal O}_c^3 &\equiv&  - \frac{1}{\sqrt 2} \frac{m_t}{M_W \sin\beta}
|Z_b^{1j}|^2 {\rm Im} \left\{   Z_{2m}^{+} L^- K^- \right\} ~, \\
{\cal O}_c^4 &\equiv& |Z_b^{1j}|^2 {\rm Im} \left\{ Z_{1m}^{-*} L^- K^-
\right\} ~, \\ 
{\cal O}_d^1 &\equiv& - {\rm Im} \left\{ K^- M^1 \right\} ~, \\ 
{\cal O}_d^2 &\equiv& {\rm Im} \left\{ K^- M^2 \right\} ~, \\
{\cal O}_d^3 &\equiv& {\rm Im} \left\{ K^+ M^2 \right\} ~, \\
{\cal O}_d^4 &\equiv& - {\rm Im} \left\{ K^+ M^1 \right\} ~, \\
{\cal O}_{b^{\prime}}^1 &\equiv& -\frac{1}{2} |Z_u^{1l}|^2 |Z_d^{1o}|^2 {\rm
Im} \left\{ L^+ L^{-*} \right\} ~, \\ 
{\cal O}_{c^{\prime}}^1 &\equiv& |Z_d^{1o}|^2 {\rm Im} \left\{ Z^-_{1m} L^{-*}
K^{+*} \right\} ~, \\ 
{\cal O}_{c^{\prime}}^2 &\equiv& |Z_d^{1o}|^2 {\rm Im} \left\{ Z^-_{1m} L^{-*}
K^{-*} \right\} ~, \\   
{\cal O}_{d^{\prime}}^1 &\equiv& |Z_u^{1l}|^2 {\rm Im} \left\{ Z^{+*}_{1m}
L^{+} K^{-*} \right\} ~, \\ 
{\cal O}_{d^{\prime}}^2 &\equiv& |Z_u^{1l}|^2 {\rm Im} \left\{ Z^{+*}_{1m}
L^{+} K^{+*} \right\} ~, \\ 
{\cal O}_e^1 &\equiv& \frac{1}{\sqrt 2} |Z_u^{1l}|^2 {\rm Im} \left\{
Z^{+*}_{1m} L^{+} M^2 \right\} ~, \\ 
{\cal O}_e^2 &\equiv& -\frac{1}{\sqrt 2} |Z_u^{1l}|^2 {\rm Im} \left\{
Z^{+*}_{1m} L^{+} M^1 \right\} ~, \\ 
{\cal O}_f^1 &\equiv& -\frac{1}{2} \frac{m_t}{M_W \sin\beta} |Z_b^{1j}|^2
|Z_d^{1o}|^2 |L^-|^2 {\rm Im} \left\{  Z_{1m}^- Z_{2m}^+ \right\} ~, \\
{\cal O}_g^1 &\equiv& -\frac{1}{\sqrt 2} |Z_d^{1o}|^2 {\rm Im} \left\{
Z_{1m}^- L^{-*} M^1 \right\} ~, \\ 
{\cal O}_g^2 &\equiv&  \frac{1}{\sqrt 2} |Z_d^{1o}|^2 {\rm Im} \left\{
Z_{1m}^- L^{-*} M^2 \right\} ~, \\ 
{\cal O}_h^1 &\equiv& \frac{1}{\sqrt 2} |Z_u^{1l}|^2 |Z_b^{1j}|^2 {\rm Im}
\left\{ L^- L^+   Z_{1m}^{-*} Z_{1m}^{+*} \right\} ~, \\
{\cal O}_h^2 &\equiv& -\frac{1}{2} \frac{m_t}{M_W \sin\beta} |Z_u^{1l}|^2
|Z_b^{1j}|^2 {\rm Im} \left\{ L^- L^+   Z_{1m}^{+*} Z_{2m}^+ \right\} ~. 
\end{eqnarray}   

\noindent In order to avoid any confusions we chose the following notation  
for the indexes of different squark mass eigenstates: $i,j,l,o=1,2$ are the
indices for the  two superpartners of the $t,b,u,d$, respectively, 
$m=1,2$ is the index for the two mass eigenstates of the charginos and $n=1-4$
is the index for the  four neutralinos mass eigenstates. Also $L^{\pm},K^+$
and $K^-$ are defined in Eqs.~A.9--A.11 and:

\begin{eqnarray} 
\xi_t^i &\equiv& Z_t^{1i*} Z_t^{2i} ~, \\
M^1 &\equiv& \frac{1}{\sqrt 2} \frac{m_t}{M_W \sin\beta} \left( Z_{2m}^{+} 
L^{+*} \xi_t^{i*} - \sqrt 2 Z_{1m}^{+} Z_N^{4n*} \xi_t^i \right) \nonumber \\
&& + \frac{1}{\sqrt 2} \left( \left(  \frac{m_t}{M_W \sin\beta} \right)^2
|Z_t^{2i}|^2 Z_{2m}^{+} Z_N^{4n*} -  
\sqrt 2 |Z_t^{1i}|^2 Z_{1m}^{+} L^{+*} \right) ~, \\
M^2 &\equiv& \frac{1}{\sqrt 2} \frac{m_t}{M_W \sin\beta} \left( \frac{4}{3}
\tan\theta_W |Z_t^{2i}|^2 Z_{2m}^{+}  
Z_N^{1n} + \sqrt 2 |Z_t^{1i}|^2 Z_{1m}^+ Z_N^{4n} \right) \nonumber \\
&& - \left( \frac{1}{\sqrt 2} \left(  \frac{m_t}{M_W \sin\beta} \right)^2
Z_{2m}^{+} Z_N^{4n} \xi_t^{i*} +  
\frac{4}{3} \tan\theta_W  Z_{1m}^{+} Z_N^{1n} \xi_t^i \right) ~.
\end{eqnarray}
\pagebreak

\begin{center}
{\bf Appendix C}\\
\end{center}

In this appendix we give the coefficients $C^k_{pq}$, $D^k_{pq}$ and $C_0^k$,
$D^k_0$ which appear in Eqs.~\ref{aalr}-\ref{cdtagll},
Eqs.~\ref{b1e}-\ref{b2h} and Eqs.~\ref{aalrd}-\ref{dalrd}
($k$ stands for the corresponding diagram in Fig.~2,
 i.e.\ $k=a,b,c,d,a^{\prime},b^{\prime},c^{\prime},d^{\prime},e,f,g$
and $h$ and $p=1,2$ , $q=1-7$). 
These coefficients which are functions of masses and momenta
are defined by the one-loop momentum integrals as follows \cite{olden}:

\setcounter{num}{2}
\setcounter{equation}{0}
\def\theequation{\Alph{num}.\arabic{equation}}

\begin{eqnarray}
&&{\rm C}_0; {\rm C}_\mu; {\rm
C}_{\mu\nu}(m^2_1,m^2_2,m^2_3,p^2_1,p^2_2,p^2_3)\equiv\nonumber 
\\ 
&& ~~~\equiv \int \frac{d^4k}{i\pi^2}
\frac{1;k_\mu;k_\mu k_\nu}{[k^2-m^2_1][(k+p_1)^2-m^2_2][(k-p_3)^2-m^2_3]} \ ,
\end{eqnarray}
\begin{eqnarray}
&&{\rm D}_0; {\rm D}_\mu; {\rm D}_{\mu\nu}(m^2_1,m^2_2,m^2_3,m^2_4,
p^2_1,p^2_2,p^2_3,p^2_4,(p_1+p_2)^2,(p_2+p_3)^2)\equiv\nonumber \\
&& ~~~\equiv \int \frac{d^4k}{i\pi^2}
\frac{1;k_\mu;k_\mu k_\nu}{[k^2-m^2_1][(k+p_1)^2-m^2_2][(k+p_1+p_2)^2-m^2_3]
[(k-p_4)^2-m_4^2]} ~, \nonumber \\
&&~~
\end{eqnarray}

\noindent where $\sum_i p_i = 0$ is to be understood above.

The coefficients are then defined through the following relations
\cite{veltman}: 

\begin{eqnarray}
&&{\rm C}_\mu=p_{1\mu}{\rm C}_{11}+p_{2\mu}{\rm C}_{12} \ ,\\
&&{\rm C}_{\mu\nu} = p_{1\mu}p_{1\nu}{\rm C}_{21} + p_{2\mu} p_{2\nu}{\rm
C}_{22} +\{p_1p_2\}_{\mu\nu}{\rm C}_{23}+g_{\mu\nu}{\rm C}_{24} \ ,\\
&&{\rm D}_\mu = p_{1\mu}{\rm D}_{11} + p_{2\mu}{\rm D}_{12}+p_{3\mu}{\rm
D}_{13} \ ,\\ 
&&{\rm D}_{\mu\nu} = p_{1\mu}p_{1\nu}{\rm D}_{21} + p_{2\mu}p_{2\nu}{\rm
D}_{22}+p_{3\mu}p_{3\nu}{\rm D}_{23}+ \{p_1p_2\}_{\mu\nu}{\rm D}_{24} +\nonumber
\\ 
&&~~~~~~~~~~~~~~~~ +\{p_1p_3\}_{\mu\nu}{\rm D}_{25}+\{p_2p_3\}_{\mu\nu}{\rm
D}_{26} +g_{\mu\nu}{\rm D}_{27} \ ,
\end{eqnarray}

\noindent where $\{ab\}_{\mu\nu}\equiv a_\mu b_\nu +a_\nu b_\mu$.

With the above definitions and notation, the coefficients $C_0^k$'s and
$C^k_{pq}$'s for the triangle diagrams in the production amplitude which
appear in Eqs.~\ref{aalr}-\ref{cdtagll}, are given by: 

\begin{eqnarray}
&&C_0^a;C^{a}_{pq} = {\rm Im} \left\{{\rm C_0;C_{pq}} (m^2_G,m^2_{\tilde{b}_j},
m^2_{\tilde{t}_i},m^2_b,\hat{s},m^2_t) \right\} ~, \\
&&C_0^b;C^{b}_{pq} = C_0^a;C^{a}_{pq}(m_G\to m_{\tilde{\chi}^0_n}) ~, \\
&&C_0^c;C^{c}_{pq} = {\rm Im} \left\{{\rm C_0;C_{pq}}
(m^2_{\tilde{b}_j},m_{\tilde{\chi}^0_n}^2,m_{\tilde{\chi}_m}^2, 
m^2_b,\hat{s},m^2_t) \right\} ~, \\
&&C_0^d;C^{d}_{pq} = C_0^c;C^{c}_{pq}(m_{\tilde{b}_j} \to m_{\tilde{t}_i}, 
m_{\tilde{\chi}^0_n} \leftrightarrow m_{\tilde{\chi}_m}) ~, \\
&&C_0^{a^{\prime}};C^{a^{\prime}}_{pq} = {\rm Im} \left\{{\rm C_0;C_{pq}}
(m^2_G,m^2_{\tilde{u}_l}, m^2_{\tilde{d}_o},m^2_u,\hat{s},m^2_d) \right\}~, \\
&&C_0^{b^{\prime}};C^{b^{\prime}}_{pq} = C_0^{a^{\prime}};C^{a^{\prime}}_{pq}
( m_G\to m_{\tilde{\chi}^0_n}) ~, \\ 
&&C_0^{c^{\prime}};C^{c^{\prime}}_{pq} = {\rm Im} \left\{{\rm C_0;C_{pq}}
(m^2_{\tilde{d}_o},m_{\tilde{\chi}_m}^2,m_{\tilde{\chi}^0_n}^2, 
m^2_u,\hat{s},m^2_d) \right\}~, \\
&&C_0^{d^{\prime}};C^{d^{\prime}}_{pq} = C_0^{c^{\prime}};C^{c^{\prime}}_{pq}
(m_{\tilde{d}_o} \to m_{\tilde{u}_l},   
m_{\tilde{\chi}^0_n} \leftrightarrow m_{\tilde{\chi}_m}) ~.
\end{eqnarray}

\noindent The $D_0^k$'s and $D^k_{pq}$'s, for the box diagrams in the
production amplitudes which appear in Eqs.~\ref{b1e}-\ref{b2h}, are given by: 

\begin{eqnarray}
&&D_0^e;D^{e}_{pq} = {\rm Im} \left\{{\rm
D_0;D_{pq}}(m_{\tilde{\chi}^0_n}^2,m^2_{\tilde{t}_i},m_{\tilde{\chi}_m}^2,
m_{\tilde{u}_l}^2, m_t^2,m_b^2,m_d^2,m_u^2,\hat s,\hat t) \right\} ~, \\
&&D_0^f;D^{f}_{pq} = D_0^e;D^{e}_{pq}( m_{\tilde{t}_i} \to m_{\tilde{b}_j},
m_{\tilde{u}_l} \to m_{\tilde{d}_o} ,  
m_{\tilde{\chi}^0_n} \leftrightarrow m_{\tilde{\chi}_m}) ~, \\
&&D_0^g;D^{g}_{pq} = D_0^e;D^{e}_{pq}( m_{\tilde{u}_l} \to m_{\tilde{d}_o}
, m_u \leftrightarrow m_d, \hat t \to \hat u ) ~, \\  
&&D_0^h;D^{h}_{pq} = D_0^g;D^{g}_{pq}( m_{\tilde{t}_i} \to m_{\tilde{b}_j}
, m_{\tilde{d}_o} \to m_{\tilde{u}_l},  
m_{\tilde{\chi}^0_n} \leftrightarrow m_{\tilde{\chi}_m}) ~,
\end{eqnarray}

\noindent and for the top decay amplitudes, the coefficients $C_0^k$'s and
$C^k_{pq}$'s which appear in Eqs.~\ref{aalrd}-\ref{dalrd}, are given
by: 

\begin{eqnarray}
&&C_0^a;C^{a}_{pq} = {\rm Im} \left\{{\rm C_0;C_{pq}} (m^2_{\tilde{b}_j},
m^2_{\tilde{t}_i},m_G^2,m_W^2,m_t^2,m^2_b) \right\} ~, \\
&&C_0^b;C^{b}_{pq} = C_0^a;C^{a}_{pq}(m_G\to m_{\tilde{\chi}^0_n}) ~, \\
&&C_0^c;C^{c}_{pq} = {\rm Im} \left\{{\rm C_0;C_{pq}}
(m_{ \tilde{\chi}^0_n}^2 ,m_{\tilde{\chi}_m}^2,m^2_{\tilde{b}_j}, 
m^2_W,m_t^2,m^2_b) \right\} ~, \\
&&C_0^d;C^{d}_{pq} = C_0^c;C^{c}_{pq}(m_{\tilde{b}_j} \to m_{\tilde{t}_i}, 
m_{\tilde{\chi}^0_n} \leftrightarrow m_{\tilde{\chi}_m}) ~.
\end{eqnarray}

For the numerical evaluation of the above form factors see \cite{olden}.
\pagebreak

\pagebreak

\begin{center}
{\bf Figure Captions}
\end{center}

\begin{description}

\item{Fig. 1:} The tree-level Feynman diagram contributing to
$u \bar{d} \to t\bar{b}$.

\item{Fig. 2:} The CP-violating, SUSY induced, one-loop diagrams for the
processes $u \bar{d} \to t\bar{b}$ and $t \to W^+ b$.  

\item{Fig. 3:} The allowed regions in the $\sin\alpha_u - \sin\alpha_d$ plane
for the NEDM not to exceed a) $1.1\times 10^{-25}$ e-cm and b) $3\times
10^{-25}$ e-cm. $M_s=400$ GeV and $m_G=500$ GeV is used. The shaded areas
indicate the allowed  regions. 
  
\item{Fig. 4:} The SUSY induced partially integrated cross-section asymmetry
as a function of $\mu$, for $M_s=400$ GeV, $m_l=50$ GeV and for $\sqrt{s}=2$
TeV\null. With a) $\tan\beta=1.5$ and b) $\tan\beta=35$. See Ref.~24. 

\item{Fig. 5:} The SUSY induced partially integrated cross-section asymmetry
as a function of $m_l$, for several values of $\mu$, $M_s=400$ GeV, $m_G=450$
GeV and for $\sqrt{s}=2$ TeV\null. With a) $\tan\beta=1.5$ and b)
$\tan\beta=35$. 

\item{Fig. 6:} The SUSY induced partially integrated cross-section asymmetry
as a function of $m_G$, for several values of $\mu$, $M_s=400$ GeV, $m_l=50$
GeV and for $\sqrt{s}=2$ TeV\null. With a) $\tan\beta=1.5$ and b) $\tan\beta=35$.

\item{Fig. 7:} The SUSY induced partially integrated cross-section asymmetry
as a function of $\tan\beta$, for several values of $\mu$, $M_s=400$ GeV,
$m_l=50$ GeV and for $\sqrt{s}=2$ 
TeV\null. With a) $m_G=350$ GeV and b) $m_G=350$ GeV\null.  

\item{Fig. 8:} The SUSY induced partial rate asymmetry in the top decays
as a function of $\mu$, $M_s=400$ GeV, $m_G=350$ GeV and for $\sqrt{s}=2$
TeV\null. With a) $\tan\beta=1.5$ and b) $\tan\beta=35$. See Ref.~24. 

\item{Fig. 9:} The SUSY induced partial rate asymmetry in the top decays
as a function of $m_l$, for several values of $\mu$, $M_s=400$ GeV and for
$\sqrt{s}=2$ TeV\null. With 
a) $\tan\beta=1.5$ and $\mu=-160$ GeV, and b) $\tan\beta=35$ and $\mu=230$
GeV\null. 

\item{Fig. 10:} The SUSY induced partial rate asymmetry in the top decays
as a function of $m_G$, for several values of $\mu$, $M_s=400$ GeV, $m_l=50$
GeV and for $\sqrt{s}=2$ TeV\null. With a) $\tan\beta=1.5$ and b) $\tan\beta=35$.

\item{Fig. 11:} The SUSY induced partial rate asymmetry in the top decays
as a function of $\tan\beta$, for several values of $\mu$, $M_s=400$ GeV,
$m_l=50$ GeV and for $\sqrt{s}=2$ 
TeV\null. With a) $m_G=350$ GeV and b) $m_G=350$ GeV\null.

%\newpage

\begin{table}
\begin{center}
\caption[first entry]{The charginos and neutralinos masses for different
values of $\left\{ \tan\beta,\mu,m_G \right\}$. 
The entries in the table are rounded to the nearest integer. 

\bigskip

\protect\label{oropttab}}
\begin{tabular}{|r||r|r|r|r|r|r|r|r||r|} \cline{3-9}
%\multicolumn{2}{c||}{~~} & \multicolumn{6}{c|}{$e^+e^- \to t \bar t Z$ (Model~II
%with Set~II)} 
\hline \hline
$\tan\beta$ & $\mu $
& \multicolumn{6}{c|}{charginos and neutralinos masses} & $m_G $
\\ \cline{3-8}
$ \Downarrow$ & $ ({\rm GeV}) \Downarrow$ & $m_{{\tilde \chi}_1} $ & $m_{{\tilde
\chi}_2} $ & $m_{{\tilde \chi}_1^0} $ & $m_{{\tilde \chi}_2^0} $ & $m_{{\tilde
\chi}_3^0} $ & $m_{{\tilde \chi}_4^0} $ & $\Leftarrow ({\rm GeV}) \Downarrow$
\\ \hline 
\hline
& & $ 138 $ & $94 $ & $ 56$ & $68$ & $111$ & $138$ & 350 \\ 
&-70& $160 $ & $ 94 $ & $ 65$ & $75$ & $107$ & $161$ & 450 \\
& & $ 185 $ & $ 92 $ & $ 67$ & $88$ & $104$ & $186$ & 550 \\ \cline{2-9}
& & $ 140 $ & $ 107 $ & $ 56$ & $85$ & $128$ & $137$ & 350 \\ 
&-90& $ 160 $ & $ 110 $ & $ 70$ & $88$ & $124$ & $160$ & 450 \\ 
& & $ 184 $ & $ 110 $ & $ 81$ & $92$ & $122$ & $184$ & 550 \\ \cline{2-9}
& & $ 161 $ & $ 118 $ & $ 56$ & $112$ & $162$ & $145$ & 350 \\ 
1.5 &-130& $ 167 $ & $ 136 $ & $ 70$ & $122$ & $160$ & $161$ & 450 \\
& & $ 185 $ & $ 143 $ & $ 84$ & $126$ & $158$ & $183$ & 550 \\ \cline{2-9}
& & $ 202 $ & $ 40 $ & $ 16$ & $67$ & $142$ & $209$ & 350 \\ 
&140& $ 214 $ & $ 56 $ & $ 29$ & $86$ & $141$ & $220$ & 450 \\ 
& & $ 229 $ & $ 70 $ & $ 41$ & $103$ & $141$ & $234$ & 550 \\ \cline{2-9}
& & $ 276 $ & $ 65 $ & $ 35$ & $75$ & $241$ & $282$ & 350 \\
&240& $ 282 $ & $ 89 $ & $ 49$ & $98$ & $241$ & $287$ & 450 \\ 
& & $ 288 $ & $ 110 $ & $ 63$ & $120$ & $241$ & $294$ & 550 \\ \hline
\hline
& & $ 176 $ & $ 65 $ & $ 41$ & $70$ & $132$ & $172$ & 350 \\ 
&-110& $ 189 $ & $ 77 $ & $ 50$ & $86$ & $130$ & $187$ & 450 \\ 
& & $ 206 $ & $ 86 $ & $ 59$ & $100$ & $128$ & $205$ & 550 \\ \cline{2-9}
& & $ 212 $ & $ 82 $ & $ 47$ & $83$ & $187$ & $208$ & 350 \\ 
&-170& $ 219 $ & $ 102 $ & $ 60$ & $103$ & $185$ & $216$ & 450 \\
& & $ 229 $ & $ 118 $ & $ 73$ & $121$ & $184$ & $227$ & 550 \\ \cline{2-9}
& & $ 228 $ & $ 85 $ & $ 48$ & $86$ & $206$ & $223$ & 350 \\ 
35 &-190& $ 233 $ & $ 107 $ & $ 61$ & $107$ & $204$ & $229$ & 450 \\ 
& & $ 241 $ & $ 125 $ & $ 74$ & $127$ & $203$ & $238$ & 550 \\ \cline{2-9}
& & $ 177 $ & $ 60 $ & $ 37$ & $69$ & $130$ & $175$ & 350 \\ 
&110& $ 191 $ & $ 72 $ & $ 47$ & $85$ & $128$ & $189$ & 450 \\ 
& & $ 208 $ & $ 82 $ & $ 56$ & $100$ & $126$ & $207$ & 550 \\ \cline{2-9}
& & $ 214 $ & $ 78 $ & $ 45$ & $80$ & $185$ & $211$ & 350 \\ 
&170& $ 221 $ & $ 97 $ & $ 58$ & $100$ & $184$ & $219$ & 450 \\
& & $ 231 $ & $ 114 $ & $ 70$ & $119$ & $183$ & $230$ & 550 \\ \hline
\hline
\end{tabular}
\end{center}
\end{table}

\end{description}

\newpage
\pagestyle{empty}

\begin{figure}
\centering
\leavevmode
\epsfysize=350pt
\epsfbox{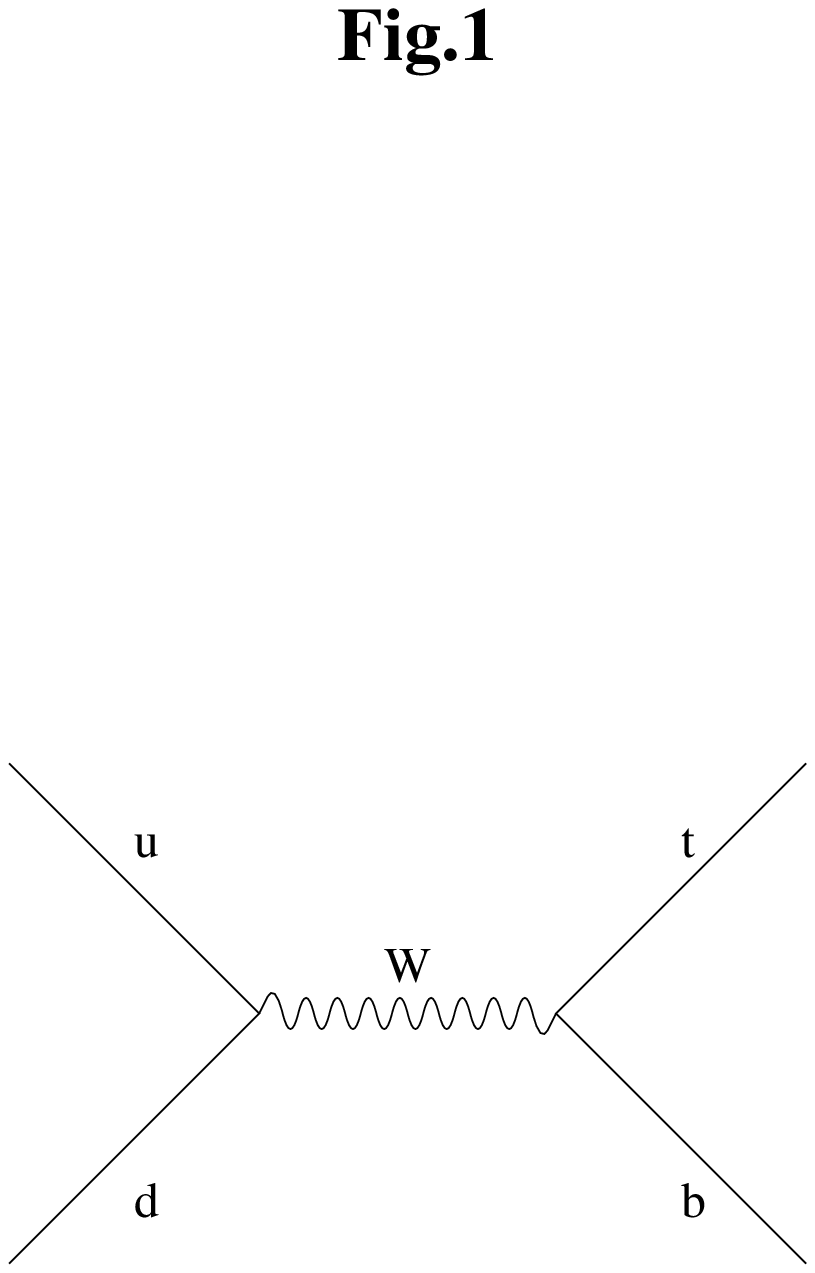}
%\caption{Feynman diagrams for the process (a) $q\bar q\to Q \bar Q$,
%and (b) $gg\to Q\bar Q$ with $Q=t$.}
\end{figure}

\begin{figure}
\centering
\leavevmode
\epsfysize=350pt
\epsfbox{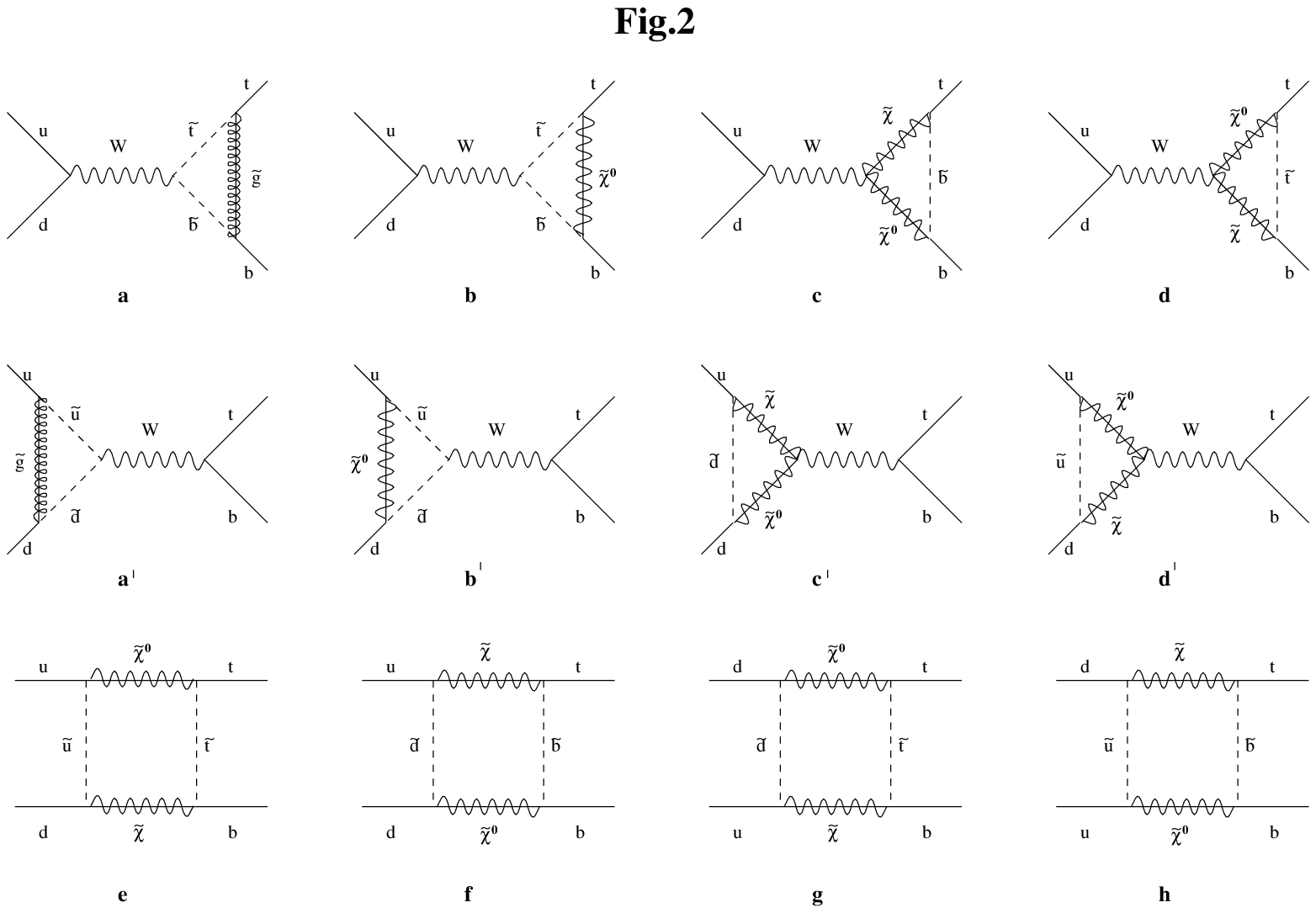}
\end{figure}

\begin{figure}
\centering
\leavevmode
\epsfysize=350pt
\epsfbox[0 0 612 792]{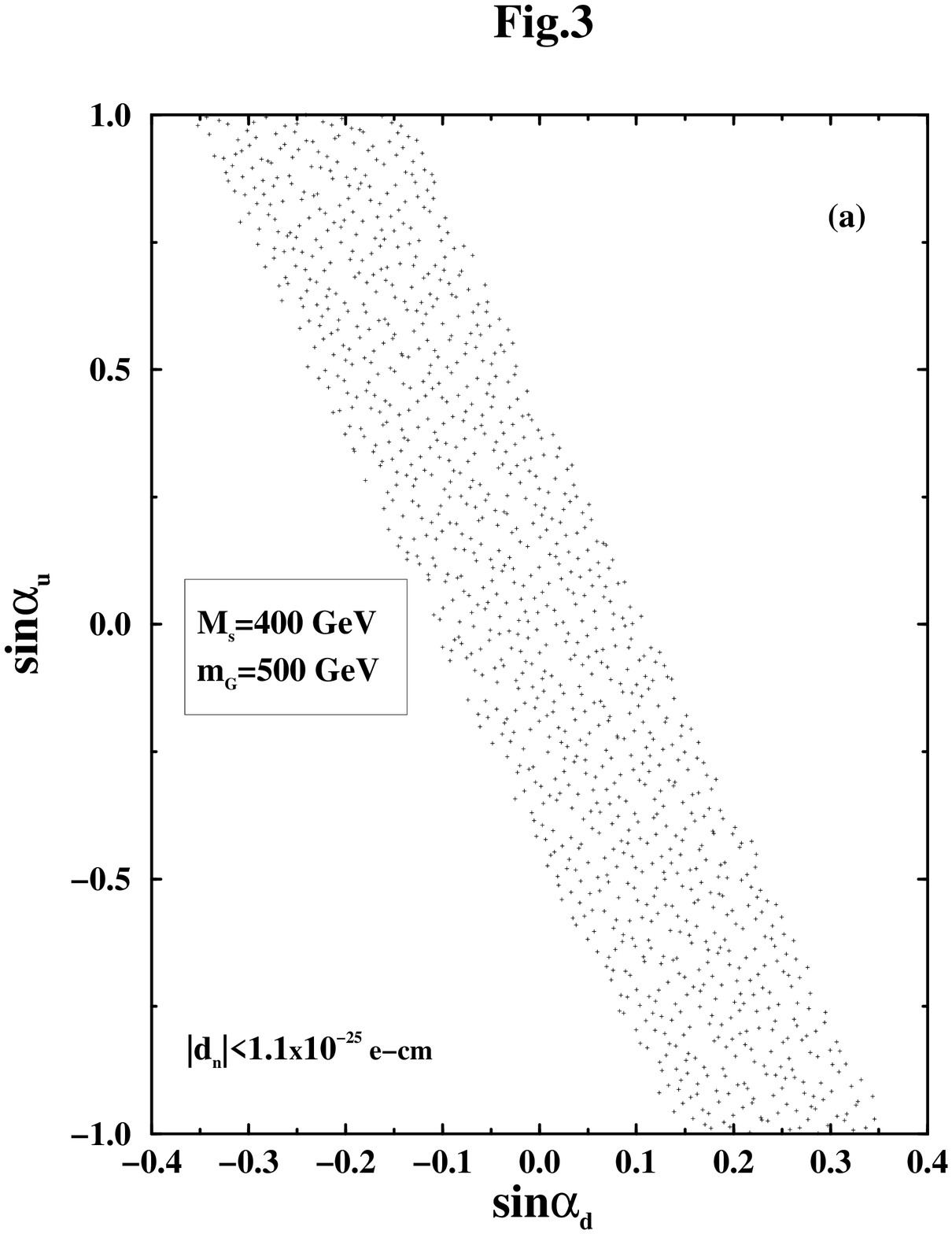}
\epsfysize=350pt
\epsfbox[0 0 612 792]{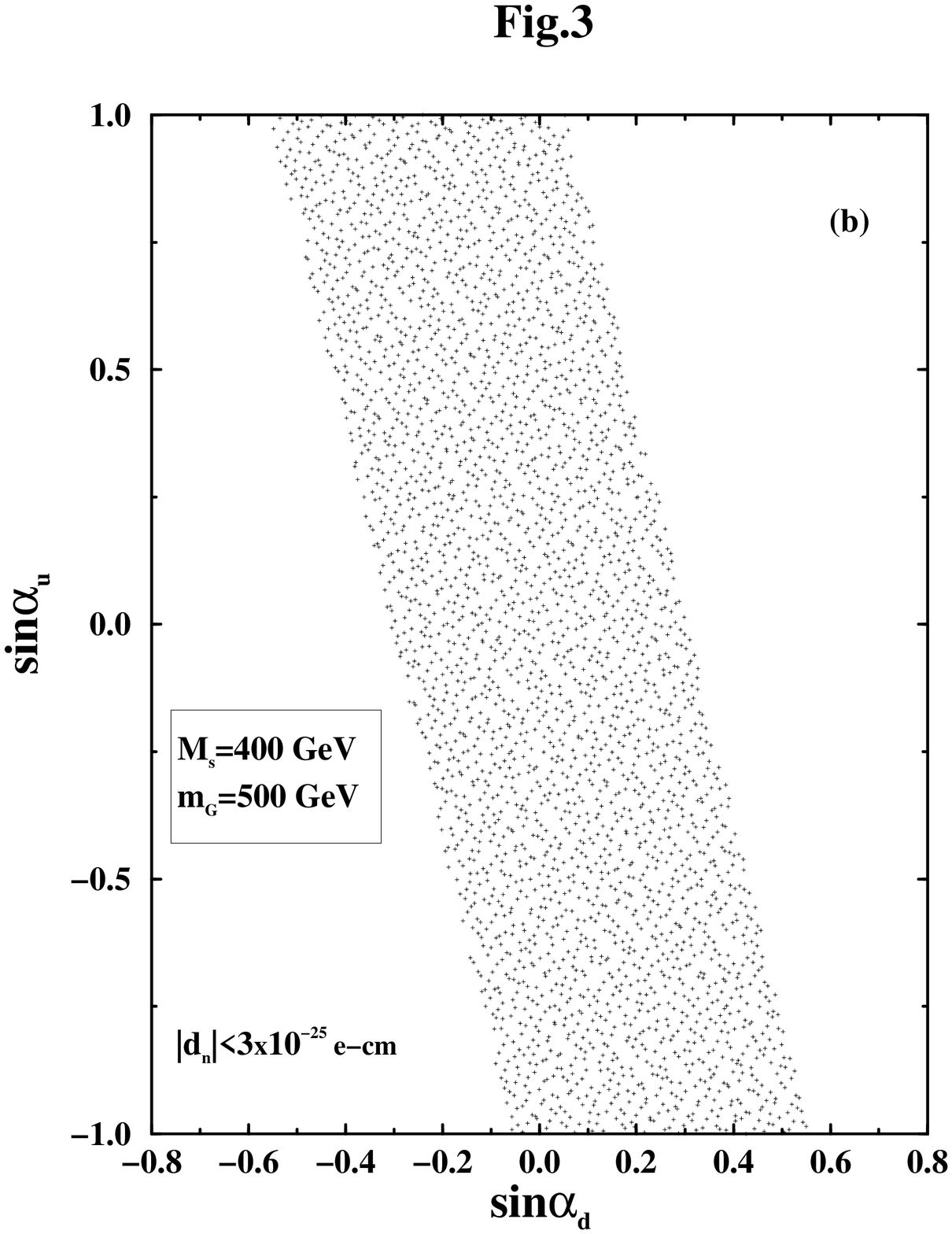}
\end{figure}

\begin{figure}
\centering
\leavevmode
\epsfysize=350pt
\epsfbox[0 0 612 792]{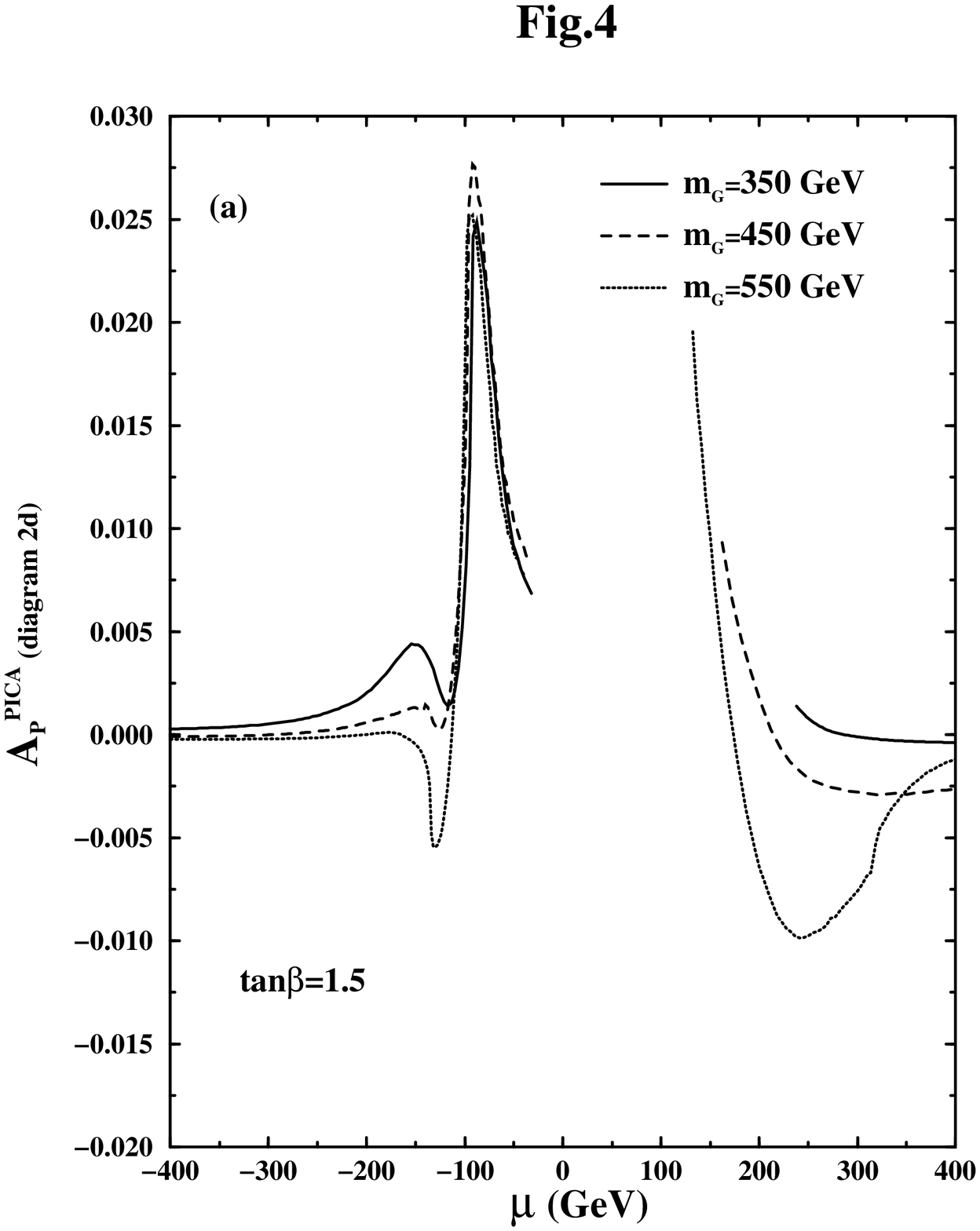}
\epsfysize=350pt
\epsfbox[0 0 612 792]{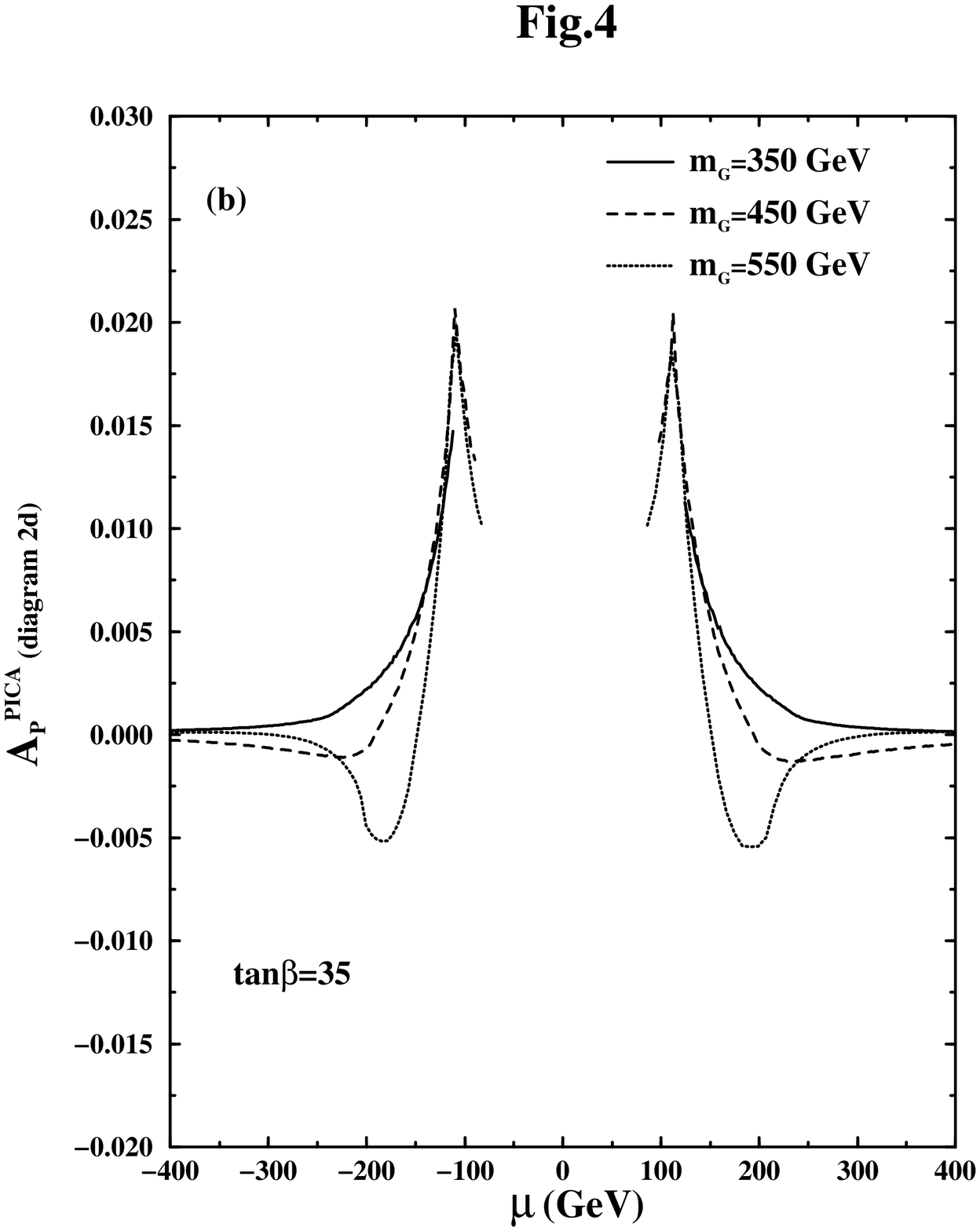}
\end{figure}

\begin{figure}
\centering
\leavevmode
\epsfysize=350pt
\epsfbox[0 0 612 792]{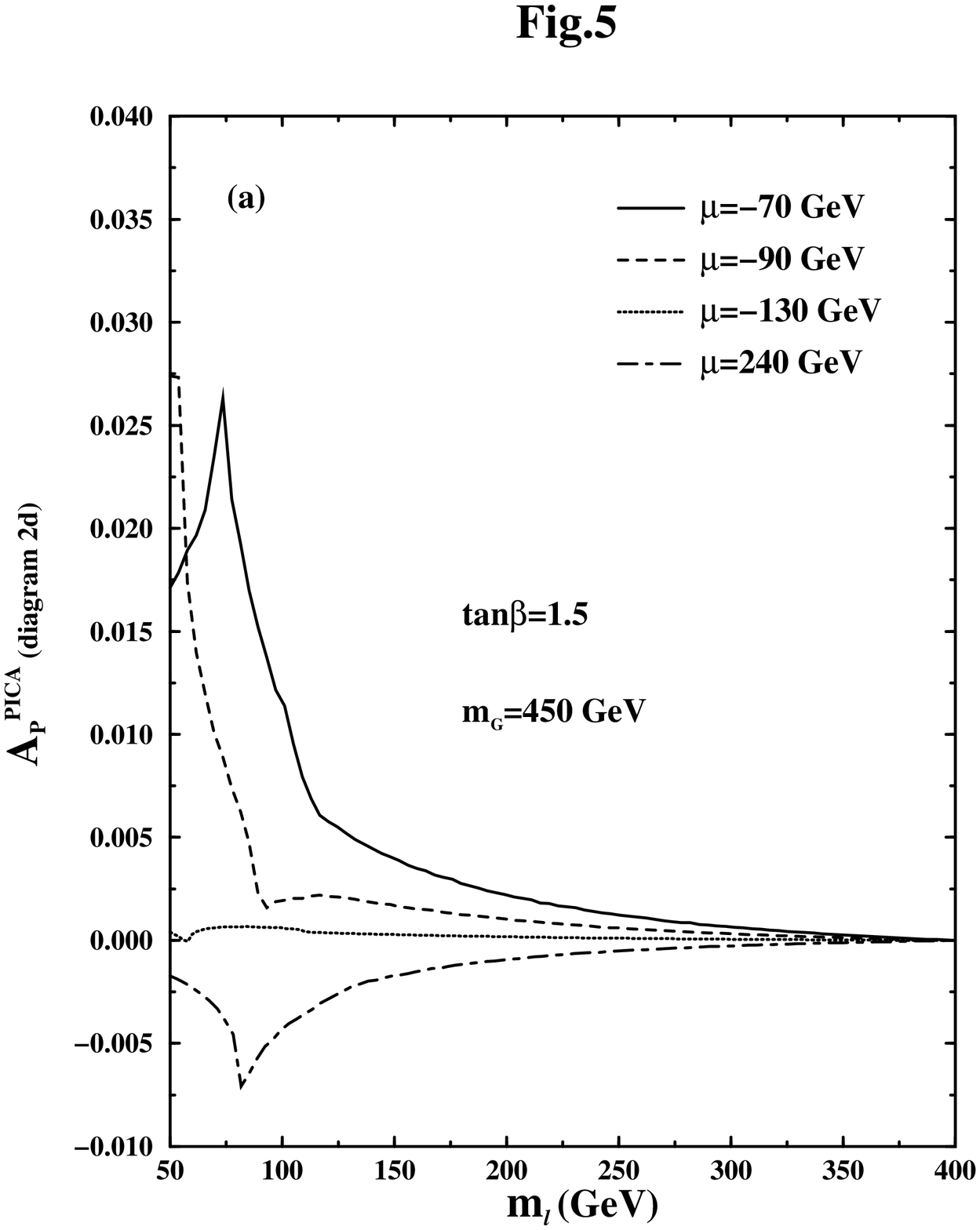}
\epsfysize=350pt
\epsfbox[0 0 612 792]{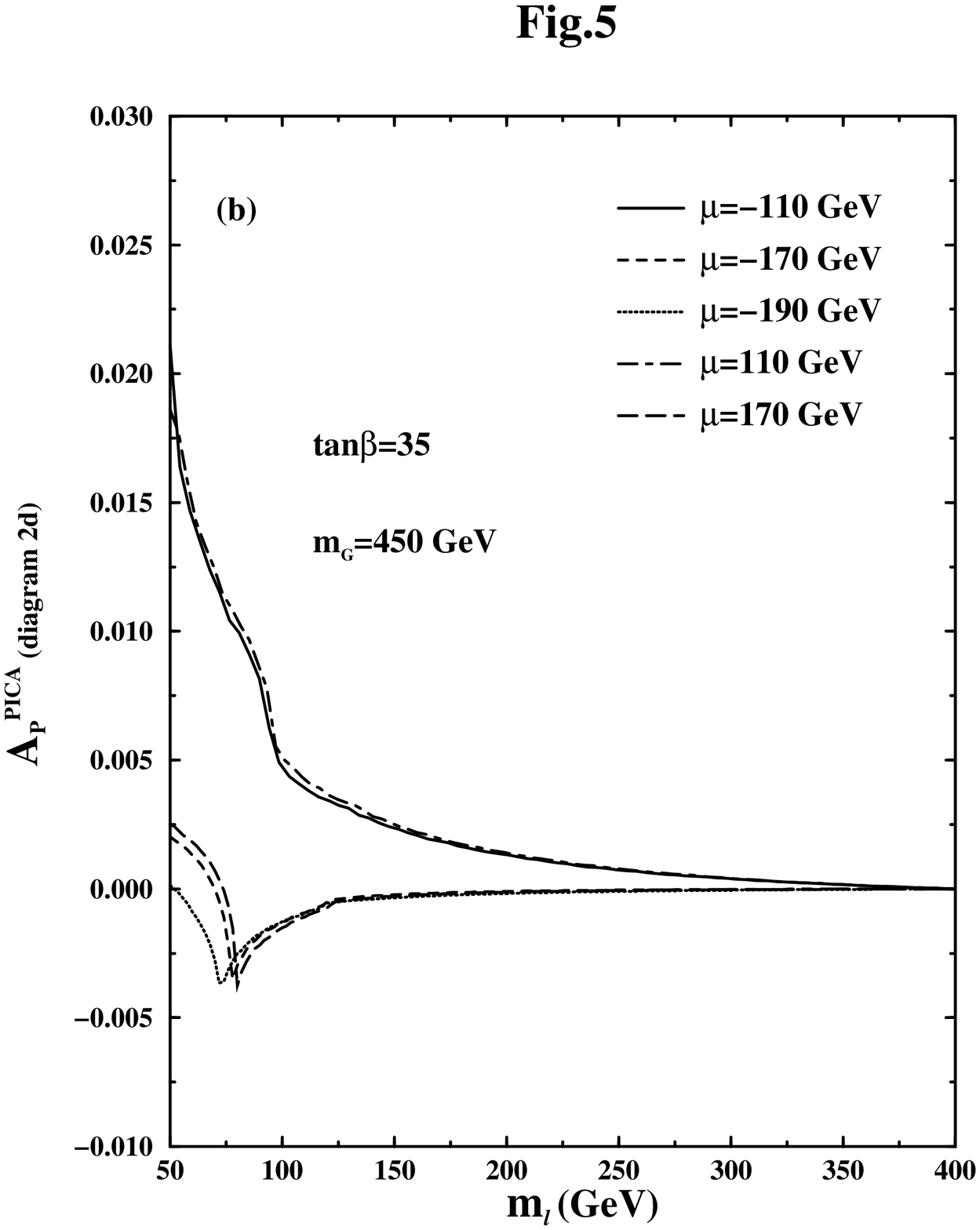}
\end{figure}

\begin{figure}
\centering
\leavevmode
\epsfysize=350pt
\epsfbox[0 0 612 792]{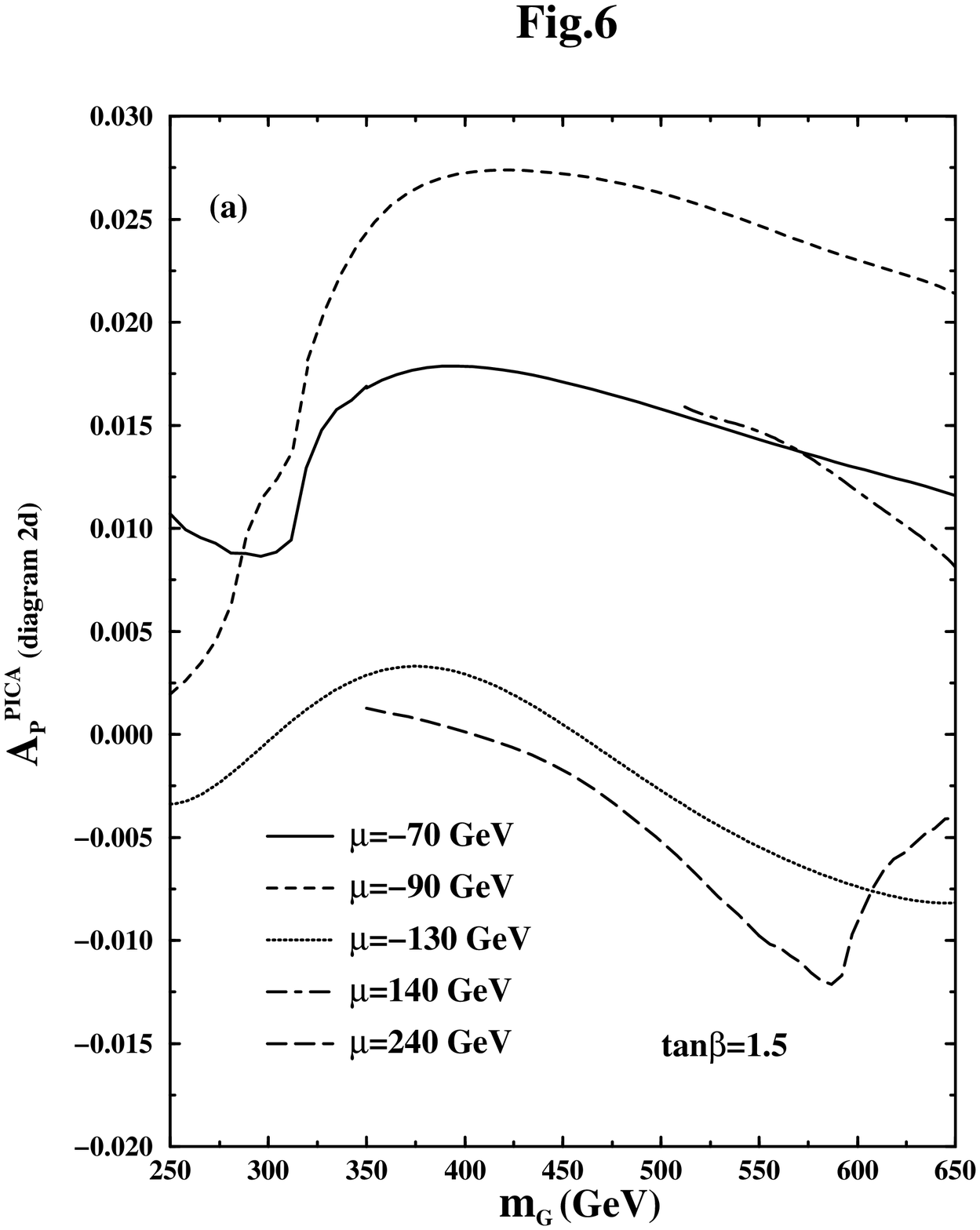}
\epsfysize=350pt
\epsfbox[0 0 612 792]{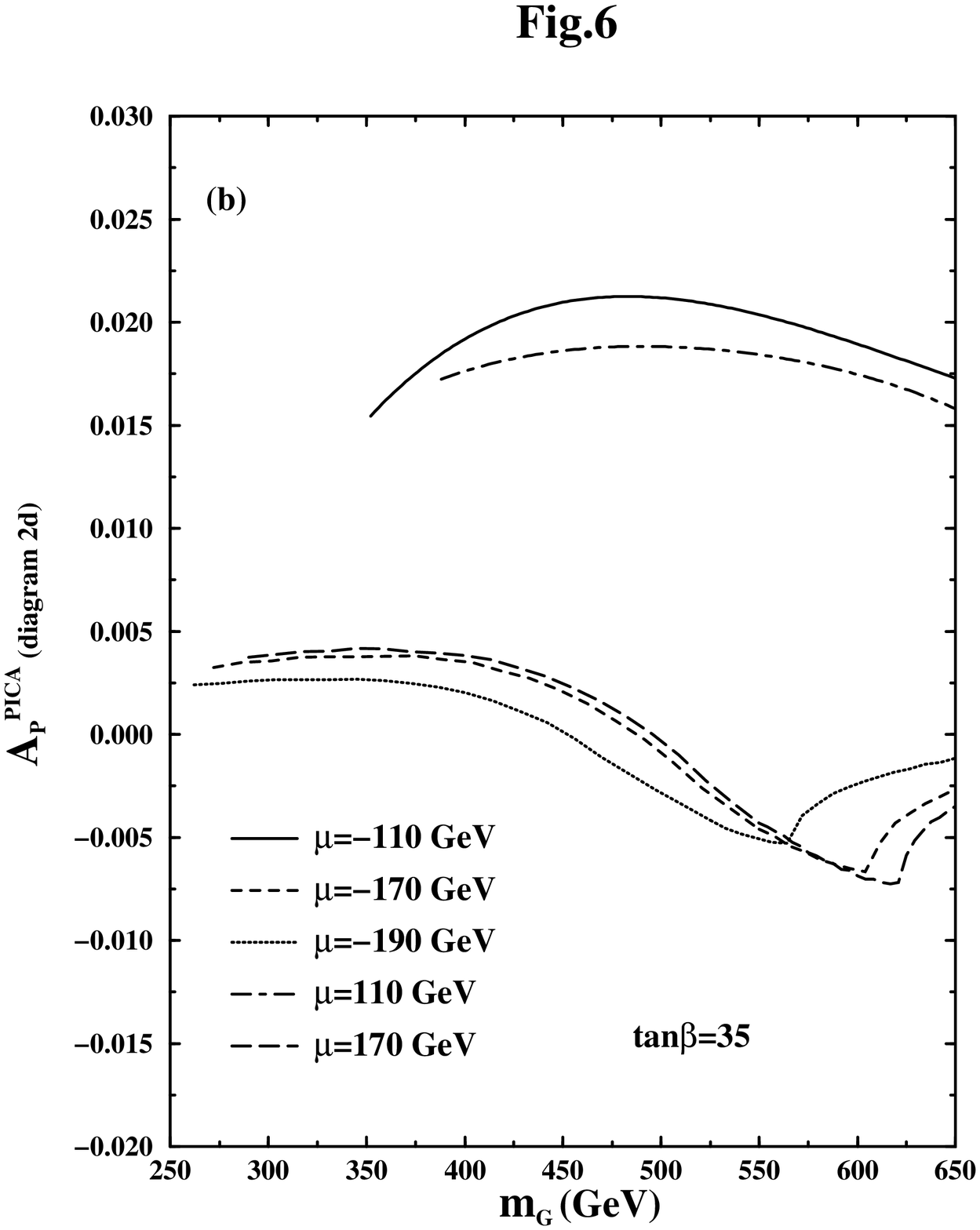}
\end{figure}

\begin{figure}
\centering
\leavevmode
\epsfysize=350pt
\epsfbox[0 0 612 792]{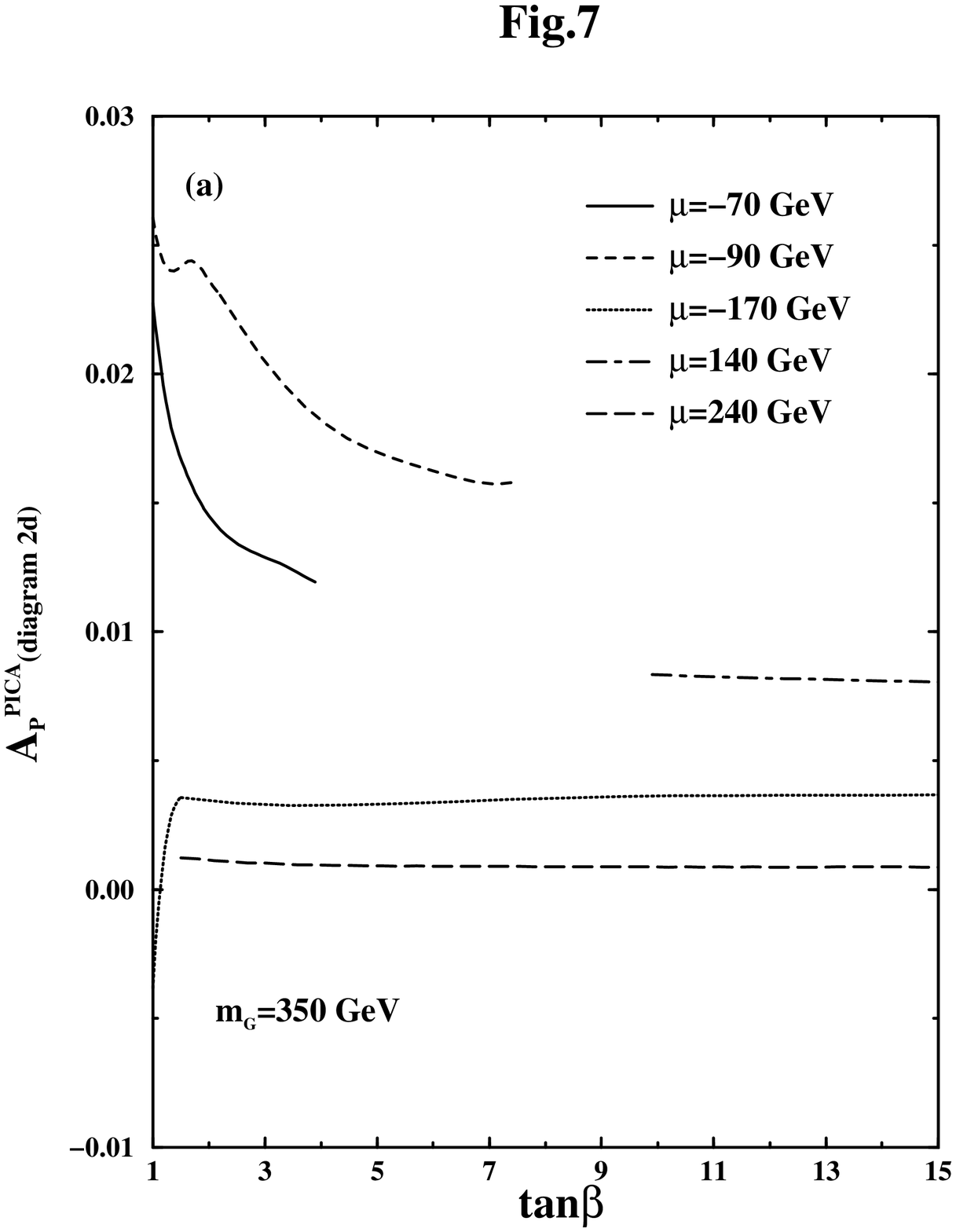}
\epsfysize=350pt
\epsfbox[0 0 612 792]{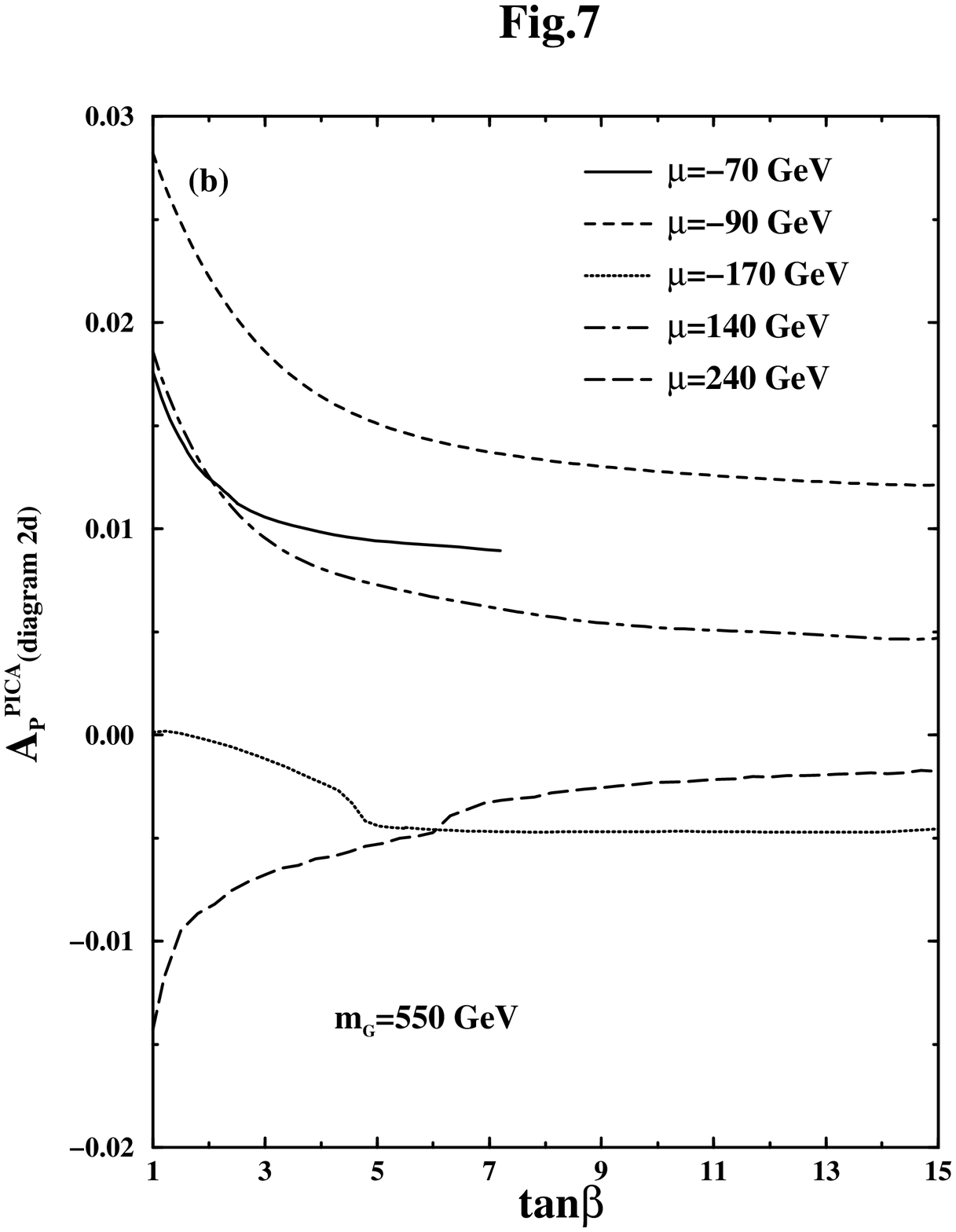}
\end{figure}

\begin{figure}
\centering
\leavevmode
\epsfysize=350pt
\epsfbox[0 0 612 792]{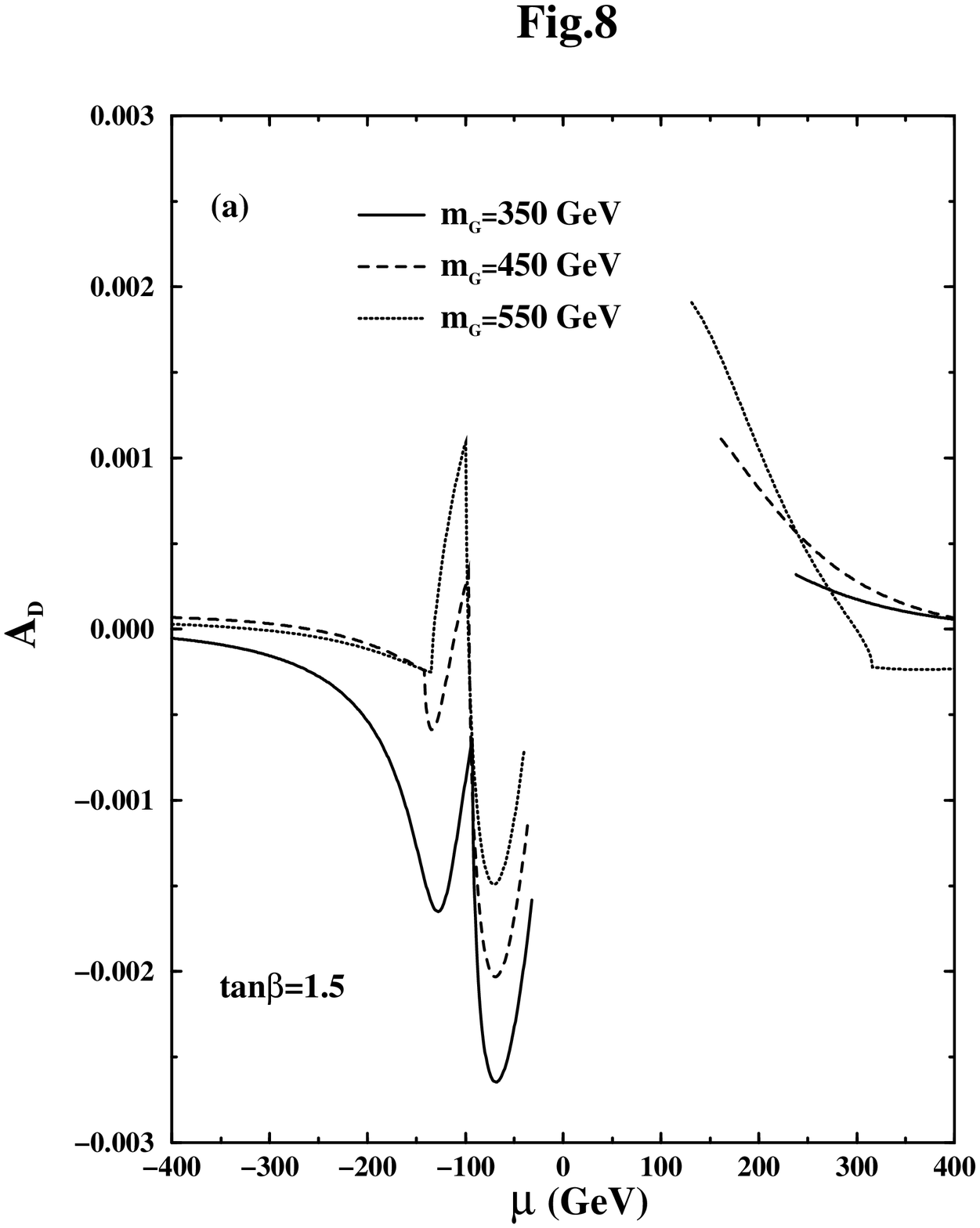}
\epsfysize=350pt
\epsfbox[0 0 612 792]{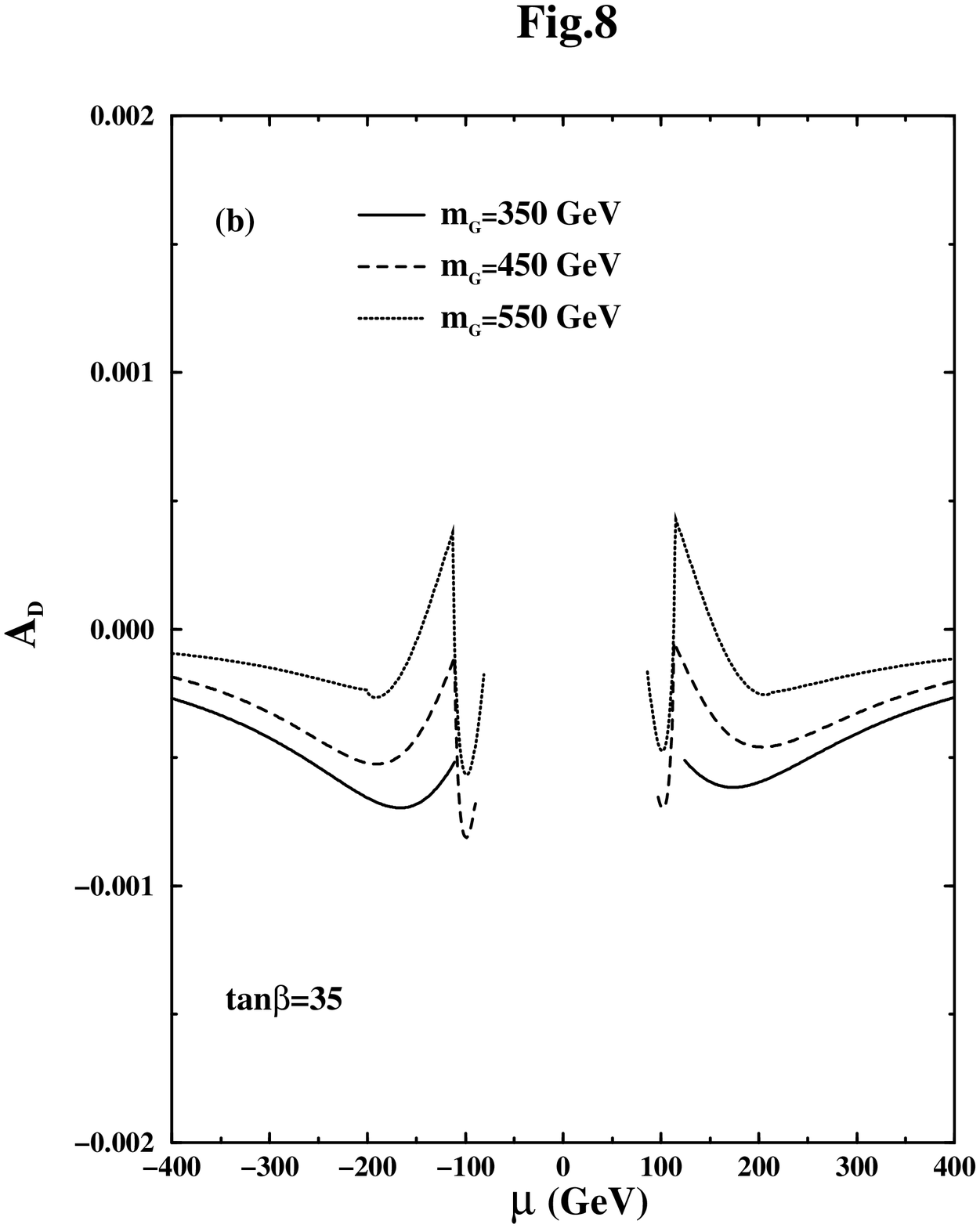}
\end{figure}

\begin{figure}
\centering
\leavevmode
\epsfysize=350pt
\epsfbox[0 0 612 792]{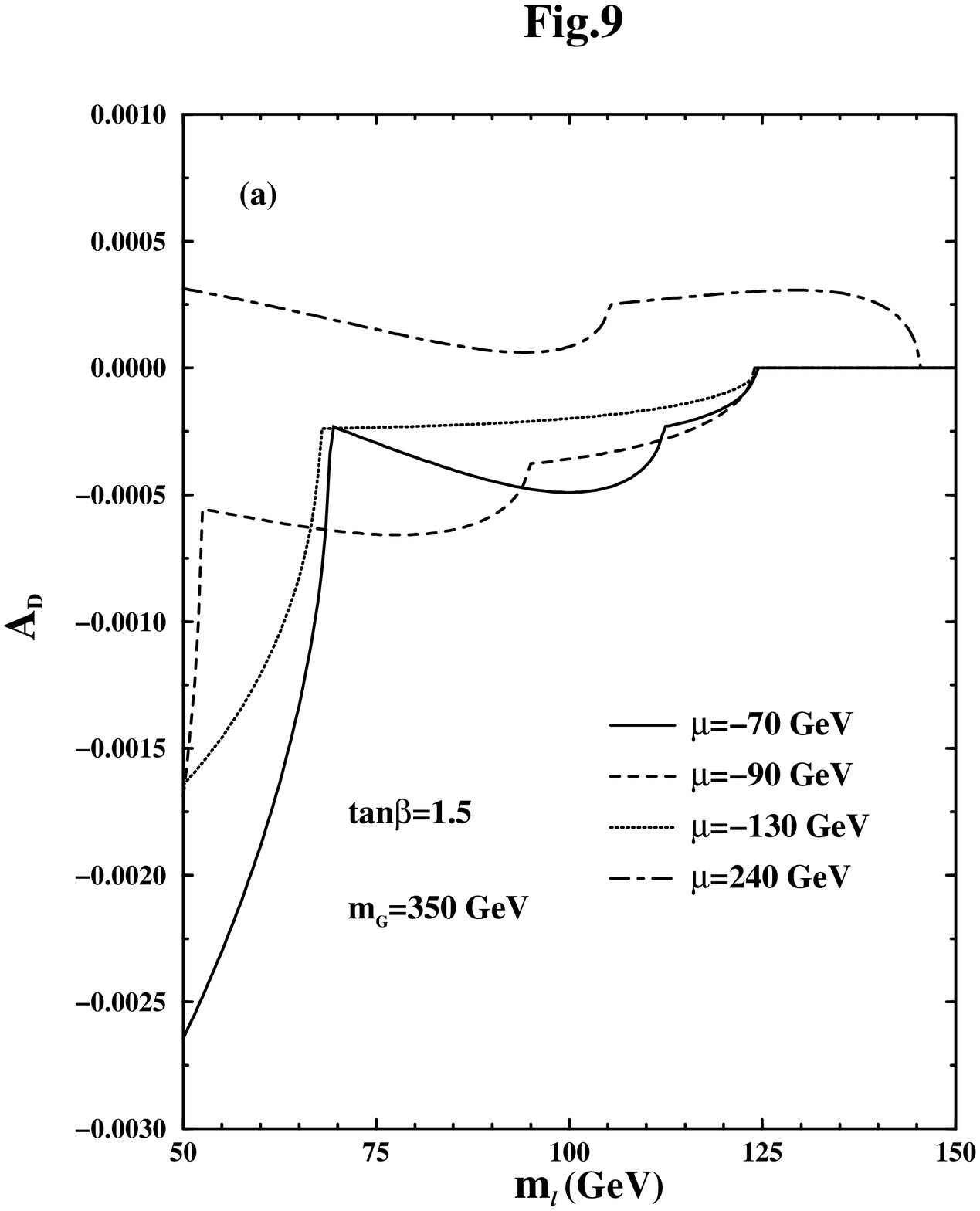}
\epsfysize=350pt
\epsfbox[0 0 612 792]{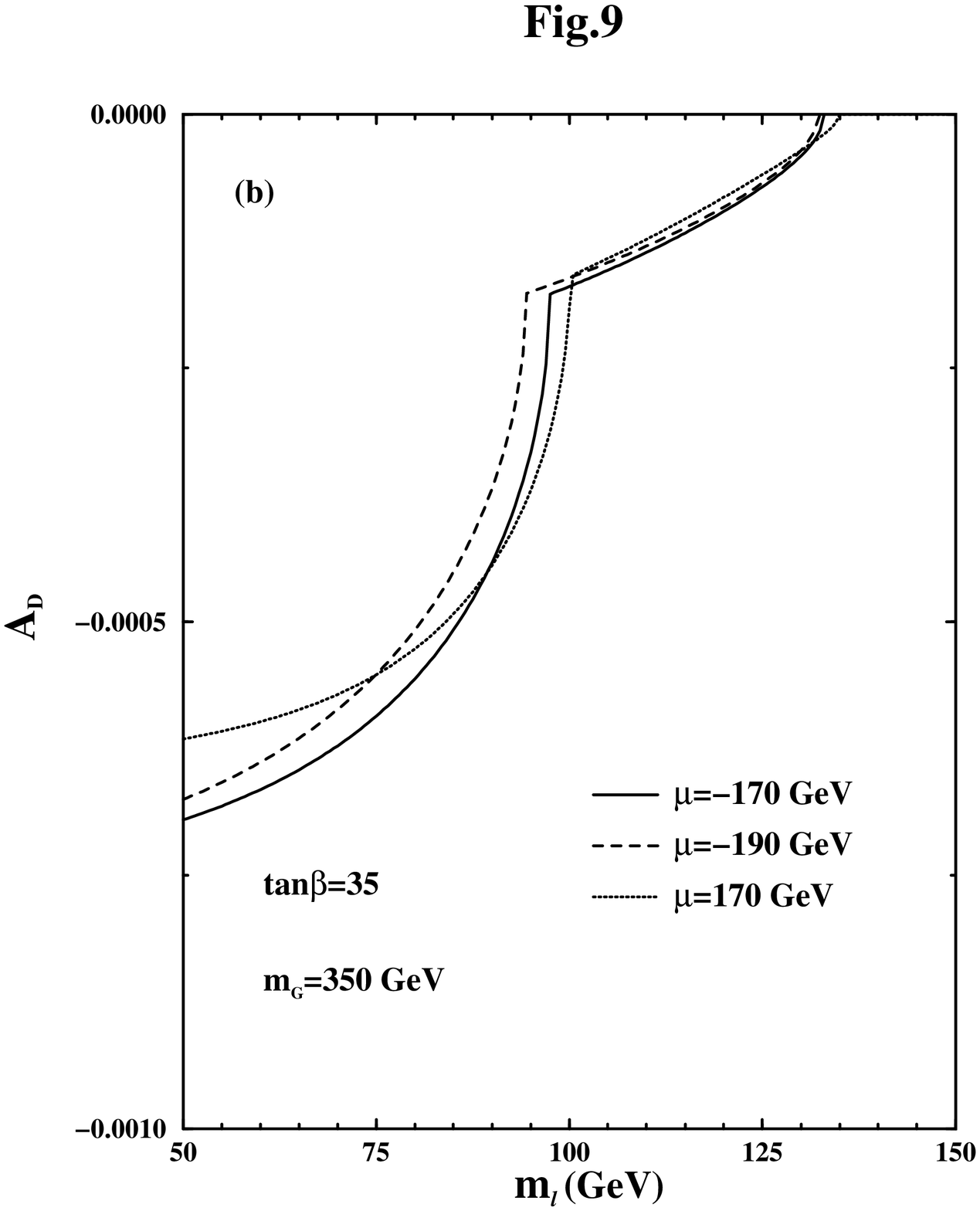}
\end{figure}

\begin{figure}
\centering
\leavevmode
\epsfysize=350pt
\epsfbox[0 0 612 792]{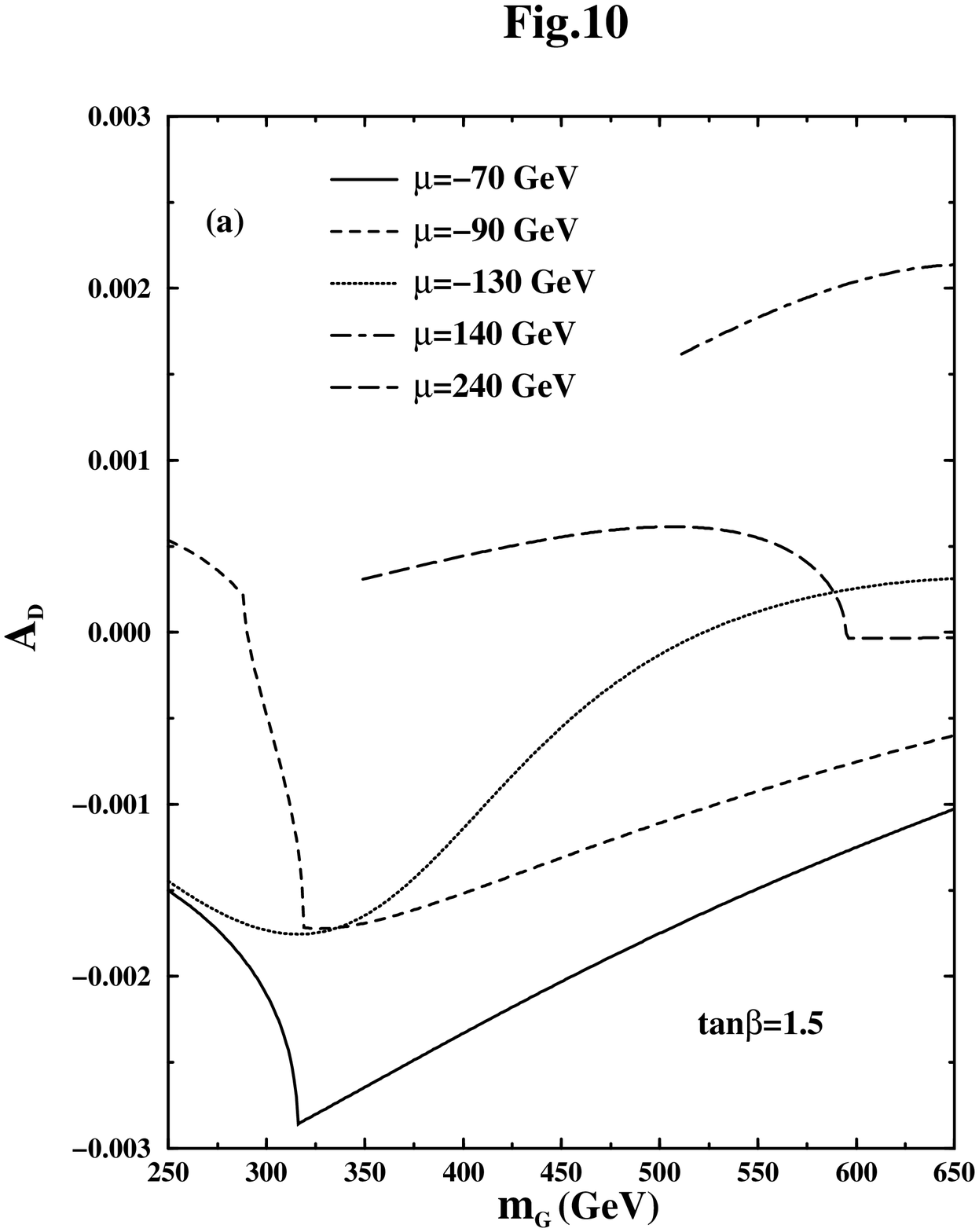}
\epsfysize=350pt
\epsfbox[0 0 612 792]{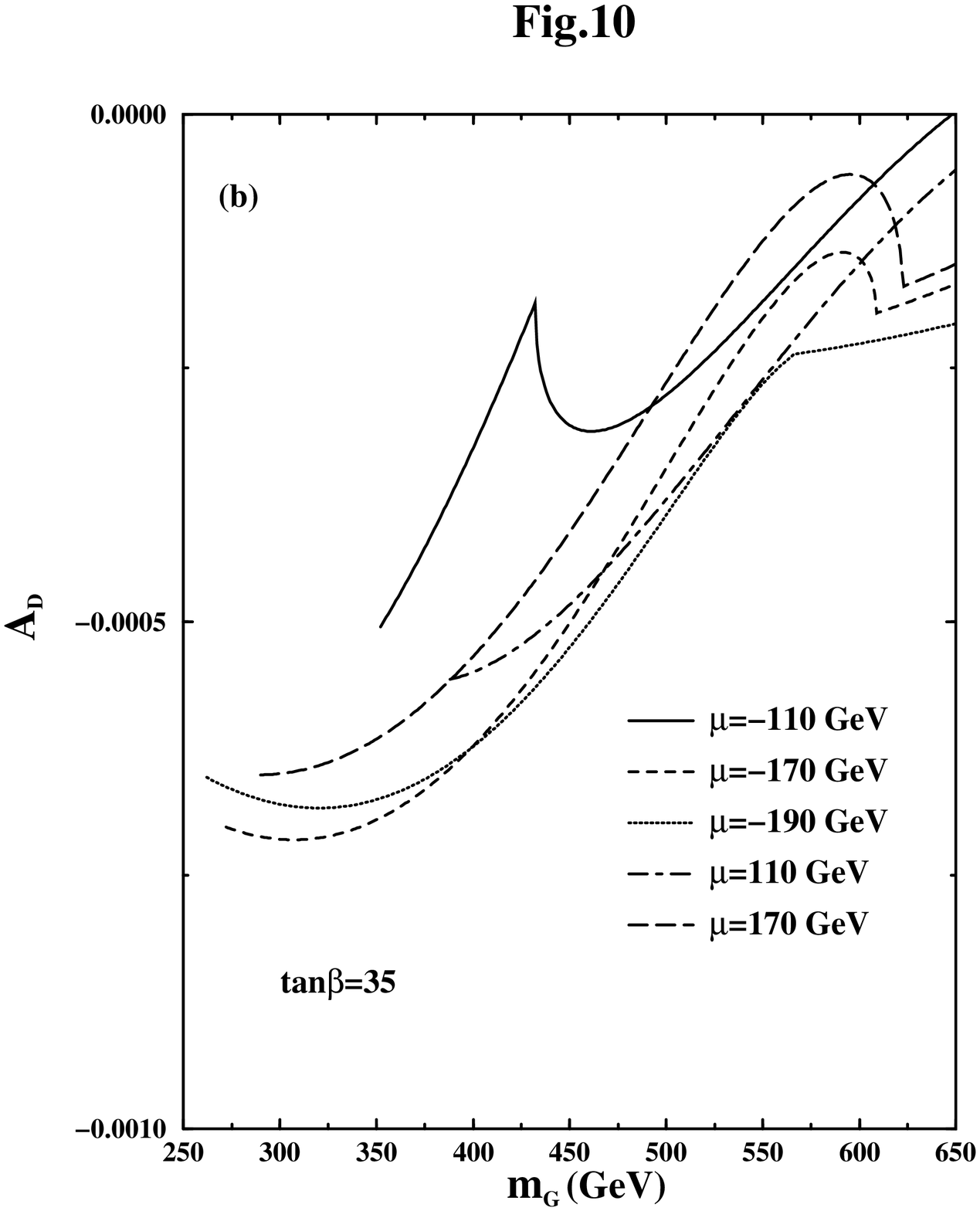}
\end{figure}

\begin{figure}
\centering
\leavevmode
\epsfysize=350pt
\epsfbox[0 0 612 792]{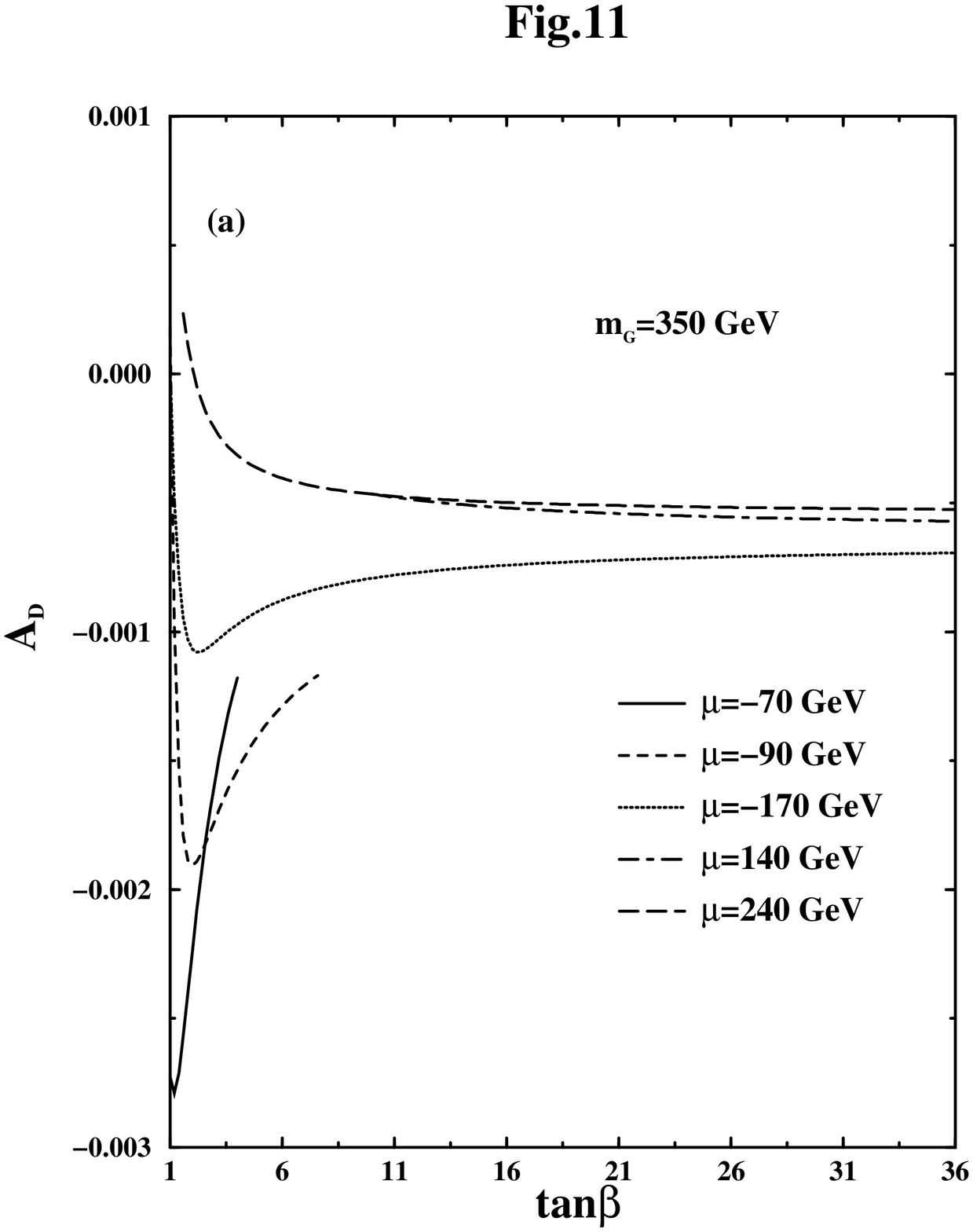}
\epsfysize=350pt
\epsfbox[0 0 612 792]{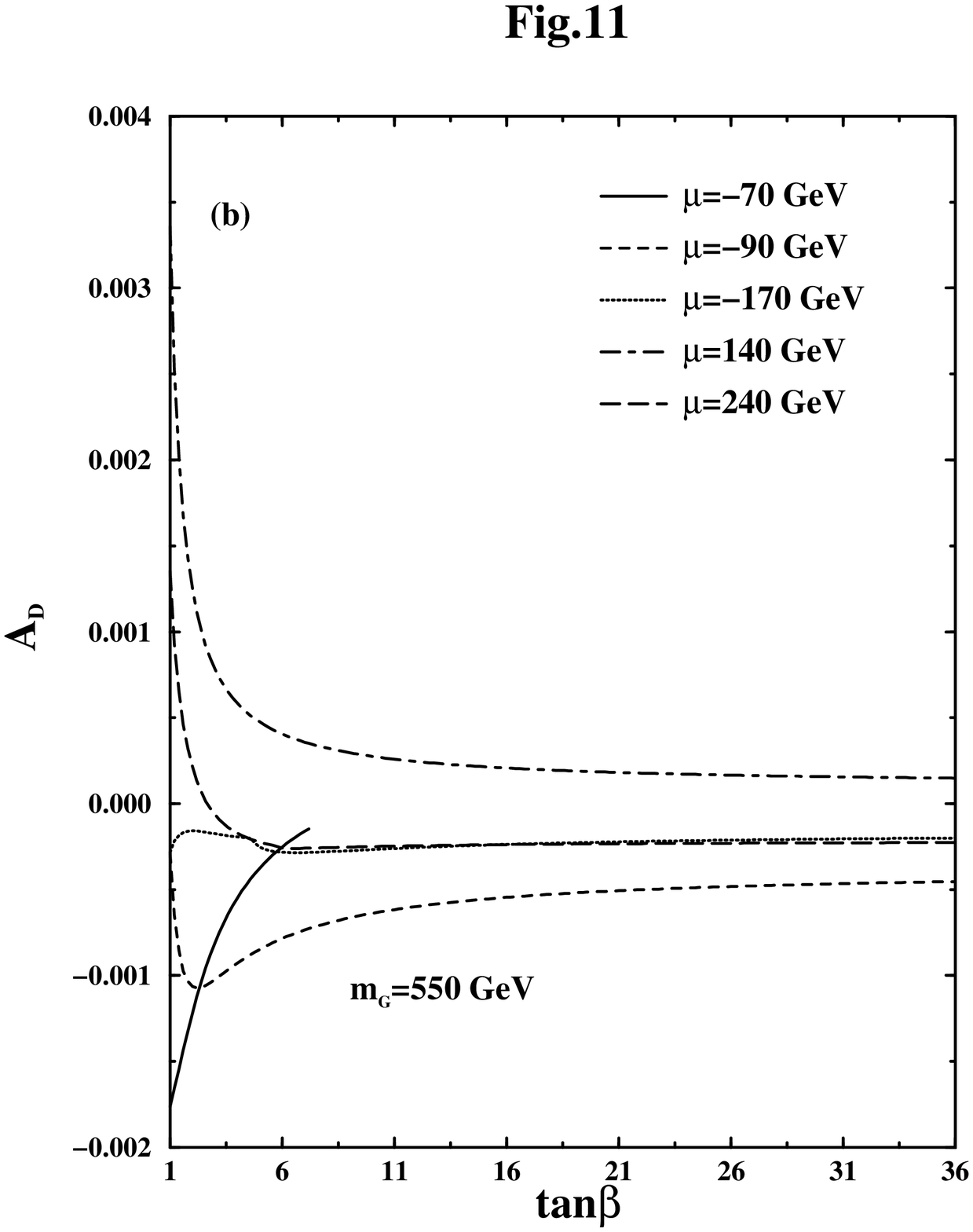}
\end{figure}


\begin{thebibliography}{99}
\bibitem{footresonant} We neglect one-loop CP-violating exchanges of
supersymmetric particles in which the ``resonant'' (on-shell) top quark is
absent, e.g., 
%ASf
between the colliding partons and the decay products of the top. Those class
of diagrams are expected to yield a much smaller CP-asymmetry than the on-shell
top process. Furthermore, in the narrow width approximation the interference
between those should be very small and also they can be experimentally
eliminated in principle by simply demanding that the $W^+ b$ invariant mass
reconstructs the top.
\bibitem{footminus} We want to mention that  the minus sign in front of the
amplitudes of the 
box diagrams. This appears as a consequence of the odd number of interchanges
of the anticommuting fermion spinors needed to go from the  ordering of
the spinors in the tree-level and one-loop triangle amplitudes to that of
the box amplitudes.
\bibitem{footnedm} The function $K(r)$ in Eq.~\ref{kr} is slightly
different from that obtained in Ref.~5. However, we find that numerically
the difference is insignificant and does not change our predictions
below.
\bibitem{footlines} Note that for some values of $\mu$ and $m_G$, some of
the lines in Figs.~4--11 end rather abruptly. This occurs due to the fact
that the corresponding combination of the set of
values $\left\{ \tan\beta,\mu,m_G \right\}$ is forbidden experimentally,
that is, either the LSP is below 20 GeV or the lighter chargino is below
65 GeV\null.
\bibitem{footcuts} We want to comment here on  the sharp peaks in some of
the Figs.~ 4--11. Recall that the asymmetries discussed in this paper are
proportional to absorptive cuts which 
are sensitive to the masses of the SUSY particles in the loops. Thus, whenever
such an absorptive threshold is satisfied (i.e., two SUSY particles in the
loop can go on their mass-shell) these asymmetries will exhibit a drastic
change in their behavior.
\bibitem{footgk} Note that an asymmetry smaller than $0.25\%$ was found by
GK in Ref.~9 for the more likely choice of $m_{\tilde b}=300$
GeV\null.
\bibitem{footmixing} We neglect generation
mixing, thus shrinking $Z_u$ and $Z_d$ from $6\times6$ to $2\times2$ matrices.
\end{thebibliography}
\end{document}